\documentclass[amsmath,amssymb,aps,prd,letterpaper,nofootinbib,notitlepage,showkeys,showpacs,10pt,twocolumn]{revtex4-1} 
\usepackage{graphicx}		

\usepackage{bm}

\usepackage[sort&compress]{natbib}	 		
	\setcitestyle{square,numbers,comma}	
\usepackage[colorlinks=true,urlcolor=blue,linkcolor=blue,citecolor=red]{hyperref}	

\newcommand{\figref}[1]{Fig.\ \ref{fig:#1}}		
\newcommand{\secref}[1]{Sec.\ \ref{sec:#1}}	

\renewcommand{\Re}[1]{\text{Re}\left[ #1 \right]}	
\renewcommand{\Im}[1]{\text{Im}\left[ #1 \right]}		
\newcommand{\eten}[1]{\times 10^{#1}}			
\newcommand{\lb}{\left}						
\newcommand{\rb}{\right}

\newcommand{\ok}{\omega_k}

\newcommand{\nck}{n^{\text{c}}_k}
\newcommand{\np}{n^{\text{p}}}
\newcommand{\bks}{|\beta_k|^2}

\begin{document}
\title{Irruption of massive particle species during inflation}
\author{Michael A.\ Fedderke}
\affiliation{Department of Physics and Kavli Institute for Cosmological Physics, The University of Chicago, 5640 S. Ellis Ave., Chicago, Illinois, 60637, USA}
\email{mfedderke@uchicago.edu}
\author{Edward W.\ Kolb} 
\affiliation{Department of Astronomy and Astrophysics, Enrico Fermi Institute, and Kavli Institute for Cosmological Physics, The University of Chicago, 5640 S. Ellis Ave., Chicago, Illinois, 60637, USA}
\email{Rocky.Kolb@uchicago.edu} 
\author{Mark Wyman}
\email{markwy@gmail.com}
\noaffiliation

\date{\today}

\begin{abstract}
All species of (non-conformally-coupled) particles are produced during inflation so long as their mass $M$ is not too much larger than $H$, the expansion rate during inflation.  It has been shown that if a particle species that is normally massive ($M\gg H$) couples to the inflaton field in such a way that its mass vanishes, or at least becomes small ($M < H$), for a particular value of the inflaton field, then not only are such particles produced, but an irruption of that particle species can occur during inflation.  In this paper we analyze creation of a massive particle species during inflation in a variety of settings, paying particular attention to models which realize such an irruptive production mechanism.
\end{abstract}


\maketitle 

\section{Introduction\label{sec:introduction}}

The epoch of inflation \cite{Guth:1980zm,Linde:1981mu,Albrecht:1982wi} is the highest-energy physical process to which we are ever likely to have observational or experimental access. Recent observational evidence for degree-scale $B$-mode polarization of the cosmic microwave background radiation \cite{Ade:2014xna}\footnote{See, e.g., Refs.\ \cite{Mortonson:2014yq,Flauger:2014uq} for a critical discussion of galactic dust foregrounds in connection with these results.} suggests that inflation occurred at or near the Grand Unified Theory scale with an expansion rate during inflation of $H\sim 10^{14}$ GeV, and that the inflaton field traversed a field space distance much larger than the Planck scale during the inflationary epoch (e.g., Refs.\ \cite{Turner:1993su,Lyth:1996im}). 

For physicists, this situation is fortuitous, because very high-energy-scale inflation provides us with a significant opportunity to uncover new physical laws. This is because as the inflaton traverses a great (super-Planckian) distance in field space, it is possible to uncover new ultraviolet physical effects and probe the couplings of the inflaton to other particles. The inflaton must be coupled to other particle species since, after inflation ends, the inflaton energy density must be converted to entropy by reheating or preheating.  Presumably, this is related to the coupling of the inflaton to ``light'' particle species.\footnote{Unless otherwise specified, ``light'' and ``heavy'' particle species refers to the mass of a species compared to the mass of the inflaton, which is approximately the expansion rate of the Universe during inflation in the inflation models we study.}  However, the inflaton might also be coupled to additional ``heavy'' particle species.  If the inflaton couples to a particle species of mass that is always much larger than the inflaton mass or the expansion rate of the Universe, the heavy field can be integrated out to form an effective field theory describing the inflaton field and its coupling to light degrees of freedom.  In this case, the heavy field will not be dynamically important during inflation.

In this paper we investigate the possibility that the inflaton couples to another particle species in such a way that the mass of the additional species depends on the value of the inflaton field.  Thus, as the inflaton field evolves, the mass of the other particle species will change with the value of the inflaton field.  We will consider several toy models in which the mass of the additional field vanishes (or at least becomes ``light'') for a particular value of the inflaton field.  For that particular value of the inflaton field, and only for that value, the additional field becomes dynamically important.  One consequence of the fact that for a particular value of the inflaton field the new particle is dynamically important, is that while the field is light it can be produced via the usual mechanism of particle creation during inflation.  This may result in a sudden growth, or irruption, in the population of the new particle species, which we dub the ``\emph{irrupton}.''

String theory provides a calculable framework for articulating the general statement that the inflaton might be coupled to heavy fields, so let us describe some characteristic examples from the literature of string cosmology.  One of the most successful early models for inflation within string theory was brane inflation \cite{Dvali:1998pa}, wherein the separation of two higher-dimensional (mem)branes played the role of the inflaton. After extensive study of this framework, it was realized that the string-theoretic context of brane inflation excluded the possibility of super-Planckian field excursions and hence observable gravitational waves \cite{Baumann:2006cd}. Intuitively, this is because the higher-dimensional Calabi-Yau geometries in which brane inflation was thought to operate cannot be enlarged well beyond the string (and, hence, four-dimensional Planck) scales without their constituting elements generating an overwhelming back-reaction.  Since this phenomenon was general (if not universal), the ``model space'' for inflation in string theory was generally thought to involve small volumes and field excursions; see, e.g., \citet{Burgess:2013sla} for a review.

Nonetheless, the breadth of the string landscape (and theorists' capacity for imagination) is also large (super-Planckian, in fact). A significant breakthrough in this direction was the model of axion monodromy \cite{Silverstein:2008sg}, a proof-of-principle calculation that geometries could exist in string theory that permitted super-Planckian field excursions. Although the well-studied models of monodromy are in some conflict with recent gravitational wave discoveries (because their gravitational wave production is \emph{too small}), they offer hope that string-theoretic constructions can agree with the cosmological data we have. Moreover, it suggests that the higher-dimensional geometry necessary to embed high-energy-scale inflation within string theory will have to be something very different from the simple geometries most often studied in the past.

Summing up, we are in a remarkable situation: if string theory is the correct description of quantum gravity, then the observation of gravitational waves tells us that the inflationary epoch can in principle affect, and hence potentially generate, many ultra-UV phenomena previously thought of chiefly as technological arcana of higher-dimensional geometries. Put more simply, if inflation is near the GUT scale and the inflaton field traverses super-Planckian distances, it can shake and rattle fields (e.g., moduli that describe and shape that geometry) as it rolls.   

In this context, then, it becomes of intense interest to understand the full range of observable phenomena that may result from such an inflationary epoch. A relatively less well studied possibility in this research space is the creation of new particles \emph{during} the inflationary period. Within the picture we have outlined, following the seminal work of \citet{Kofman:2004yc}, this can be thought of as the physical manifestation of enhanced symmetry points within the geometrical space of fields that characterize the extra dimensions (see also Refs.\ \cite{Watson:2004aq} and \cite{Green:2009ds}). In such a set-up, the vanishing of effective masses corresponds to a momentary enhancement of symmetry, which we expect on general grounds to be a dynamical attractor.  By the same token, it is worth going beyond the relatively simple effective theory described in \citet{Kofman:2004yc}, where a dynamically varying mass was captured by a simple $g \phi^2 \chi^2$ potential (where $\phi$ is the inflaton and $\chi$ the representative extra modular field), to richer dynamical systems. 

To that end, we extend the study of particle creation beyond the canonical $g \phi^2 \chi^2$ model, which has been a familiar friend since its introduction in the context of preheating after inflation \cite{Kofman:1994rk}. In particular, although we continue to work in the context of a simple two-scalar-field model, we will study in detail two models that encapsulate the phenomenon of a field that is heavy (and thus dynamically unimportant) before, during, and after inflation {\em except} at a particular value of the inflaton field during inflation. 

The two models we investigate in detail have different starting points for coupling a new field to the inflaton. In our first model, the new field is coupled to the inflaton field through a potential term with a simple Yukawa-type coupling of the new field to the inflaton.  We will refer to this as the ``potentially-coupled'' case.  For the second model we study the coupling of the new field to the inflaton traces to the kinetic interactions of the inflaton field and the new field.  We will refer to this as the ``kinetically-coupled'' case.  We do not suggest that the two models we investigate span the space of all possibilities for massive particle species irruption during inflation; indeed, there are many other possibilities one might consider.  We concentrate on the two models we have chosen in order to understand the issues that we anticipate will be generic to any model of massive species irruption.

While previous studies have considered potentially-coupled models \cite{Chung:1999ve,Elgaroy:2003hp,Romano:2008rr,Barnaby:2009mc,Barnaby:2010sq}, this work will also focus on kinetic interactions between the two fields.  Such interactions are characteristic of the supergravity limit of string theory (see, e.g., \citet{GrootNibbelink:2001qt} and references therein). Complete formal perturbative analysis of such systems has been done \cite{Lalak:2007vi,Cremonini:2010ua,Cremonini:2010sv}, but the resulting system is complex and can describe a surprising range of nontrivial phenomena, including reduced speed-of-sound effective dynamics \cite{Tolley:2009fg}, step-like features in the effective potential for the inflaton \cite{Achucarro:2012fd}, and temporarily non-adiabatic evolution of the inflaton itself \cite{Shiu:2011qw,Baumann:2011su,Avgoustidis:2012yc}.

In the original model for super-heavy dark matter production, the simple fact of the highly energetic (yet still adiabatic) inflationary background was exploited to generate a tiny number density of ultra-heavy particles that could play the role of dark matter. The mechanism we will describe is a generalization of that approach, where the existence of non-trivial multi-field dynamics will allow the prospective dark matter particle to become effectively light (or even temporarily tachyonic) during inflation, thus allowing it to be produced. Once created, the particle's mass then varies strongly with the value of the inflaton, turning the former into an end-of-inflation very massive particle, with a mass greater than $10^{13}$ GeV.

In the next section we discuss the criteria for species irruption.  We then review a model for creation of massive particles during creation under the condition that the particle mass is unaffected by the value of the inflaton field. While there is no species irruption in this model it serves as a useful baseline in understanding irruption in models where the species mass does depend on the inflaton field.  We then discuss the adiabatic conditions that must be violated for particle creation to occur.  Also in \secref{irrup} we describe the potentially-coupled and kinetically-coupled models.  Finally, in this section we discuss the expressions for the number density of the particles in terms of the Bogoliubov coefficient and the issue of initial conditions.

In \secref{abundance} we review the relationship between the Bogoliubov coefficient and the present number density assuming the produced particle is stable.  Section \ref{sec:numerics} discusses the numerical methods we employ as well as the limitations of our numerical study.  In \secref{results} we present the numerical results in the three models we consider.  We  comment on possible implications of irruption of massive particle species and conclude in \secref{applications}.  Appendices A and B contain longer technical derivations of some results which we will have occasion to refer to multiple times in the text.

\section{Irruption of Particle Species During Inflation \label{sec:irrup}}

In this section we discuss the creation of particles during inflation. When we refer to a  ``massive'' particle, we mean a particle species with a mass \emph{at the end of inflation} larger than the expansion rate of the Universe \emph{during inflation}.  

The idea of creation of particles during any phase of the expansion of the Universe traces back to the (largely forgotten) 1939 paper of Schr\"{o}dinger, \emph{The proper vibrations of the expanding universe} \cite{alarming}.  Here, we briefly summarize what we have learned in the 75 years since that paper about cosmological particle creation:
\begin{enumerate}
\item For a particle species to be created during the expansion of the Universe it must participate in the breaking of conformal invariance.  This is usually accomplished by a mass for the field and/or a non-conformal coupling of a scalar field to the Ricci scalar.
\item Particle (in this case, the inflaton) creation during inflation is the origin of the temperature and density perturbations seen as temperature anisotropies in the background radiation.
\item Particle (in this case, the graviton) creation during inflation is the origin of the gravitational waves (tensor modes) deduced from the background radiation polarization pattern. 
\item In the inflationary phase the expansion rate of the Universe $H$ is nearly constant, and with the assumption of adiabatic initial conditions for each quantum mode, creation of particles of mass $m$ larger than $H$ is suppressed by a factor of $\exp(-m/H)$.
\item Gravitationally created particles of mass comparable to (or slightly larger than) $H$, if stable, would be a candidate for dark matter \cite{Chung:1998ua,Chung:1998zb,Kuzmin:1998kk}.
\item Particles of mass larger than $H$ can only be created during inflation if one is willing to accept some sort of trans-Planckian particle creation \cite{Kolb:2007vd}.
\end{enumerate}

The above considerations assume that the mass, the couplings, and the kinetic term of the field are constant.  The situation changes if the particle couples to the inflaton, which evolves during inflation.  

In the first proposal studying creation of massive particles during inflation, \citet{Chung:1999ve} assumed the existence of a fermion field $\psi$ that has a Yukawa coupling to the inflaton $\phi$ of the form ${\cal L}_Y = \lambda \phi \bar{\psi}\psi$ and a Lagrangian mass term of the form ${\cal L}_M= -M_0\bar{\psi}\psi$.  For nonzero values of $\phi$, the mass of the $\psi$ would be $M(\phi)= M_0 - \lambda\phi$, where $M_0$ is the mass at $\phi=0$.  This leads to a critical value of the inflaton field, $\phi_*=M_0/\lambda$, where the mass vanishes.  Even if $M_0\gg H$, there will be a resonant production of $\psi$ during inflation when the inflaton field is around $\phi_*$.  Note that in large-field models of inflation $\phi$ is of order $M_{\text{Pl}}$, so one can have resonant production even if $M_0\gg H$.  Several papers extended this work to the analogous situation for a scalar field $\chi$ coupled to the inflaton via a Lagrangian effective mass term of the form  ${\cal L}^{\text{eff}}_M = -\tfrac{1}{2}g^2(\phi-\phi_*)^2\chi^2$ \cite{Elgaroy:2003hp,Romano:2008rr,Barnaby:2009mc}. Again, there is a critical value of the inflaton field $\phi=\phi_*$ where the mass vanishes and particle creation can occur.

\subsection{A Simple Non-Irruptive Model \label{sec:simplemodel}}

Before turning to the complicated cases of $\phi$-dependent mass terms, let us consider the simple case of production of particles of fixed mass $M$ in the expanding Universe; such a model has been used in, e.g., the context of inflationary production of superheavy dark matter (e.g., Refs.\ \cite{Chung:1998zb,Kuzmin:1998kk}).  Start with a particle of mass $M$ with action 
\begin{equation}
S = \int d^4x\sqrt{-g}\left[\frac{1}{2} g^{\mu\nu} \partial_\mu\chi\partial_\nu\chi - \frac{1}{2}M^2\chi^2  - \frac{1}{2}\xi \mathcal{R}\chi^2 \right] .
\label{eq:LagM}
\end{equation}
Here $\xi$ is a constant in the coupling term of the scalar field to the Ricci scalar $\mathcal{R}$. In this paper we will assume a flat Friedmann-Robertson-Walker (FRW) spacetime with mostly minus signature: $ds^2 = dt^2 - a^2(t)d\textbf{x}^2 = a^2(\eta)(d\eta^2 - d\textbf{x}^2)$, where $\eta$ is conformal time.  In the flat FRW background $\mathcal{R}=6a''/a^3$, where $'$ denotes $d/d\eta$ (dot will denote $d/dt$).  The equation of motion for $\chi$ is
\begin{equation}
\ddot{\chi}+3H\dot{\chi}+\frac{\eta^{ij}\partial_i\partial_j\chi}{a^2(t)} + M^2\chi + \xi\mathcal{R}\chi= 0 .
\end{equation}
Now we mode expand $\chi = \sum_k \hat{a}_k u_k + \hat{a}_k^\dagger u_k^*$ and make the plane wave ansatz
\begin{equation}
u_k(\textbf{x},t) = \frac{e^{i\textbf{k}\cdot\textbf{x}}}{(2\pi)^{3/2}a(t)}\chi_k(t) \label{eq:plane_wave}
\end{equation}
where $\bf{k}$ is the comoving momentum.\footnote{The choice of normalization is such that $\chi_k \cdot (\chi_{k}^*)' - {\chi_k}'\cdot \chi_{k}^* = i$ gives modes normalized with respect to the usual inner product $(u_k,u_{k'}) = -i\int_\Sigma d\Sigma^\alpha \lb[-g_{\Sigma} \rb]^{1/2} u_k\! \stackrel{\leftrightarrow}{\partial}_{\!\alpha}\! u_{k'}^* = \delta^{(3)}(\bm{k}-\bm{k'})$, etc. We take the usual creation/annihilation algebra $[\hat{a}_k,\hat{a}_{k'}^\dagger]=\delta^{(3)}(\bm{k}-\bm{k'})$, etc. (see, e.g., pp.\ 44--45 of Ref.\  \cite{BirrellDavies1982}).\label{footnote:norm}} This gives the mode equation $\chi_k$ (here and below $k\equiv|\bm{k}|$, not the four-momentum)
\begin{align}
\ddot{\chi}_k + H \dot{\chi}_k + \left[ \frac{k^2}{a^2} - \frac{a''}{a^3}\left(1-6\xi\right) + M^2 \right] \chi_k &= 0 ,
\end{align}
where we used $a^{-1} a'' =a\ddot{a} + \dot{a}^2$. Passing to conformal time using $a^{-2} \chi_k'' = \ddot{\chi}_k + H \dot{\chi}_k$, the mode equation takes the form 
\begin{equation}
\chi_k'' + \omega_k^2(\eta) \chi_k = 0,
\label{modesimple}
\end{equation}
as is indeed expected for a sensible mode expansion.  In Eq.\ (\ref{modesimple}), $\omega_k^2(\eta)$ is given by
\begin{align}
\omega_k^2(\eta)& = k^2 - \frac{a''}{a}(1-6\xi) + a^2 M^2  \nonumber \\
&= a^2H^2\left[\left(\frac{k}{aH}\right)^2+H^{-2}\left[M^2-(1-6\xi)a^{-3}a''\right]\right].
\label{eq:deSitter_omega}
\end{align}
Note that if $a^{-3}a''(1-6\xi)=M^2$ is a non-zero constant, the total coupling of the field is conformal, although the mass term and the Ricci scalar term individually break conformal symmetry.  Since we will eventually encounter sufficient complexity, we will assume henceforth that $\xi=0$, and make the choice that $\chi$ is a ``minimally-coupled'' scalar field. 

In this simple model we can see the underlying cause of particle creation.  For a static case, $a$ is constant (so $a''$ vanishes) and $M$ is constant as well, so $\omega_k^2$ is constant in conformal time.  If we choose at some initial time a pure outgoing wave (positive frequency mode),
\begin{equation}
\chi_k(\eta) = \frac{1}{\sqrt{2\omega_k}}e^{-i\omega_k \eta},
\end{equation}
then it will remain a solution without admixture onto incoming waves (negative frequency modes).  If $\omega_k(\eta)$ is not constant (in our simple case due to the $a''/a$ term \emph{and} the mass term), the previous statement need not be true.  One might try an adiabatic approximate solution of the form
\begin{equation}
\chi_k(\eta) = \frac{1}{\sqrt{2\omega_k(\eta)}}e^{-i \int\omega_k(\eta) d\eta}.
\end{equation}
This (zeroth-order) adiabatic solution (see, e.g., Ref.\ \cite{BirrellDavies1982} for a discussion of the adiabatic approximation in this context) is constructed to satisfy the equation of motion (EOM) Eq.\ \eqref{modesimple} up to terms of ${\cal O}\lb( | \omega_k' / \omega_k^2 |^2 , | \omega_k''/\omega_k^3| \rb)$, and so the conditions for the adiabatic solution to be a good approximation are that 
\begin{align}
{\cal A} &\equiv \left|\frac{\omega_k'}{\omega_k^2}\right|^2 \ll 1 &&\text{and}&
{\cal B} &\equiv \left|\frac{\omega_k''}{\omega_k^3}\right| \ll 1.
\label{nonadiabatic}
\end{align}
If either condition fails, the adiabatic solution is not a good approximation. When the solution is non-adiabatic the incoming and outgoing modes (Schr\"odinger's ``proper vibrations'') are mixed: particles are created.\footnote{Here we are glossing over the fact that one can only really speak of particles if the evolution is adiabatic.}  Notice that in the inflationary Universe $a''$ is \emph{positive}.  This means that it is possible to have a ``tachyonic'' mass for $\chi$: i.e., $\omega_k^2<0$.

Using the simple scalar-field model above, let us see where the evolution is non-adiabatic.  The problem is very simple if we consider the evolution of the scalar field in a de Sitter background.  In de Sitter space, $H$ is constant and conformal time and the scale factor are related by $a=-1/\eta H$ where $\eta$ is in the range $-\infty\leq\eta\leq0$.  With the definitions $x\equiv k\eta$ ($0<x^2<\infty$) and $\gamma=M^2/H^2-2$, where $-2<\gamma<\infty$,
\begin{align}
\left|\frac{\omega_k'}{\omega_k^2}\right|^2 &= \left|\frac{\gamma^2}{(x^2+\gamma)^3} \right|, \\
\left|\frac{\omega_k''}{\omega_k^3}\right| &=\left| \frac{\gamma^2}{(x^2+\gamma)^3} +\frac{3\gamma}{(x^2+\gamma)^2}\right|  .
\end{align}   
Note that $\gamma=0$ is a special point where there would not be particle creation.  This happens if $M^2/H^2=2$. As mentioned above, this is where $(a''/a^3)(1-6\xi)=M^2$ (for $\xi = 0$, as we consider here).  The condition for non-adiabatic particle creation, violation of Eq.\ (\ref{nonadiabatic}), is satisfied when $x^2\simeq-\gamma$, or $k^2\eta^2 = k^2/a^2H^2\simeq-\gamma=2-M^2/H^2$.  For $M^2/H^2\ll2$ the evolution is non-adiabatic at $k/aH \sim 1$, i.e., when a momentum mode crosses the Hubble radius.  But for $M^2/H^2\gg1$, the evolution is always very nearly adiabatic and particle creation is suppressed.  The two lessons we have learned from this model are
\begin{enumerate}
\item for $M\gg H$, one should not expect significant production of particles
\item for $M\ll H$, particle creation is continuous, since for any value of conformal time (or, equivalently the scale factor) there is a comoving momentum mode $k$ that satisfies $k^2\eta^2=2$.   
\end{enumerate}

This simple model assumes that the mass of the $\chi$ field remains constant.  However, if the mass varied during inflation, one might imagine that today the mass of the species is much larger than $H$ during inflation, but for some period during inflation the mass vanished (or at least became much less than the value of $H$) due to the species coupling to the inflaton.  If this occurs, there can be an irruption of the particle species, but only while the mass is less than ${\cal O}(H)$.  This is exactly what occurs in the models discussed in the next two subsections.

We remark that for the remainder of this paper, we employ a simple ``chaotic'' inflation model rather than this fixed de Sitter background.  Production of massive states during inflation has also been considered for hybrid inflation models as well as natural inflation models \cite{Chung:2001cb} and the results seem to imply that the phenomenon of massive particle production is generic.

\subsection{Massive Particle Species Irruption from a Potential Term \label{sec:potentialirruption}}

For the first irruption model consider the following simple two-field model:
\begin{align}
S &= \int d^4x\sqrt{-g}\bigg[ \frac{1}{2} g^{\mu\nu} \partial_\mu\phi\partial_\nu\phi - V(\phi) \nonumber \\ 
& \qquad\qquad\qquad \quad + \frac{1}{2} g^{\mu\nu}\partial_\mu\chi\partial_\nu\chi - U(\chi,\phi) \bigg]
\label{eq:LagP}
\end{align}
The $\phi$ field will serve as the inflaton. As the purpose of this paper is to study the the sudden increase in the population of the particle species (the irruption of the particle species) $\chi$, rather than referring to $\chi$ as ``the second scalar field'' we will refer to it as ``the \emph{irrupton}.''

We will assume the simplest inflaton potential and a simple inflaton--irrupton interaction term that will serve our purpose (we refer to this as the ``potentially-coupled'' irrupton):
\begin{align}
V(\phi) & = \frac{1}{2}m^2\phi^2,  \\
U(\chi,\phi) & = \frac{1}{2}g^2\left(\phi -\phi_* \right)^2 \chi^2   .
\end{align}  

During inflation we will assume $\phi>0$, $\dot{\phi}<0$, and $\phi_*>0$.  The recent BICEP2 determination of tensor modes in the background \cite{Ade:2014xna} suggests that the inflaton mass and the expansion rate during inflation are of order $10^{14}$ GeV, and that inflaton field excursions during inflation are super-Planckian (although, see Refs.\ \cite{Flauger:2014uq,Mortonson:2014yq} for critical analysis of galactic dust foregrounds in connection with this result).  The effective mass of the irrupton is $M(\phi)= g\left|\phi-\phi_*\right|$; today, we have $\phi=0$ and $M(0)=g\phi_*$.  We wish to choose $\phi_*$ so that $U(\chi,\phi)$ vanishes during inflation: since the inflaton field excursion in this model is super-Planckian, $\phi_*$ may be chosen to be in excess of $M_{\text{Pl}}$, and the mass of the irrupton today may be of the scale of the Planck mass. We will discuss possible implications of this observation in the final section of the paper.

The equation of motion for the (spatially homogeneous) inflaton field is
\begin{equation}
\ddot{\phi} + 3H\dot{\phi} + m^2\phi =-g^2\left(\phi -\phi_* \right) \chi^2 \approx 0,
\end{equation}
where the approximate equality reflects the fact that we will ignore the back-reaction on the classical inflaton field (and hence on the metric) induced by the irrupton.\footnote{One can, and we will, justify this \emph{a posteriori} for the present purposes by showing that the energy density extracted by irrupton production is a negligible fraction of the inflaton energy density. See, however, Refs.\ \cite{Barnaby:2010sq,Chung:1999ve} for a discussion of observational effects in the cosmic microwave background which can arise in this model, or its fermionic cognate, when the back-reaction is not ignored.}  For the chosen potential, inflation ends at $\phi \simeq 0.2 M_{\text{Pl}}$, and 50 $e$-folds before the end of inflation corresponds to $\phi \simeq 2.8 M_{\text{Pl}}$. 

The equation of motion for the irrupton is 
\begin{equation}
\ddot{\chi}+3H\dot{\chi}+\frac{\eta^{ij}\partial_i\partial_j\chi}{a^2(t)} 
+ g^2 \left(\phi -\phi_* \right)^2 \chi =0 .
\end{equation}
Employing the same mode expansion as before and again making the plane wave ansatz as in Eq.\ \eqref{eq:plane_wave}, the mode equation for $\chi_k$ is 
\begin{align}
\ddot{\chi}_k + H \dot{\chi}_k + \left[ \frac{k^2}{a^2} - \frac{a''}{a^3} + g^2(\phi-\phi_*)^2 \right] \chi_k = 0 .
\label{eq:timemode}
\end{align}
Passing to conformal time the mode equation again takes the form 
\begin{equation}
\chi_k'' + \omega_k^2(\eta) \chi_k = 0,
\label{modeold}
\end{equation}
but now $\omega_k^2(\eta)$ is given by
\begin{align}
\omega_k^2(\eta) &= k^2 - \frac{a''}{a} + a^2 M_\mathrm{eff}^2 \nonumber \\
& = a^2H^2 \left[ \left(\frac{k}{aH} \right)^2 + H^{-2}\left( M_\textrm{eff}^2 -a^{-3}a''\right) \right],  \label{square} \\
M_\mathrm{eff}^2 &= g^2(\phi-\phi_*)^2 \equiv M_g^2 ( \nu - \nu_* )^2 , \label{eq:Meffpotential}
\end{align}
where $M_g \equiv g M_{\text{Pl}}$, and we have defined the variable $\nu \equiv \phi / M_{\text{Pl}}$, where $M_{\text{Pl}}$ is the Planck mass, for future convenience.

\begin{figure*}
\includegraphics[width = \textwidth]{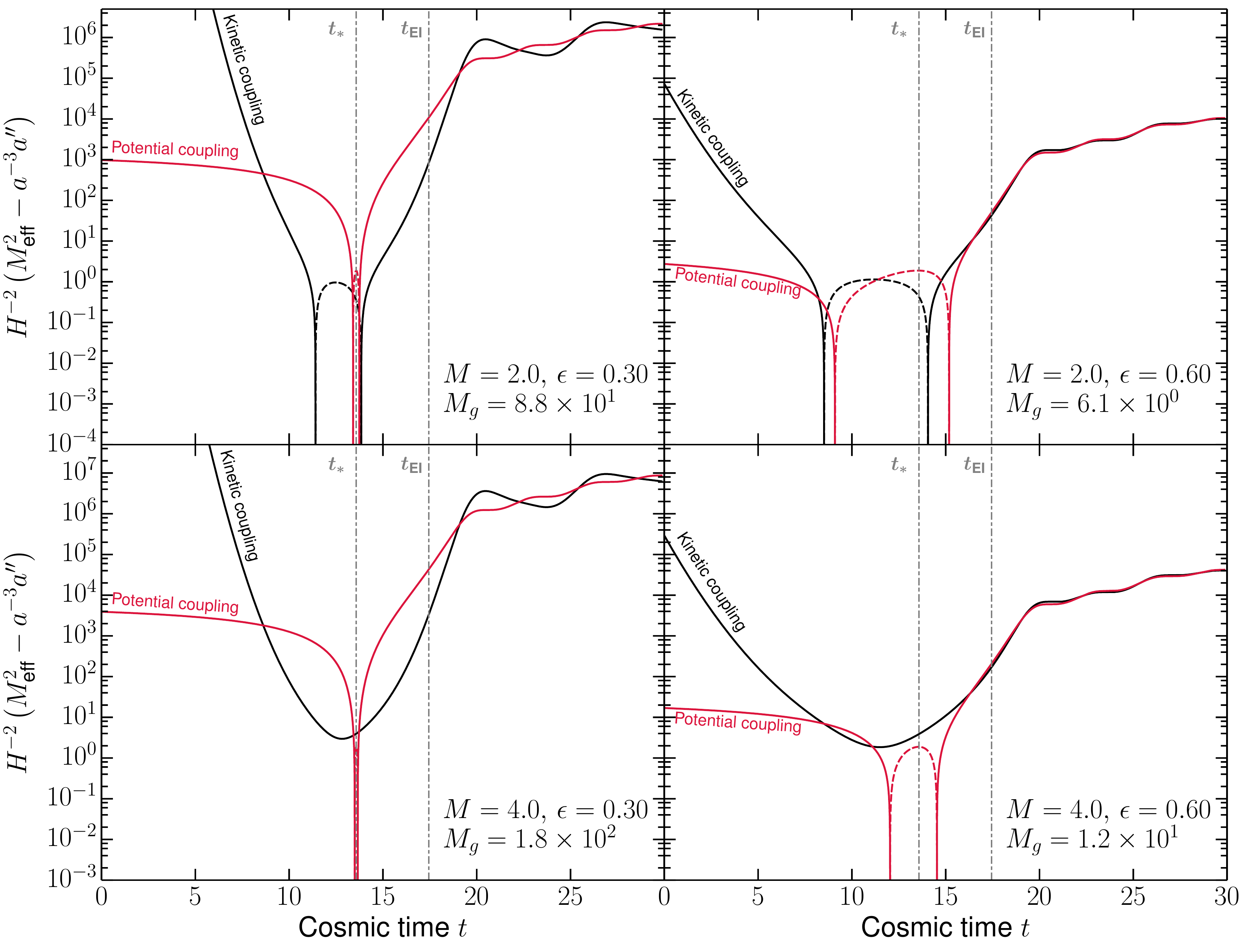}
\caption{ \label{fig:mass_plot_comp_t}(Color online) A graph of $H^{-2}(M_\textrm{eff}^2-a^{-3}a'')$ as a function of time in two models for four choices of parameters.  The models are for inflaton--irrupton coupling through either the potential term (discussed in \secref{potentialirruption}) or through the kinetic term (discussed in \secref{kineticirruption}).  Where the lines are dashed, $H^{-2}(M_\textrm{eff}^2-a^{-3}a'') < 0$. The dashed vertical lines denoted ``$t_{\text{EI}}$'' mark the end of the inflationary phase.  Time is in units of $m^{-1}$, where $m$ is the inflaton mass, and $t=0$ corresponds to the time when $\phi=3M_{\text{Pl}}$. The inflaton field has the value $\phi=\phi_*=0.8M_{\text{Pl}}$ at $t=13.6$, indicated by the dashed vertical lines marked ``$t_*$.''  Model parameters are chosen such that in each individual plot the irrupton masses are the same at $\phi\equiv0$ and $a''=0$ (i.e., at late time) for the two models.}
\end{figure*}

Just as before, species irrupton will occur when one of the conditions in Eq.\ (\ref{nonadiabatic}) is violated.  For sure, $\omega_k'$ and $\omega_k''$ are more complicated in this model than the model of \secref{simplemodel}, but if we make use of what we learned in \secref{simplemodel} we expect that irruption will occur when $\omega_k^2$ passes through zero.  Since $(k/aH)^2$ is positive, for $\omega_k^2$ to pass through zero we must have $H^{-2}\left( M_\textrm{eff}^2 -a^{-3}a''\right)<0$.  A graph of $H^{-2}\left( M_\textrm{eff}^2 -a^{-3}a''\right)$ as a function of cosmic time (not conformal time) is given in Fig.\ \ref{fig:mass_plot_comp_t}. Regions in the evolution where the above quantity is negative are shown by the dashed part of the curve.  Irruption will occur in and around the dashed regions.

A basic understanding of the results may be obtained if we make a couple of simple approximations.  First, recall that 
\begin{equation}
\frac{a''}{a^3} = \frac{4\pi G}{3}\left(\rho-3p\right) = \frac{8\pi G}{3} \left(2\rho-\frac{3}{2}\dot{\phi}^2\right) ,
\label{appoa3}
\end{equation}
where the first equality holds for any FRW model, and the second equality holds if the energy density is dominated by the inflaton (regardless of inflaton potential).  One could numerically solve the inflaton field equation for $\dot{\phi}$, or use the slow-roll approximation for $\dot{\phi}$; however, to get a rough idea of what is expected we can make an even cruder (but still reasonable) approximation and ignore the $\dot{\phi}$ term in Eq.\ (\ref{appoa3}) altogether with the result $a^{-3}a'' \sim 2H^2$.  

Since for this inflaton model $\phi \sim 10^5 H$ during inflation, unless $g$ is quite small, $M_\textrm{eff}$ will be much larger than $H$ except in a very narrow range near $\phi=\phi_*$.  So, for irruption we may write $M_\textrm{eff}^2=g^2\delta\phi^2$ where $\delta\phi =\phi-\phi_*$. 

Using $M_\textrm{eff}^2=g^2\delta\phi^2$ and $a^{-3}a''=2H^2$, the square bracket in Eq.\ (\ref{square}) becomes $(k/aH)^2-2+g^2\delta\phi^2/H^2$.  We expect irruption when $(k/aH)^2 \sim 2-g^2\delta\phi^2/H^2 \sim 2-g^2\delta\phi^2/m^2$, where we have made the further approximation that $H\sim m$, the inflaton mass.  

We can draw a couple of expected results from these crude approximations:
\begin{enumerate}
\item The condition for irruption, $\omega_k^2\sim 0$, obtains for only for a rather narrow range of $|\delta\phi| \lesssim  m/g$.  Since $m\sim 10^{13}$ GeV, $M_{\text{Pl}}\sim 10^{19}$ GeV, and $\phi\sim M_{\text{Pl}}$, unless $g\lesssim 10^{-6}$ irruption occurs for $\delta\phi\ll \phi$.
\item The duration of irruption will increase as $g$ decreases, roughly as $|\delta\phi|\sim m/g$.
\item Since $g^2\delta\phi^2/H^2$ is positive definite, the largest $k/aH$ can be during irruption is of order unity. 
\item The spectrum of produced particles is peaked; potentially strongly peaked at large $g$.  This is different than the model of the previous section.
\item Particles with present mass much larger than $H$ can be created during inflation if the particle species couples to the inflaton in such a way that its effective mass vanishes during inflation.
\item In this model, if $g$ is not much smaller than about $0.1$ or so, the present mass of the irrupton can be larger than the Planck mass.  We will speculate on the implications and complications of this fact in the concluding section.
\item We have assumed the irrupton is stable. Again, we will discuss this in the concluding section.
\end{enumerate}

In the next subsection we will consider an even more complicated irruption model.  It will prove very useful to understand the results of that model on the basis of the results of these first two simpler models.

\subsection{Massive Particle Species Irruption from Non-Canonical Kinetic Term \label{sec:kineticirruption}}

Now consider the following two-field model where the inflaton is coupled to the irrupton through the irrupton kinetic term (we refer to this as the ``kinetically-coupled'' irrupton):
\begin{align}
S &= \int d^4x\sqrt{-g}\bigg[ \frac{1}{2} g^{\mu\nu} \partial_\mu\phi\partial_\nu\phi - V(\phi) \nonumber \\ 
&\qquad\qquad\qquad\quad + \frac{1}{2} e^{2f(\phi)}g^{\mu\nu}\partial_\mu\chi\partial_\nu\chi - U(\chi,\phi)  \bigg],
\label{eq:Lag}
\end{align}
where $\phi$ is the inflaton and $\chi$ is the irrupton.  Here, $U$ and $V$ are potentials defined such that the $U$ contains only potential terms depending on $\chi$ and $\chi-\phi$ interaction terms, and $V$ contains all terms depending on $\phi$ only. 

We again take $\phi$ to be a spatially homogeneous classical field, governed by the EOM
\begin{align}
\ddot{\phi} + 3H\dot{\phi} + V_{,\phi} = f_{,\phi}e^{2f(\phi)}(\partial\chi)^2 - U(\phi,\chi)_{,\phi} \approx 0,
\end{align}
where the approximate equality above reflects the fact that we will again ignore the back-reaction on the classical inflaton field (and hence the metric) induced by the $\chi$ field.  In terms of the dimensionless field $\nu \equiv \phi / M_{\text{Pl}}$, we have 
\begin{align}
\ddot{\nu} + 3H\dot{\nu} + m^2\nu = 0,
\label{eq:nu_EOM}
\end{align}
where we have specialized to $V(\phi) = \frac{1}{2} m^2 \phi^2$.  Ignoring the back-reaction on the metric is equivalent to ignoring the contribution of the $\chi$ field to the total energy density, so from the Friedmann equation it follows that
\begin{align}
H(t) \equiv \frac{\dot{a}(t)}{a(t)} = \sqrt{\frac{4\pi}{3}} \lb( \dot{\nu}^2 + m^2 \nu^2 \rb)^{1/2}. \label{eq:Hubble_nu}
\end{align}
For our numerical work in this paper we will solve the classical field equation for $\nu$, assuming that $\nu(t_0 = 0) = 3$ and that the inflaton field undergoes initial slow-roll $\dot{\nu}(t_0 = 0) = - 1 / \sqrt{12\pi}$, yielding roughly 57 $e$-foldings of inflation.  We also fix the normalization of the scale parameter to be $a(t_{\text{EI}}) = 1$ where $t_{\text{EI}} \approx 17.5$ is the end of inflation (defined to be the moment when $\ddot{a}(t_{\text{EI}})=0$, or $w=-1/3$) which gives $H(t_{\text{EI}}) \approx 0.50$.\footnote{We denote all parameters at the end of inflation by the subscript ``EI''.} 

The irrupton field $\chi$ is governed by the field equation 
\begin{align}
\ddot{\chi} + 3H\dot{\chi} + \frac{\eta^{ij}\partial_i\partial_j\chi}{a^2(t)} +  2f_{,\phi}\dot{\phi}\dot{\chi} + e^{-2f(\phi)}U_{,\chi} =0.
\end{align}
Specializing to $U(\chi,\phi) \equiv \frac{1}{2}\cdot \mathcal{U}(\phi)\cdot \chi^2$, passing to the $\nu$ variable and making a field redefinition $\mu = e^{f(\nu)}\chi$ to eliminate the mixed derivative term,\footnote{At the level of the action, this re-definition canonically normalizes the $\mu$ kinetic term: $S \supset \int d^4x \sqrt{-g} \frac{1}{2} \dot{\mu}^2.$} this becomes
\begin{align}
\ddot{\mu} + 3H \dot{\mu}  + \lb[ \begin{array}{l} \dfrac{\eta^{ij}\partial_i\partial_j}{a^2} - f_{,\nu} \lb( \ddot{\nu} + 3H \dot{\nu} \rb) \\[2ex] - \dot{\nu}^2\lb( f_{,\nu\nu} + f_{,\nu}^2 \rb)  + e^{-2f(\nu)} \mathcal{U}(\nu) \end{array} \rb]\mu = 0 .
\end{align}

We again mode-expand $\mu = \sum_k \hat{a}_k u_k + \hat{a}_k^\dagger u_k^*$ and make the plane wave ansatz
\begin{equation}
u_k(\textbf{x},t) = \frac{e^{i\textbf{k}\cdot\textbf{x}}}{(2\pi)^{3/2}a(t)}\mu_k(t)
\end{equation}
where $\bf{k}$ is the comoving momentum. This gives the mode equation for $\mu_k$,
\begin{align}
\ddot{\mu}_k + H \dot{\mu}_k + \lb[ \begin{array}{l} \dfrac{k^2}{a^2} - \dfrac{a''}{a^3} + m^2 \nu f_{,\nu} \\[2ex]  - \dot{\nu}^2 \lb( f_{,\nu\nu} + f_{,\nu}^2 \rb) + e^{-2f(\nu)} \mathcal{U}(\nu)\end{array} \rb] \mu_k &= 0,
\label{eq:modek}
\end{align}
where we used Eq.\ \eqref{eq:nu_EOM} to simplify. 

In this paper we will specialize to the potential $\mathcal{U}(\nu) = M^2$, and we assume that the function $f(\phi)$ takes the form $f(\phi) = - (\phi-\phi_*)^2 / 2 \epsilon^2 M_{\text{Pl}}^2$, which implies  $f(\nu) = - (\nu-\nu_*)^2 / 2\epsilon^2 $.  We then have $f_{,\nu} = -(\nu-\nu_*)/\epsilon^2$ and $f_{,\nu\nu} = -1/\epsilon^2$, so that the mode equation can be written in the form
\begin{widetext}
\begin{align}
\ddot{\mu}_k + H \dot{\mu}_k + \frac{\omega_k^2(\eta)}{a^2} \mu_k &= 0 \qquad \Leftrightarrow \qquad  \mu_k'' + \omega_k^2(\eta) \mu_k = 0 \label{eq:mode} \\ 
\omega_k^2(\eta) &= k^2 - \frac{a''}{a} + a^2 M_{\text{eff}}^2 = a^2H^2\left[\left(\frac{k}{aH}\right)^2 + H^{-2}(M_\mathrm{eff}^2-a^{-3}a'')\right] \label{eq:omegak2} \\
M_{\text{eff}}^2 &= M^2 \exp\lb[ \frac{(\nu-\nu_*)^2}{\epsilon^2} \rb] - \frac{m^2}{\epsilon^2} \nu(\nu-\nu_*) + \frac{\dot{\nu}^2}{\epsilon^2} \lb[ 1- \frac{1}{\epsilon^2}(\nu-\nu_*)^2 \rb]. \label{eq:Meff2}
\end{align}
\end{widetext}
which implies that the late-time $(\nu,\dot{\nu}\approx0)$ effective mass of the $\mu$ field is given by  $M^\infty_{\text{eff}} = Me^{\nu_*^2/2\epsilon^2}$, which may be much larger than $M$ if $\epsilon \ll \nu_*$. In all our numerical work, we will take $\nu_* = 0.8$; the solution of the mode equation becomes increasingly numerically intractable as $\nu_*$ is increased or $\epsilon$ is decreased.

Note that as $\epsilon \rightarrow \infty$, $M_\textrm{eff}\rightarrow M =$ constant.  Thus in the limit of large $\epsilon$, the kinetically-coupled irrupton model approaches the simple model of \secref{simplemodel}, albeit in a different inflationary background, as we already noted.

Before we discuss the numerical calculation of species irruption in this model, we can observe similarities and differences between this model and potentially-coupled irruption model.  In Fig.\ \ref{fig:mass_plot_comp_t} we show the function $H^{-2}(M_\textrm{eff}^2-a^{-3}a'')$ for a variety of parameter choices. There are some general observations we can draw:
\begin{enumerate}
\item One very general result is that the kinetically-coupled irrupton mass is very large at early time. 
\item For some model parameters the function can appear similar to that for the potentially-coupled irrupton model (see, e.g., the upper-right plot).
\item For some model parameters the function never becomes negative (see, e.g., the lower plots).
\item The region where the function is negative is not centered on the time when $\phi=\phi_*$.
\end{enumerate}

\subsection{Irrupton number density and irrupton initial conditions \label{sec:betak}}

Now let us turn to the numerical calculation of irrupton irruption.  As noted, the form of the mode equation for $\mu_k$ on the right in Eq.\ \eqref{eq:mode} is an ``harmonic oscillator'' equation with $\eta$-dependent frequency in conformal time; however, we find it more convenient\footnote{\label{foot:eta_vs_t}The relation $dt = a\, d\eta$ implies that for exponentially small $a$, as occurs in the early inflationary epoch, a very large range of $\eta$ must be covered to cover even a small range of $t$. Since the 57 $e$-folds of inflation in our computation last $\Delta t \approx 18$ in the units in which we perform the computation, whereas $a(0) \sim 10^{-25}$, this would be a significant problem.} for our numerical work to use the form on the left of Eq.\ \eqref{eq:mode} in terms of cosmic time $t$.\footnote{Here, and throughout the remainder of the paper, we use $\mu_k$ generically to mean ``the mode function.'' Whenever the constant-$M$ or potentially-coupled models are under discussion, $\mu_k$ should of course be read instead as $\chi_k$: cf.\ Eq.\ \eqref{eq:mode} and Eqs.\ \eqref{modesimple} and/or \eqref{modeold}.} We also emphasize that if $\omega_k^2$ anywhere runs negative (as it may do owing to the presence of the $-a''/a$ term, which would be absent if the irrupton were coupled conformally to the metric, rather than minimally), the mode becomes ``tachyonic'' and the mode function diverges exponentially.\footnote{Despite this exponential growth, the evolution does always exactly preserve the norms defined in footnote \ref{footnote:norm}.}

During the non-adiabatic phase, the notion of particle number is ambiguous \cite{BirrellDavies1982}, but in both the early-time and late-time regimes, where the expansion is adiabatic with respect to any given mode, the notion of a particle number regains physical validity. In order to extract the number of particles produced in mode $k$, we make use of the method of Bogoliubov coefficients. We specify some initial conditions (equivalent to the in-vacuum choice; see below) then numerically integrate the mode equation for $\mu_k$ to some late time after inflation has ceased ($\ddot a < 0$), and use this solution to extract the Bogoliubov coefficient $\beta_k$ giving the overlap of the exact solution to the mode equation subject to the early-time-vacuum initial conditions with the exact solution which is pure negative-frequency at late-time. This gives the differential comoving number density of particles present in mode $\bm{k}$ in the asymptotic late time regime as
\begin{align}
(2\pi)^3 \frac{ d n }{d^3k} \equiv \bks 
&= \frac{\omega_k}{2} \lb[ \lb| \mu_k \rb|^2 + \omega_k^{-2} \lb| \mu_k' \rb|^2 \rb] - \frac{1}{2} \nonumber \\
&=  \frac{\omega_k}{2}\left[ |\mu_k|^2 + \frac{a^2}{\omega_k^2} |\dot{\mu}_k|^2 \right]  - \frac{1}{2}.
\label{eq:nk}
\end{align}
Clearly the identification of the rhs of Eq.\ \eqref{eq:nk} as the absolute value squared of a complex quantity is only sensible if the rhs is real and positive. If $\omega^2_k<0$ (the tachyonic regime), the rhs is imaginary and cannot be identified as $|\beta_k|^2$.  While the identification of the rhs as $|\beta_k|^2$ is sensible if $\omega^2_k>0$, it can only be interpreted as a particle density in the adiabatic regime, $\mathcal{A}=\lb| \omega_k'/\omega_k^2 \rb|^2 \ll 1$ and $\mathcal{B}= \lb| \omega_k''/\omega_k^3 \rb| \ll 1$. We will frequently have recourse to refer to the differential comoving number density $\nck$ of particles per logarithmic interval of $k \equiv |\bm{k}|$ which is given by
\begin{align}
\nck \equiv  \frac{d n }{d\ln k} \equiv \frac{k^3}{2\pi^2} |\beta_k|^2,
\label{eq:nck}
\end{align}
as well as the total physical number density $\np$ of particles, which is given by
\begin{align}
\np  =  \frac{1}{a^3}\int \frac{d^3k}{(2\pi)^3} \bks = \frac{1}{a^3} \int_{-\infty}^\infty \nck \, d\ln k.
\label{eq:number}
\end{align}
If $\omega_k^2$ does run negative anywhere, the resulting tachyonic evolution of the mode function will show up as an exponential enhancement in the particle production since $\mu_k$ and $\dot{\mu}_k$ will be exponentially larger at late time.\footnote{While this statement is true, the issue is a bit more subtle than it appears.  It will be discussed in more detail in \secref{results}.} On the other hand, if $\omega_{k}^2 \gg 0$ is everywhere too large (say, we choose $k$ much larger than other parameters), then there is never appreciable particle creation as the mode will suffer neither tachyonic evolution nor a strongly non-adiabatic background space-time expansion with respect to that mode.

The question of which initial conditions to specify for the mode function $\mu_k$ is subtle since their choice defines which basis of solutions are used for the mode expansion, making the question equivalent to the deep issue of the choice of vacuum in a non-static spacetime \cite{BirrellDavies1982}.  For practical reasons of numerical stability, we choose initial conditions such that we start in an instantaneous Minkowski vacuum,\footnote{These initial conditions would obtain by requiring the exact solution match onto the zeroth-order adiabatic approximate solution at $t= t_0^k$, correct up to terms of zeroth adiabatic order. That is, these initial conditions specify a ``zeroth-order adiabatic vacuum.'' See, e.g., Sec.\ 3.5 of Ref.\ \cite{BirrellDavies1982}. }
\begin{align}
\mu_k(t^k_0) &= \frac{e^{i\pi/4}}{\sqrt{2\omega_k(t^k_0)} },   &\text{and} && \dot{\mu}_k(t^k_0) &= -i\frac{\omega_k(t^k_0)}{a(t^k_0) } \mu_k(t_0^k),
\label{eq:ICs}
\end{align}
for mode $k$ at a time $t=t^k_0$ chosen such that, for the mode in question, $\omega_k(t^k_0)/a(t^k_0)$ is equal to a very large threshold value (at least $5\eten{3}m$). That is, we pick the vacuum state $|0\rangle$ such that $\hat{a}_k(t_0^k)|0\rangle=0$, which implies that $n^{\text{c}}_k(t_0^k)=0$. We immediately note that this choice of initial conditions does indeed specify a different (zeroth-order adiabatic) vacuum for every mode as $t_0^k$ is mode-dependent; however, since up to and including this time, the adiabaticity parameters $|\omega_k'/\omega_k^2|^2 \ll 1$ and $|\omega_k''/\omega_k^3| \ll 1$ are very small, and the bases of mode functions specified by the imposition of the initial conditions at $t_0^k$ or ${t_0^{k}}'$ differ from each other only by terms of adiabatic order greater than zero, the family of vacua are all approximately equivalent. Indeed, we have explicitly verified that by increasing the threshold value of $\omega_k(t^k_0)/a(t^k_0)$, the amount of particle production we find does not change. There is one subtlety involved here, mainly relevant for the kinetically-coupled case: if $M_{\text{eff}}$ is very large at early time (and/or $k$ is very small), $\omega_k^2/a^2$ may cross the large threshold value while dominated by the $M_{\text{eff}}^2$ term, which could possibly lead to a situation where $t_0^k$ comes after the mode in question crosses outside the Hubble radius at $t_{\text{HC}}$ (see Figs.\ \ref{fig:mass_plot_1} and \ref{fig:mass_plot_2}).  This would be potentially problematic on conceptual grounds, and we thus choose to present results only for modes which satisfy $t_0^k< t_{\text{HC}}$.  

\section{Present-Day Abundance for Stable Irruptons \label{sec:abundance}}

Before we turn to a more detailed discussion of our numerical work, we specify how the comoving number density of irruptons produced in the final $e$-foldings of inflation is translated into a physical present-day relic abundance assuming the irrupton is stable. We follow the discussion of Ref.\ \cite{Chung:1998zb}.

The irruptons are produced mainly during the final stages of inflation, which we assume to be followed by a brief matter-dominated phase characterized by coherent inflaton oscillations about the potential minimum $\phi=0$. At the onset of oscillations, the Universe is in a low-entropy frozen state owing to the inflationary expansion; however, after some number of inflaton oscillations, during which time the overwhelming majority of the energy density of the Universe is contained in the inflaton field, the inflaton energy density is converted to radiation, heating the Universe to some high temperature $T_{\text{RH}}$ following which, in the standard thermal history of the Universe, there is no significant further entropy production and the Universe in the large simply undergoes adiabatic expansion to the present time.

We consider first the epoch after the inflaton energy density has been converted to radiation. The inflaton is non-relativistic (NR) so that the irrupton energy density, $\rho_I = M_{\text{eff}}^\infty n^{\text{p}}$ with $n^{\text{p}}$ given by Eq.\ \eqref{eq:number} and where $M_{\text{eff}}^\infty$ is the late-time effective irrupton mass (see Eq.\ \eqref{eq:Meff_inf} below). Under adiabatic expansion the comoving entropy density is constant, so it follows that the ratio 
\begin{align}
\frac{\rho_I}{\rho_{R}} \frac{g_{\text{eff}}}{h_{\text{eff}}} T
\end{align}
is constant, where $\rho_R$ is the radiation energy density, and $g_{\text{eff}}$ ($h_{\text{eff}}$) is the effective number of relativistic degrees of freedom relevant for the computation of $\rho_R$ (entropy density $s$). The present-day relic abundance of the heavy irrupton species is thus
\begin{align}
\lb. \Omega_I h^2 \rb|_0 = \lb. \Omega_{R} h^2 \rb|_0  \lb(\frac{T_{\text{RH}}}{T_0}\rb) \lb( \frac{ h_{\text{eff}} }{ g_{\text{eff}} } \rb)_0 \lb( \frac{ \rho_I}{\rho_R} \rb)_{\!\text{RH}},
\label{eq:topresent}
\end{align}
where ``RH'' denotes quantities evaluated at the moment of matter-radiation equality at the completion of (p)reheating, $T_0 = 2.7255\,$K \cite{Fixsen:2009gs}, $\Omega_{R} h^2 = 4.149\eten{-5},\ g_{\text{eff},0} = 3.38$ and $h_{\text{eff},0} = 3.91$ \cite{Kolb:1990vq} and we have taken $g_{\text{eff}}=h_{\text{eff}}$ at the completion of (p)reheating.

It remains to determine $\lb(\rho_I/ \rho_R \rb)_{\text{RH}}$. To do this, we note that during the matter-dominated inflaton-oscillation epoch, the overwhelming majority of the energy density is in the NR inflaton coherent oscillations, so the Friedmann equation yields $ \rho_{\phi} \approx \rho_{\text{tot}} = 3M_{\text{Pl}}^2 H^2/8\pi$. The irruptons carry a subdominant component of the energy density (see Sec.\ \ref{sec:applications}) and are already NR at this epoch, so we still have $\rho_I = M_{\text{eff}}^\infty n^{\text{p}}$ with $n^{\text{p}}$ given by Eq.\ \eqref{eq:number} (the mass of the irrupton is fairly well approximated by its asymptotic large-time value $M_{\text{eff}}^\infty$ after a few inflaton oscillations). The ratio $(\rho_I/\rho_\phi)_\text{osc.} \sim H^{-2}a^{-3} \sim t^{-2w}$ is thus constant for a pure-matter era ($w=0$). This means that we may extract $\rho_I$ at any point during the matter-dominated era once $\rho_I$ has stabilized. In order to relate $(\rho_I/ \rho_\phi)_\text{osc.}$ to $\lb(\rho_I / \rho_R \rb)_{\text{RH}}$ we must now make two assumptions: 
\begin{enumerate}
\item the entire energy density of the inflaton coherent oscillations ends up in radiation after the inflaton oscillations decay
\item the transition between the matter-dominated oscillation epoch and the radiation-dominated epoch happens fairly quickly so that (a) the Universe does not expand significantly during the transition which would cause the ratio $\rho_I/\rho_\text{total} = \rho_I / (\rho_R + \rho_\phi)$ to change non-trivially, and (b) there is no significant further entropy production once in the radiation-dominated era. 
\end{enumerate}
Under these assumptions, it is a good approximation to set $( \rho_I / \rho_{\phi} )_\text{osc.} \approx \lb(\rho_I / \rho_R \rb)_{\text{RH}}$ and to apply Eq.\ \eqref{eq:topresent}. Putting this all together, we have
\begin{align}
\lb. \Omega_I h^2 \rb|_0 & \approx \lb. \Omega_{R} h^2 \rb|_0  \lb(\frac{T_{\text{RH}}}{T_0}\rb) \lb( \frac{ h_{\text{eff}} }{ g_{\text{eff}} } \rb)_0 \times \lb(  \frac{3M^2_\text{Pl}}{8\pi}   H^2(\tilde{t}\,)  \rb)^{-1} \nonumber \\[1ex] &\quad \times \lb[  \frac{M_{\text{eff}}^\infty}{a^3(\tilde{t}\,)} \int_{-\infty}^\infty \nck(\tilde{t}\,) \,d\ln k \rb] 
\end{align}
where $\tilde{t}$ is some reference time during the matter-dominated inflaton oscillation era at which we choose to extract the particle spectrum from our numerical solutions (see \secref{numerics}). In order to make contact with our numerical simulations in which we work in units for $k$, $H$, and $t$ which are based on $m=1$, we can re-write the above result as
\begin{align}
\lb. \Omega_I h^2 \rb|_0 &\approx \frac{8\pi}{3} \lb( \Omega_{R} h^2 \rb)_0  \lb(\frac{T_{\text{RH}}}{10^{9}\text{GeV}} \rb) \lb(\frac{10^{9} \text{GeV}}{T_0}\rb)  \nonumber \\[1ex] &\quad \times \lb( \frac{ h_{\text{eff}} }{ g_{\text{eff}} } \rb)_0  \lb(  \frac{ m }{ 10^{13} \text{GeV} }\rb)^2 \lb( \frac{10^{13} \text{GeV} } { M_{\text{Pl} } } \rb)^2 \nonumber \\[1ex] &\quad \times \frac{ {M_{\text{eff}}^\infty}' }{ H'^2(\tilde{t}\,) a^3(\tilde{t}\,) } \int_{-\infty}^\infty n_{k'}^c(\tilde{t}\,) \,d\ln k'   
\end{align}
or
\begin{align}
& \lb. \Omega_I h^2 \rb|_0 \times \lb(\frac{T_{\text{RH}}}{10^{9}\text{GeV}} \rb)^{-1} \!\! \times \lb(  \frac{ m }{ 10^{13} \text{GeV} }\rb)^{-2} \nonumber \\
&\approx 1.1 \eten{6}  \times \lb[ \frac{ {M_{\text{eff}}^\infty}' }{ H'^2(\tilde{t}) a^3(\tilde{t}) } \times  \int_{-\infty}^\infty n_{k'}^c(\tilde{t}\,) \,d\ln k' \rb],
\label{eq:relic_abundance}
\end{align}
where the values of $H'$, ${M_{\text{eff}}^\infty}'$, and $n_{k'}$ are all extracted from the numerical solution of the $\nu$ EOM and the mode equation for $\mu_k$ in the units where $m=1$. For the three models we have discussed in this paper, $M_{\text{eff}}^\infty$ is given by
\begin{equation}
{M_{\text{eff}}^\infty}' = \left\{ 
\begin{array}{ll}
M/m \qquad & \textrm{constant\ } M  \\[0.8ex]
(M_g/m)\nu_*=0.8M_g \qquad & \textrm{potentially-coupled} \\[1ex]
(M/m)\exp(\nu_*^2/2\epsilon^2)\qquad  & \textrm{kinetically-coupled}
\end{array} \right.
\label{eq:Meff_inf}
\end{equation}
Also, in the matter-dominated era after inflation, $H'^2(\tilde{t}) a^3(\tilde{t})\simeq 0.18$.

We shall henceforth drop the primes with the understanding that all of these quantities are measured in units of $m$.

\section{Numerical Methods and Issues \label{sec:numerics}}

In order to calculate the final value of the Bogoliubov coefficient, $\beta_k$, and hence the number density of the irrupton, in principle we require the late-time values of $|\mu_k|^2$ and $|\dot{\mu}_k|^2$ [see Eq.\ \eqref{eq:nk}]. The straightforward procedure is to integrate the seven-dimensional system of first-order coupled ordinary differential equations (odes) for the scale factor $a$, the inflaton field value $\nu$ and its derivative $\dot{\nu}$, and the real and imaginary components of the irrupton mode function and its first derivatives
\begin{align}
\bm{F}[t] \equiv \lb\{\ a,\ \nu,\ \dot{\nu},\ \Re{\mu_k},\ \Im{\mu_k},\ \Re{\dot{\mu}_k},\ \Im{\dot{\mu}_k}\ \rb\}.
\end{align}
In some cases we will consider, this straightforward procedure is impractical.  One issue is for large irrupton mass the irrupton oscillation frequency may be much, much larger than the inflaton oscillation frequency.  We will also see that for some parameters we require integration of the system very deep into the matter-dominated era after inflation.  Integrating the seven-dimensional system is occasionally unwieldy and computationally limited.  Therefore, we will have occasion to employ a different computational strategy that can be used in the regimes where there is no tachyonic phase.

\subsection{Integration of the Full Seven-Dimensional System \label{sec:seven}}

For the numerical integration of the seven-dimensional system of odes we utilized the the Runge-Kutta Dormand-Prince 8(53) algorithm \texttt{dop853} natively implemented in the \texttt{scipy v0.13} module for \texttt{python v2.7}, and have cross-checked our results against at least one other solver (\texttt{scipy}'s native \texttt{lsoda} algorithm) for a large subset of parameter values, finding agreement for most input parameter choices, and noting that it is fairly obviously the \texttt{lsoda} algorithm which fails when the two numerical solutions disagree. 

In implementing these algorithms it was necessary to impose very strict error tolerances and small maximum step sizes $\delta t$ as a variety of egregious numerical issues arise in solving the system of equations owing to the magnitude of some terms which enter. For example, an incomplete cancellation at early times between the positive and negative terms in Eq.\ \eqref{eq:nk} arising from tolerances which are too loose can cause $|\beta_k|^2$ to jump nonphysically to a value many orders of magnitude larger than it should be in the first few time steps, and this erroneously large value places a floor on how small a value of $|\beta_k|^2$ can be probed in our simulations at late times, which causes complications in tracking certain features in the particle spectra.

\begin{figure*}
\includegraphics[width = \textwidth]{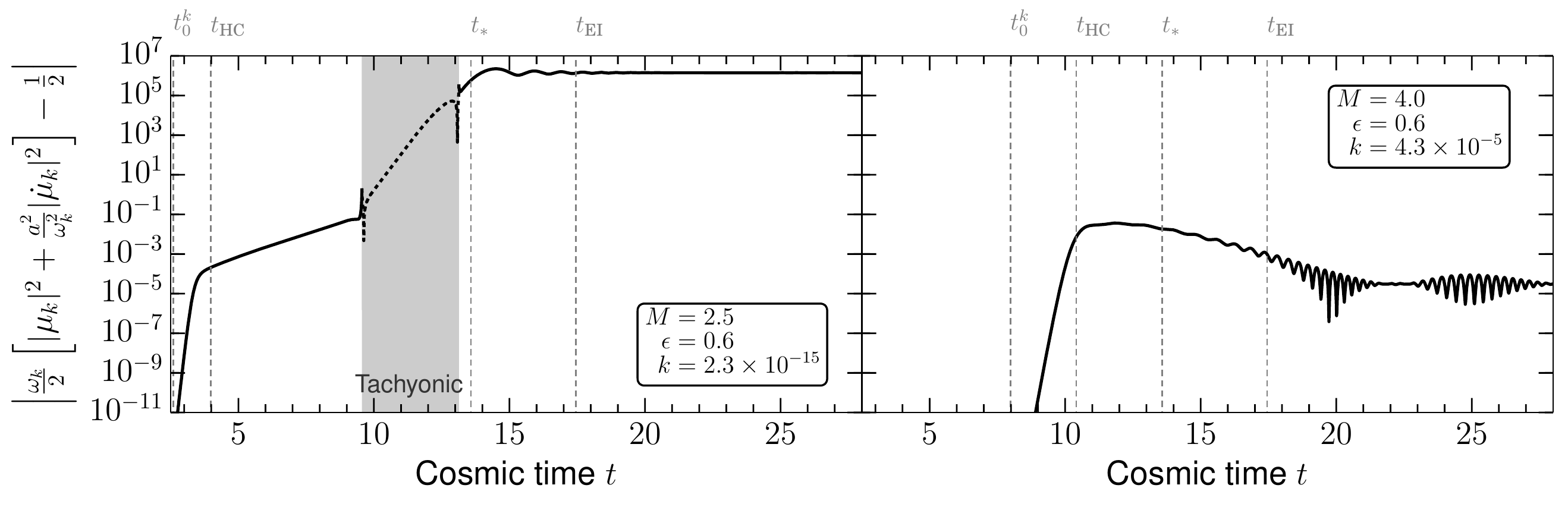} 
\caption{ \label{fig:beta_k_example}  The time evolution of $\left| \frac{\omega_k}{2}\left[ |\mu_k|^2 + \frac{a^2}{\omega_k^2} |\dot{\mu}_k|^2 \right]  - \frac{1}{2} \right|$ for two representative cases for the kinetically-coupled irrupton as extracted from the numerical solution for the choices of parameters $M = 2.5, \ \epsilon = 0.6$ and $k = 2.3\times10^{-15}$ (left plot) and $M = 4.0, \ \epsilon = 0.6$ and $k = 4.3\times10^{-5}$ (right plot). If $\omega_k^2>0$, $\frac{\omega_k}{2}\left[ |\mu_k|^2 + \frac{a^2}{\omega_k^2} |\dot{\mu}_k|^2 \right]  - \frac{1}{2}$ is real and equal to $|\beta_k|^2$.  The evolution during the tachyonic phase  is presented for demonstrative purposes to indicate the rapid increase during this period; we caution, however, that there is no fashion in which it can be interpreted as a particle number during the tachyonic phase. At both early- and late-times, the background space-time evolution is sufficiently adiabatic with respect to this mode that $|\beta_k|^2$ can indeed be interpreted as the occupation number for mode $\bm{k} = k \bm{\hat{k}}$; note that the late-time behavior (i.e., after the end of inflation) shows that $|\beta_k|^2$ undergoes damped oscillation about a constant non-zero value, indicating particle production has occurred. The times $t^k_0, t_{\textrm{HC}}, t_*$ and $t_{\textrm{EI}}$ indicate, respectively, the times when the initial conditions Eq.\ \eqref{eq:IC} were imposed, when the mode crosses the Hubble radius ($k = aH$), when $\nu=\nu_*$, and when inflation ends ($\ddot{a}<0$). }
\end{figure*}

Furthermore, the cosmic-time oscillation frequency $\omega_k/a$ can become very large in both the early and late evolution of the inflationary epoch making it computationally unfeasible (given the number of solutions required to scan over parameter space) to perform sufficiently small $\delta t$ time steps when the oscillation frequency $\omega_k/a$ of the mode function is larger than about $\mathcal{O}(10^4)$.\footnote{As was alluded to above (see footnote \ref{foot:eta_vs_t}), the opposite side of the same coin is that by going to conformal time, the domain of the $\eta$ integration becomes unfeasibly long to step forward any reasonable amount of cosmic time.}  This necessitated a subdivision of  the computation. As we are ignoring the back-reaction due to the irrupton on the space-time metric, as a first step we solved the restricted three-dimensional problem for $\bm{\tilde{F}}[t] = \lb\{\ a\ ,\ \nu\ ,\  \dot{\nu}\ \rb\}$ specifying initial conditions $\bm{\tilde{F}}[0] = \lb\{\ 1.60928\eten{-25}\ ,\ 3\ ,\ - 1 / \sqrt{12\pi}\ \rb\}$ where the value of $a(0)$ was set by the requirement that $a(t_{\text{EI}}) = 1$. We used this solution to find the time $t_0^k$ (or more precisely, the nearest sampled time-step earlier than this) where $\omega_k/a \sim 5\eten{3}$ and is decreasing.  As discussed above, this is the point at which we specify the mode function $\mu_k$ to match the zeroth-order approximate adiabatic solution (i.e., specify the in-vacuum). Having identified this point, we then switched to integrating the full seven-dimensional system\footnote{For reasons of numerical accuracy, we prefer this approach to interpolating the already-known solution for the restricted three-dimensional problem and solving just the four-dimensional system for $\mu_k$.} $\bm{F}[t]$ starting with initial conditions
\begin{align}
\bm{F}[t_0^k]  &= \Bigg\{\ \tilde{a}(t_0^k)\ ,\ \tilde{\nu}(t_0^k)\ ,\ \dot{\tilde{\nu}}(t_0^k)\ ,\ \frac{1}{ 2\sqrt{\omega_{k}(t_0^k) }}\ , \nonumber \\ &\quad \quad \ \frac{1}{ 2\sqrt{\omega_{k}(t_0^k) }}\ ,\ \frac{\sqrt{ \omega_{k}(t_0^k) }}{2\tilde{a}(t_0^k)}\ ,\  -\frac{\sqrt{ \omega_{k}(t_0^k) }}{2\tilde{a}(t_0^k)}\ \Bigg\}
\label{eq:IC}
\end{align}
where the quantities with a tilde are obtained from the integration of the restricted three-dimensional problem, and the $\mu_k$ initial conditions are precisely those defined in Eq.\ \eqref{eq:ICs}. We obtain $|\beta_k|^2$ from this full solution via Eq.\ \eqref{eq:nk}; an example plot demonstrating the time-evolution of $|\beta_k|^2$ thus extracted is shown in Fig.\ \ref{fig:beta_k_example}.

We extract the final particle spectrum using the ``asymptotic'' value of $|\beta_k|^2$ deep in the matter-dominated inflaton oscillation era.  When $|\beta_k|^2$ has stabilized to a damped oscillation about a constant central value at large time, we extract the asymptotic value by averaging over the last few oscillations at some late time.  However, for certain parameter choices, $|\beta_k|^2$ does not stabilize to oscillations about a constant central value even by $t=500$, but rather is still executing oscillation about a downward-drifting central value. Although for such modes a stable asymptotic value for $|\beta_k|^2$ is achieved if the mode equation is integrated for sufficiently long (see Appendix \ref{app:late_time_asymptotics}), this is not computationally feasible via straightforward solution of the seven-dimensional system, and below we describe the method we use to obtain the asymptotic value of $|\beta_k|^2$ at very late time in such cases.

\subsection{The Iterative Solution Approach \label{sec:integro}}

Now we describe another approach to the calculation of $|\beta_k|^2$ that can be used only in the absence of a tachyonic phase.  This approach is amenable to a very useful and computationally much less demanding iterative solution approach when $|\beta_k|$ remains small.  It also is the formalism we use to extract the late-time asymptotic value of $|\beta_k|^2$ for model parameters where it has not stabilized to its asymptotic value by $t$ of a few hundred, which generically only occurs for modes without a tachyonic phase.

If $\omega_k$ is real (no tachyonic behavior), the solution to the mode equation may be written as (see, e.g., Ref.\ \cite{Kofman:1997yn})
\begin{equation}
\mu_k = \frac{\alpha_k(t)}{\sqrt{2\omega_k}} e^{-i\Phi(t)} +  \frac{\beta_k(t)}{\sqrt{2\omega_k}} e^{+i\Phi(t)} 
\label{eq:albe}
\end{equation}
where the accumulated phase $\Phi(t)$ is given by [obviously $\Phi_0\equiv\Phi(t_0^k)$]
\begin{equation}
\Phi(t) \equiv \int_{t_0^k}^{t} \frac{\omega_k(t')}{a(t')}dt' + \Phi_0.
\end{equation}
Equation \eqref{eq:albe} is a solution to the mode equation \eqref{eq:mode} if $\alpha(t)$ and $\beta(t)$ satisfy the coupled equations of motion 
\begin{align}
\dot{\alpha}_k & = \frac{\dot{\omega}_k}{2\omega_k} \beta_k e^{2i\Phi}     \nonumber \\ 
\dot{\beta}_k  & = \frac{\dot{\omega}_k}{2\omega_k} \alpha_k e^{-2i\Phi},
\end{align}
which also implies that 
\begin{align}
\dot{\mu}_k = -i \frac{\omega_k}{a} \lb[ \frac{\alpha_k(t)}{\sqrt{2\omega_k}} e^{-i\Phi(t)} -  \frac{\beta_k(t)}{\sqrt{2\omega_k}} e^{+i\Phi(t)} \rb].
\end{align}
In line with our previous discussion, we take initial conditions
\begin{align}
\mu_k(t_0^k) &= \frac{1}{\sqrt{2 \omega_k(t_0^k)}} e^{i\pi/4} \nonumber \\
\dot{\mu}_k(t_0^k) &= -i \frac{\omega_k(t_0^k)}{a(t_0^k)} \mu_k(t_0^k),
\end{align}
where the threshold value of $\omega_k/a$ which dictates the value of $t_0^k$ can be taken to be much larger than $\omega_k(t_0^k)/a(t_0^k)=5\eten{3}$ in this method provided we have a sufficiently accurate background solution for $\lb\{\ a,\ \nu,\ \dot{\nu}\ \rb\}$; we utilize a threshold value of $\omega_k(t_0^k)/a(t_0^k)=10^7$, and have checked explicitly (for a subset of parameter values) that the results are insensitive to this parameter provided it remains large. These initial conditions imply that ($\Phi_0$ is of course arbitrary; we simply take the value consistent with our previous discussion)
\begin{align}
\alpha_k(t_0^k) & = 1, & \beta_k(t_0^k) & = 0, & \text{and }&& \Phi_0 &= -\frac{\pi}{4}.
\end{align}
The occupancy number for the mode $\bm{k} = k \bm{\hat{k}}$ is given by 
\begin{align}
n_{\bm{k}} &= \frac{1}{2} \lb[ |\alpha_k|^2 + |\beta_k|^2 - 1 \rb] = |\beta_k|^2 = |\alpha_k|^2-1, \label{eq:nk_alpha_beta}
\end{align}
where we have used the Wronskian condition $|\alpha_k|^2 - |\beta_k|^2 = 1$ at the last two steps; this condition follows from demanding correctly normalized modes per footnote \ref{footnote:norm}.

This approach is amenable to an iterative solution when $|\beta_k|$ is small and the evolution is nearly adiabatic (we will quantify this statement shortly).  To develop the iterative solution, we introduce a formal small parameter $\epsilon$ by rescaling $\dot{\omega}_k/\omega_k$,
\begin{align}
\frac{\dot{\omega}_k}{\omega_k} \rightarrow \epsilon \frac{\dot{\omega}_k}{\omega_k},
\end{align}
and expanding the solutions in powers of $\epsilon$,
\begin{align}
\alpha_k(t) &\equiv \sum_{n=0}^{\infty} \epsilon^{2n} \alpha_k^{(2n)}(t) \nonumber \\
\beta_k(t)  &\equiv \sum_{n=0}^{\infty} \epsilon^{2n+1} \beta_k^{(2n+1)}(t),
\end{align}
in terms of which we impose the initial conditions on the $\alpha^{(n)}_k(t)$ and $\beta^{(n)}_k(t)$ as 
\begin{align}
\alpha_k^{(0)}(t) &\equiv 1, \qquad \Phi_0 = -\frac{\pi}{4} \\
\alpha_k^{(2n)}(t_0^k) = \beta_k^{(2n-1)}(t_0^k) &= 0  \quad \text{ for } \quad n \geq 1.
\end{align}
Substitution into the equations of motion for $\alpha_k$ and $\beta_k$, followed by equating coefficients of $\epsilon^n$ to zero for all $n$, then sending $\epsilon \rightarrow 1$ at the end of the process, yields the equations of motion for $\alpha^{(n)}_k$ and $\beta^{(n)}_k$,
\begin{align}
\dot{\alpha}_k^{(0)} & = 0 \nonumber \\
\dot{\beta}_k^{(2n+1)} & = \frac{\dot{\omega}_k}{2\omega_k} \alpha_k^{(2n)} e^{-2i\Phi} \quad \text{ for } \quad n \geq 0 \nonumber \\
\dot{\alpha}_k^{(2n+2)} & = \frac{\dot{\omega}_k}{2\omega_k} \beta_k^{(2n+1)} e^{+2i\Phi}  \quad \text{ for } \quad n \geq 0 \label{eq:Cone},
\end{align}
which has the following solutions consistent with the initial conditions:
\begin{align}
\alpha_k^{(0)}(t) &\equiv 1 , \qquad  \Phi_0 = -\frac{\pi}{4} \nonumber \\
\beta_k^{(2n+1)}(t) & = \int_{t_0^k}^{t} \frac{\dot{\omega}_k(t')}{2\omega_k(t')} \alpha_k^{(2n)}(t') e^{-2i\Phi(t')} \, dt'   \nonumber \\
\alpha_k^{(2n+2)}(t) & = \int_{t_0^k}^t \frac{\dot{\omega}_k(t')}{2\omega_k(t')} \beta_k^{(2n+1)}(t') e^{+2i\Phi(t')} \, dt'   \label{eq:C},
\end{align}
where the last two lines hold for $n\geq0$.
From these expressions, the iterative solution method is obvious.  Convergence is generally obtained after only a few iterations; we always truncate the series for $|\beta_k|^2$ at the tenth iterate and use this value in Eq.\ \eqref{eq:nk_alpha_beta} to obtain $n_{\bm{k}}$.

Note, however, that to utilize this method, we still need a very accurate solution for the background $\bm{\tilde{F}}[t] = \lb\{\ a\ ,\ \nu\ ,\  \dot{\nu}\ \rb\}$ out to whatever time we choose to run the solution, which must be obtained by solving the (restricted three-dimensional set of) odes for these fields per the methods discussed in the previous subsection. Although dramatically less computationally intensive than solving the full seven-dimensional set of odes, this is still time-consuming if we wish to have the solutions out to very late time to track $|\beta_k|^2$ all the way to its asymptotic value. In the next subsection we describe how in our actual numerical implementation of this iterated method, we have made a modification to the procedure just outlined which allows this problem also to be avoided.

In order to understand the conditions for the iterative solution to be a good approximation, consider the first-order solution $\beta_k^{(1)}$, which yields the lowest-order approximation for the occupancy number $n_{\bm{k}}$:
\begin{align}
n_{\bm{k}}^{(1)} = \lb|\beta_k^{(1)}\rb|^2 = \lb| \int_{t_0^k}^t \frac{\dot{\omega}_k(t')}{2\omega_k(t')} e^{-2i\Phi(t')} \, dt' \rb|^2.
\end{align}
Since in the absence of a tachyonic phase $\omega_k(t)$ is everywhere positive and real, $\dot{\Phi}(t)=\omega(t)/a(t)> 0$, and the accumulated phase is a strictly increasing function of time.  Thus, we can recast the first-order approximation using the accumulated phase as the integration variable:
\begin{align}
n_{\bm{k}}^{(1)} = \lb|\beta_k^{(1)}\rb|^2 = \frac{1}{4} \lb| \int_{\Phi_0}^{\Phi(t)} \lb. \frac{\omega'_k(t')}{\omega^2_k(t')} \rb|_{t'=t'(\Phi)} e^{-2i\Phi} \, d\Phi \rb|^2. \label{eq:beta1_phase}
\end{align}
In the asymptotic late-time regime $t\rightarrow\infty$, we have $\Phi(t) \rightarrow \infty$, so that
\begin{align}
n_{\bm{k}}^{(1)}(t\rightarrow \infty) &= \frac{1}{4} \left| \int_{\Phi_0}^{\infty} \frac{\omega'_k}{\omega^2_k} e^{-2i\Phi} \, d\Phi \right|^2 \nonumber \\ 
&\leq \int_{\Phi_0}^{\infty} \lb|  \frac{\omega'_k}{\omega^2_k} \rb|^2 \, d\Phi = \int_{\Phi_0}^{\infty}\! {\cal A}\, d\Phi, \label{eq:upper_bound}
\end{align}
where we used the Cauchy-Schwarz inequality and ${\cal A}$ is the adiabaticity parameter defined in Eq.\ \eqref{nonadiabatic}. The appearance in the integral of the ``square-root'' (with phase) of the  adiabaticity parameter $ {\cal A} $ refines our previous argument that the size of the adiabaticity parameters limits the amount of particle production.  It is in fact the total time-integrated magnitude of the adiabaticity parameter which provides a hard upper limit to the amount of particle production. Note of course that modifications to the adiabaticity parameter near its maximum clearly impact the upper bound more strongly.

It is important to note that the simple expression in Eq.\ \eqref{eq:upper_bound} is only an extremely crude upper bound to $n_{\bm{k}}^{(1)}(t\rightarrow \infty)$; we do not expect this bound to be saturated as it represents the integral of the envelope of the highly oscillatory integrand, rather than the integrand itself. \emph{Phase cancellations are important.}

\subsection{Late-time Solution in the Matter-Dominated Era \label{sec:MD}}

Well after inflation ends, the dynamics of the expansion of the Universe is that of a matter-dominated (MD) model with the energy density from the oscillating inflaton field (of course, the energy in the inflaton field eventually must be converted to radiation, so this MD phase is not of infinite duration; absent postulating a specific model for this process, we cannot assess its impact on our results). Where necessary, we exploit this fact to extend our solution for $|\beta_k|^2$ very deep into the MD era to extract its asymptotic late-time value without having to solve a set of odes.

For a MD phase the scale factor evolves as $a(t)\propto t^{2/3}$.  We will choose a reference time $t_\mathrm{ref}$ following inflation deep into the MD era (in our computations, we use $t_{\text{ref}}\sim10^3$).  Then the scale factor and expansion rate may be written as
\begin{align}
a(t) &= a_\mathrm{ref}\left(\frac{t-\tau}{t_\mathrm{ref}-\tau}\right)^{2/3} && \text{and} &
H &= \frac{2}{3}\frac{1}{t-\tau}, \label{eq:HMD}
\end{align} 
where $\tau$ is the effective bang time assuming a MD Universe all the way back to the singularity; $\tau$ has no physical significance.  The inflaton field and its time derivative are given by
\begin{align}
\nu(t)     &= A\frac{\sin(t-\tau)}{t-\tau} + B\frac{\cos(t-\tau)}{t-\tau}\ , \nonumber \\
\dot\nu(t) &= A\frac{\cos(t-\tau)}{t-\tau} - A\frac{\sin(t-\tau)}{(t-\tau)^2} \nonumber \\
          &\quad  - B\frac{\sin(t-\tau)}{t-\tau} - B\frac{\cos(t-\tau)}{(t-\tau)^2} \ , \label{eq:nudot}
\end{align}
where $A$ and $B$ are given by
\begin{align}
A & \equiv \left( \dot\nu_{\text{ref}} (t_{\text{ref}}-\tau)+ \nu_{\text{ref}} \right) \cos(t_{\text{ref}}-\tau) \nonumber \\ &\phantom{\equiv} + \nu_{\text{ref}} ( t_{\text{ref}}-\tau) \sin(t_{\text{ref}}-\tau ), \nonumber \\
B & \equiv -\left( \dot\nu_{\text{ref}} (t_{\text{ref}}-\tau)  + \nu_{\text{ref}} \right) \sin(t_{\text{ref}}-\tau) \nonumber \\ &\phantom{\equiv} + \nu_{\text{ref}} ( t_{\text{ref}}-\tau ) \cos(t_{\text{ref}}-\tau ).
\end{align}

For the late-time MD evolution we require $\dot{\omega}_k/\omega_k$.  There are three terms in the expression for $\omega_k^2(t)$: $k^2$, $a''/a$, and $a^2M^2_\mathrm{eff}$.  At late-time in the MD era the three terms scale as $t^0$, $t^{-2/3}$, and $t^{4/3}$, respectively, so at late time we will use $\omega_k^2(t)\simeq a^2(t)M^2_\mathrm{eff}$.  This leads to $\dot{\omega}_k/\omega_k = H + \dot{M}_\mathrm{eff}/M_\mathrm{eff}$.

At late time 
\begin{align}
M_\mathrm{eff} &= M & (\text{constant}\ M) \nonumber\phantom{.} \\
M_\mathrm{eff} &= M_g(\nu_*-\nu) & (\text{potentially-coupled}) \phantom{.}\nonumber \\
M_\mathrm{eff} &\simeq M\exp\left[(\nu-\nu_*)^2/2\epsilon^2\right] & (\text{kinetically-coupled}) .
\end{align}
For all cases we can write for the late-time MD era
\begin{equation}
\frac{\dot{\omega}_k}{2\omega_k} = \frac{1}{2} H - \delta\, \dot{\nu} ,
\end{equation}
where $H(t)$ is given by Eq.\ \eqref{eq:HMD}, $\dot{\nu}$ is given in Eq.\ \eqref{eq:nudot}, and for $\nu\ll \nu_*$, $\delta=0$, $\delta=1/2\nu_*$ and $\delta\simeq\nu_*/2\epsilon^2$ for the constant-$M$, potentially-coupled, and kinetically-coupled models, respectively. 

We are now positioned to find the expressions for $\alpha(t)$ and $\beta(t)$ in the MD era. First consider the accumulated phase $\Phi(t)$,
\begin{align}
\Phi(t) &\equiv \Phi_0 + \int_{t_0^k}^t \frac{\omega_k(t')}{a(t')}\, dt'  \nonumber \\ 
& \approx \Phi_0 + \int_{t_0^k}^{t_\mathrm{ref}} \frac{\omega_k(t')}{a(t')} \, dt' + M_\infty(t-t_\mathrm{ref})   \nonumber \\ 
& \equiv \Phi_\mathrm{ref} + M_\infty(t-t_\mathrm{ref}) ,
\end{align} 
where $\Phi_0$ is the accumulated phase at $t_0^k$, $\Phi_{\text{ref}}$ \emph{is evaluated in the full numerical evolution}, and $M_\infty = M_g\nu_*$ in the potentially-coupled model and $M_\infty = M e^{\nu_*^2/2\epsilon^2}$ in the kinetically-coupled model. We have assumed here that $|\nu| \ll \nu_*$.

Supposing that we have obtained $\alpha^{(n)}_{k,\text{ref}} = \alpha^{(n)}_{k}(t_{\text{ref}})$ and $\beta^{(n)}_{k,\text{ref}} = \beta^{(n)}_{k}(t_{\text{ref}})$ (e.g., by the methods of the previous subsection), we may now use our MD-era expressions to extend these to later times $t>t_{\text{ref}}$ by iterating (for $n\geq0$)
\begin{widetext}
\begin{align}
\beta^{(2n+1)}_k(t) &= \beta^{(2n+1)}_{k,\text{ref}}+\frac{1}{2} e^{-2i\Phi_\text{ref}} \int_{t_\text{ref}}^t \left[H(t_1)-2\delta\, \dot{\nu}(t_1)\right]\alpha^{(2n)}_k(t_1)\, e^{-2iM_\infty(t_1-t_\text{ref})}\ dt_1 \label{eq:beta_late}\\
\alpha^{(2n+2)}_k(t) &= \alpha^{(2n+2)}_{k,\text{ref}}+\frac{1}{2} e^{+2i\Phi_\text{ref}} \int_{t_\text{ref}}^t \left[H(t_1)-2\delta\, \dot{\nu}(t_1)\right]\beta_k^{(2n+1)}(t_1)\, e^{+2iM_\infty(t_1-t_\text{ref})}\ dt_1.
\label{eq:alpha_late}
\end{align}
\end{widetext}
In our actual numerical implementation for the alternative solution method based on $\alpha$ and $\beta$, we utilize the iterated method implicit in Eq.\ \eqref{eq:C} without change for $t\leq t_{\text{ref}}$, but for $t > t_{\text{ref}}$ we obtain $\alpha_k^{(n)}$ and $\beta^{(n)}_k$ with the iterated method implicit in Eqs.\ \eqref{eq:beta_late} and \eqref{eq:alpha_late}.

One could of course also use a hybrid method in which the full seven-dimensional system of odes is integrated through any tachyonic regions, and the iterative solution method is used to evolve the solution forward to very late time starting from some time after the tachyonic phase ends. We never find this necessary.

We develop an analytical understanding of the late-time asymptotic behavior of $\bks$ in Appendix \ref{app:late_time_asymptotics}.

\subsection{Parameters scanned and breakdown of methods employed \label{sec:scan}}

For the kinetically-coupled case, we have completed a scan over the parameters $(M,\epsilon)$ at fixed $\nu_* = 0.8$ in the ranges $M \in [ 0.2, 7]$, $\epsilon > 0.2-0.25$ (the smallest computationally feasible lower cutoff here is somewhat dependent on the choice of $M$) with the largest $\epsilon$ investigated being effectively infinite (specifically, $10^{30}$) to allow us to compare our numerical investigations to the simple model of \secref{simplemodel} in the context of the chaotic inflation background, which has been previously investigated in Ref.\ \cite{Kuzmin:1998kk}. Additionally, we investigated the $M$ dependence of the spectra up to $M=9$ at fixed $\epsilon = 0.6$. For the potentially-coupled case, we have completed a scan over $M_g$ in the range $M_g \in [ 1.2, 3.7\eten{4}]$.  In Table \ref{tab:which_method}, we summarize explicitly which numerical method was used in obtaining the various spectra we present in the next section; the general rule-of-thumb is that we use the iterative solution method wherever possible, but directly solve the full seven-dimensional set of coupled odes whenever a tachyonic phase is present.

Where we choose to present results for $\Omega h^2$, the particle spectrum extracted as detailed above is integrated over $k$ per Eq.\ \eqref{eq:relic_abundance} in the maximal numerically sampled range $k \in [10^{-20}, 20]$\footnote{We occasionally work outside this range if necessary to capture a relevant feature: for example, at $M=0.2$ and $\epsilon = 0.8$ for the kinetically-coupled case, $\nck$ peaks around $k \sim 10^{-19}$, and we wish to capture this behavior fully to get an accurate particle number, so we extend the range of integration down to $k=10^{-22}$.} to obtain a relic abundance $\Omega h^2$, provided that the numerical results we have extracted indicate that this integral converges in the infrared ($M<1$ results at large $\epsilon$ in the kinetically-coupled case are problematic in this regard) and subject to modification in the upper limit of the integral to avoid obvious numerical artifacts which enter at large $k$ (i.e., very small $|\beta_k|^2$).  

\renewcommand*\arraystretch{1.5}
\begin{table*}[t]
\begin{ruledtabular}
\caption{\label{tab:which_method} A summary of which numerical method has been used to obtain the late-time asymptotic value of $|\beta_k|^2$, organized by figure number. The end-time $t_f$ used in these solutions varies and is taken to be ``late enough'' in the sense that we can reliably extract the asymptotic value of $|\beta_k|^2$. This can be as short as $t_f = 100$ for cases which quickly go to their late-time asymptotic value (e.g., $M = 2$ at small $\epsilon$), or as long as $t_f \sim 4.5 \eten{5}$ for particularly stubborn cases which take a very long time to ``ring down'' (e.g., $M>4$ at very large or very small $k$); a full listing of the values of $t_f$ used would not be enlightening. In this table ``7D'' refers to the straightforward solution of the full seven-dimensional system of equations for the background fields and the mode equation as described in Sec.\ \ref{sec:seven}, and ``Iter.'' refers to the iterated solution method for $\alpha$ and $\beta$ described in Sec.\ \ref{sec:integro}, taken where necessary with the late-time MD-era modification described in Sec.\ \ref{sec:MD}.}

\begin{tabular}{lll}
Figure Number and Description					&	Parameter Range or Identifier			&	Method Employed \\ \hline
Fig.\ \ref{fig:eps_inf}	($\epsilon \rightarrow \infty$)	& $k \gtrsim 0.5$ and $M \geq 1.5$; or $M \geq 5.0$ 	& 	Iter. 	\\
										&	$k \lesssim 0.5$ and / or $M < 1.5$		& 	7D 	\\\hline
Fig.\ \ref{fig:potential} (potentially-coupled)		& 	---								&	7D 	\\ \hline
Fig.\ \ref{fig:eps_0.60} (left plot, $\epsilon = 0.6$)	&	$k \gtrsim 0.1$ and $M\geq 0.6$		& 	Iter. 	\\
										&	All others							& 	7D	\\ \hline
Fig.\ \ref{fig:eps_0.60} (right plot, $\epsilon = 0.6$, large $M$)	%
										& 	All, solid lines						&	Iter.	\\
										& 	Selected, open circles				&	7D	\\ \hline
Fig.\ \ref{fig:M}, $M = 2.0$ (upper plot)			&	$k \gtrsim 0.6$ and $\epsilon \geq 0.4$	&	Iter. 	\\
										&	$k \lesssim 0.6$ and $\epsilon \geq 0.4$	&	7D	\\
										&	$k \gtrsim 0.4$ and $\epsilon = 0.3$		&	Iter.	\\
										&	$k \lesssim 0.4$ and $\epsilon = 0.3$	&	7D	\\
										&	$k \gtrsim 6\eten{-2}$ and $\epsilon = 0.2$   &	Iter.	\\
										&	$k \lesssim 6\eten{-2}$ and $\epsilon = 0.2$ &	7D	\\ \hline
Fig.\ \ref{fig:M}, $M = 4.0$ (both lower plots)		&	All, solid lines						& 	Iter.	\\
										& 	Selected, open circles				&	7D	\\			
\end{tabular}
\end{ruledtabular}
\end{table*}

\section{Results \label{sec:results}}

In this section we present the numerical results for particle species irruption either in the model of inflaton--irrupton coupling through the potential term (the model discussed in \secref{potentialirruption}) or in the model of inflaton--irrupton coupling through the kinetic term (the model discussed in \secref{kineticirruption}).   The most general models of these types have either two or three free functions: an inflaton potential $V(\phi)$, an inflaton--irrupton potential $U(\chi,\phi)$, and for the kinetically-coupled case only, a function $f(\phi)$ that describes the coupling of the inflaton to the kinetic term of the irrupton.  For $V(\phi)$ we choose the simple inflaton potential $V(\phi)=\frac{1}{2}m^2\phi^2$.  In the potentially-coupled case, for $U(\chi,\phi)$ we choose the same potential as in \secref{potentialirruption}: $U(\chi,\phi)=\frac{1}{2}g^2(\phi-\phi_*)^2\chi^2 \equiv \frac{1}{2}M_g^2(\nu-\nu_*)^2\chi^2$; for the kinetically-coupled case, for $U(\chi,\phi)$ we choose the same potential as in \secref{kineticirruption}: $U(\chi,\phi)=\frac{1}{2}M^2\chi^2$.  Finally, for $f(\phi)$ in the kinetically-coupled case, we choose $f(\phi) = -(\phi-\phi_*)^2/2\epsilon^2M_{\text{Pl}}^2 \equiv -(\nu-\nu_*)^2/2\epsilon^2$.  Thus, the two or three free functions are described, respectively, in terms of either three parameters: $\left\{\ m,\ M_g,\ \nu_*\ \right\}$, or four parameters: $\left\{\ m,\ M,\ \nu_*,\ \epsilon\ \right\}$.   From background radiation measurements $m\sim 10^{13}$ GeV.  We will express $M$ and $M_g$ in units of $m$ and $t$ in units of $m^{-1}$.  For the chaotic inflation model we consider, ranges of $\phi$ in the observable region of the background radiation are $0.2\lesssim \phi/M_{\text{Pl}}\equiv \nu  \lesssim 3$. So we will make the choice $\phi_*/M_{\text{Pl}}\equiv\nu_*=0.8$ for all numerical results presented.  With these choices for $m$ and $\nu_*$, the one parameter we will vary for the potentially-coupled irrupton is $\left\{\ M_g\ \right\}$ (in units of $m$), and the two parameters we will vary for the kinetically-coupled case are $\left\{\ M,\ \epsilon\ \right\}$ (again, $M$ in units of $m$ and $\epsilon$ dimensionless).  

\begin{figure*}
\includegraphics[width = 0.71\textwidth]{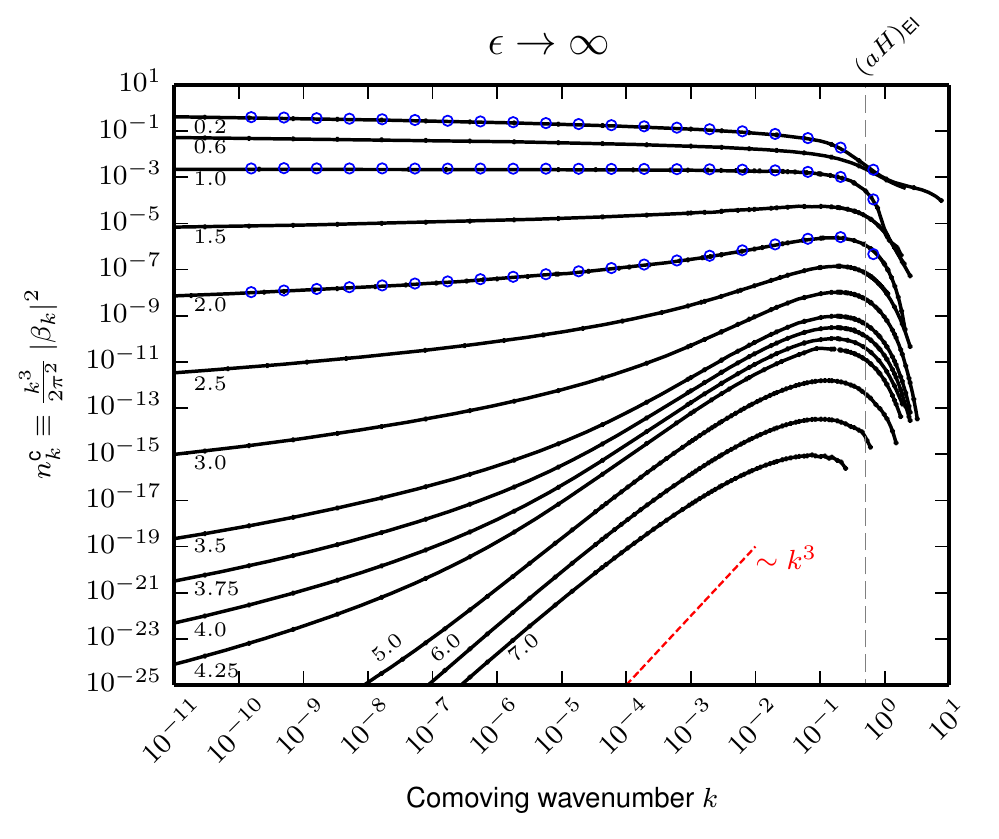}
\caption{ \label{fig:eps_inf} (Color online) Comoving particle density spectra $\nck$ as a function of comoving $k$ (in units of $m$) for a minimally-coupled scalar field with constant mass $M$ (as annotated; also in units of $m$) as discussed in \secref{simplemodel}, but using the chaotic inflation model to describe the evolution of the background (i.e., the $\epsilon \rightarrow \infty$ limit of the kinetically-coupled irrupton of \secref{kineticirruption}).  Different curves are for various heavy particle masses $M$. For comparison, the thin (red) dashed line indicates the scaling of $\nck$ with $k$ if the scaling were $k^3$ (constant $|\beta_k|^2$). The (blue) circles on the lines for $M = 0.2,\ 1$, and $2$ are sampled from spectra in Ref.\ \cite{Kuzmin:1998kk}. The vertical dashed line denotes $k = a_{\text{EI}}H_{\text{EI}}$. See Table \ref{tab:which_method} for a summary of which numerical methods were applied to obtain these results.}
\end{figure*}

We note that the $\epsilon \rightarrow \infty$ limit of the kinetically-coupled irrupton is a minimally-coupled scalar field of mass $M$ in a background chaotic-inflation model.   This model was described in \secref{simplemodel} (but for de Sitter space with constant $H$, not in chaotic inflation with a slowly evolving $H$, as for the numerical results of this section).  Our result for the comoving number density of produced particles as a function of $M$ in this limit is shown in Fig.\ \ref{fig:eps_inf}. This case has been previously considered in the literature, e.g., Ref.\ \cite{Kuzmin:1998kk}, for the same inflationary regime we utilize, and where available we reproduce their results well (the blue circles in the figure are sample points taken from Ref.\ \cite{Kuzmin:1998kk}).  The qualitative behavior of the curves in Fig.\ \ref{fig:eps_inf} are different for different $M$.  For $M<1$ there is a slow growth of $\nck$ in the infrared corresponding to $\bks \sim k^{-3-x}$ for some $x>0$ leading to an infrared (IR) divergence in the number of particles produced.\footnote{Presumably, this IR divergence is cut off if inflation has a finite duration.} For $M>1$ the spectra decrease in the far infrared corresponding to $\bks \sim k^{-3+x}$ for some $x>0$. For $M=1$, the spectrum of $\nck$ is consistent with being exactly flat in the IR. There is a sharp drop in the spectra for $k\gtrsim1$ for $M>1$; the drop is more gradual for small $M$, and a small bump is even evident in the spectrum for $M=0.2$ around $k \sim 0.5$.  These features will be explained below.

\begin{figure*}
\includegraphics[width = 0.71\textwidth]{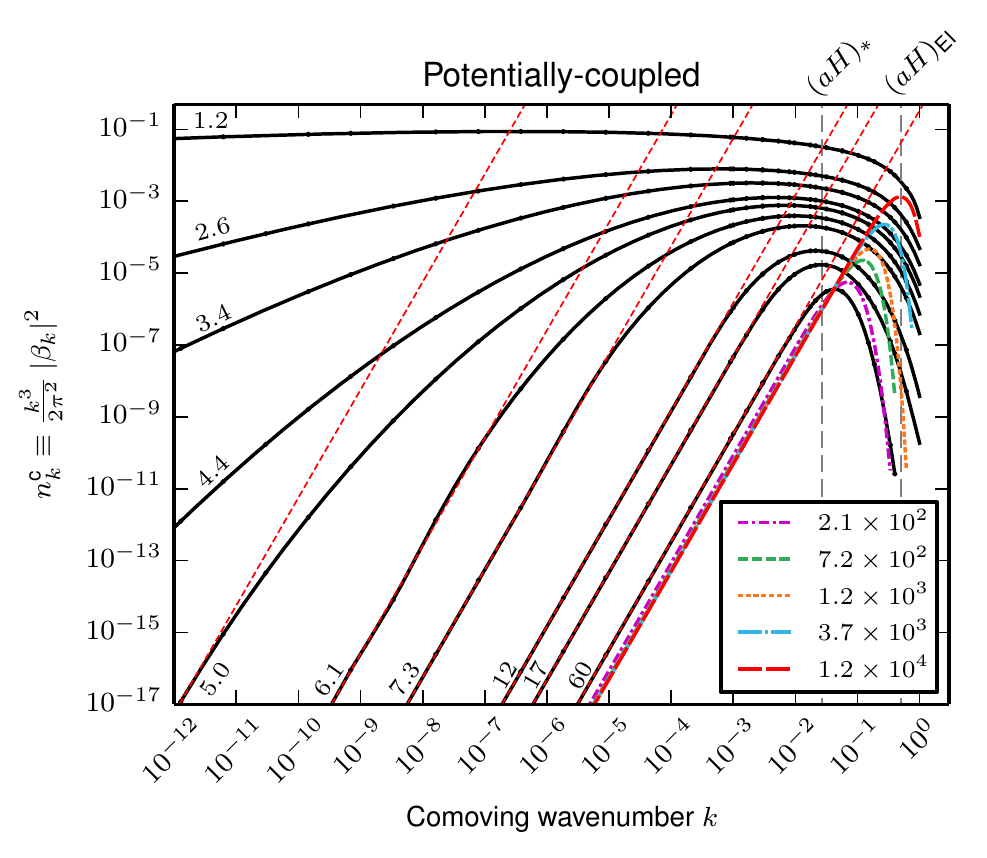}
\caption{ \label{fig:potential} (Color online) Comoving particle density spectra $\nck$ for the potentially-coupled irrupton as a function of comoving $k$ (in units of $m$) for various parameter choices $M_g \equiv g M_{\text{Pl}} = 1.22\eten{6} g$ (as annotated; in units of $m$). The choice of $M_g$ is indicated either by the black numbers annotating the lines, or by the numbers in the legend labeling the differently-styled (colored) lines. These latter cases are plotted differently to aid the reader visually, and also because these large-$M_g$ spectra show qualitatively different behavior, which is discussed in the text. The thin (red) dashed lines illustrate an $\nck$ scaling proportional to $k^3$ (constant $|\beta_k|^2$). The vertical dashed lines denote $k = a_*H_*$ and $k = a_{\text{EI}}H_{\text{EI}}$, where $*$ denotes the values at the instant where $\phi = \phi_*$. See Table \ref{tab:which_method} for a summary of which numerical methods were applied to obtain these results.}
\end{figure*}

\begin{figure*}
\includegraphics[width = 0.495\textwidth]{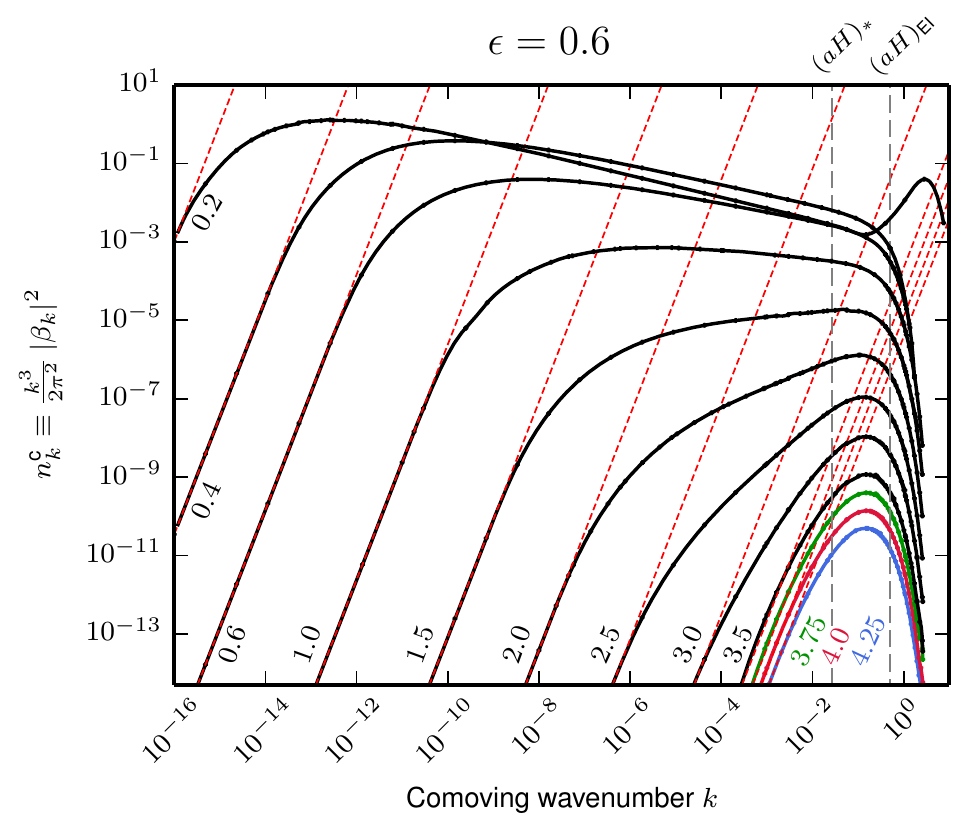}
\includegraphics[width = 0.495\textwidth]{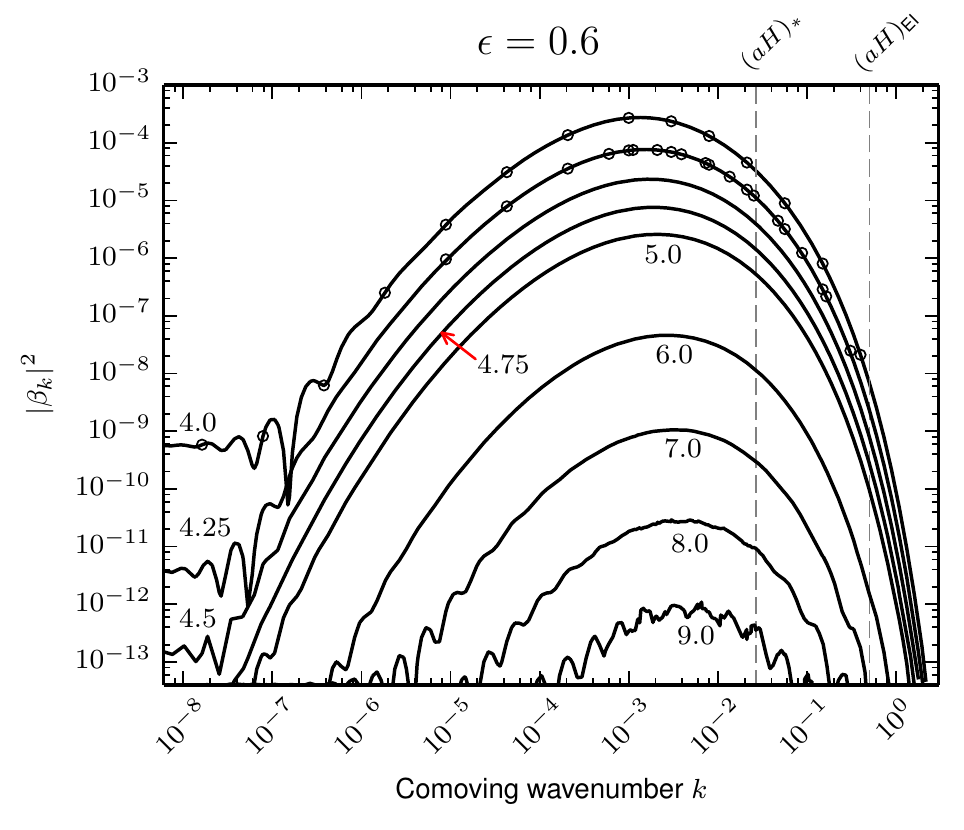}
\caption{ \label{fig:eps_0.60}  (Color online) Comoving particle density spectra $\nck$ (left plot) and the late-time Bogoliubov coefficient (i.e., mode-$\bm{k}$ occupancy number) $|\beta_k|^2$ (right plot) for the kinetically-coupled irrupton for as a function of comoving $k$ (in units of $m$) for various heavy-particle masses $M$ (as annotated; also in units of $m$) with fixed $\epsilon = 0.6$. We choose to present $|\beta_k|^2$ rather than $\nck$ in the right plot for greater clarity in these very-large-$M$ cases. In the left plot, the thin (red) dashed lines illustrate a scaling proportional to $k^3$ (constant $|\beta_k|^2$). The vertical dashed lines denote $k = a_*H_*$ and $k = a_{\text{EI}}H_{\text{EI}}$, where $*$ denotes the values at the instant where $\phi = \phi_*$. See Table \ref{tab:which_method} for a summary of which numerical methods were applied to obtain these results, and the meaning of the open circles and solid lines in the right plot.} 
\end{figure*}

\begin{figure*}[p]
\includegraphics[width = 0.71 \textwidth]{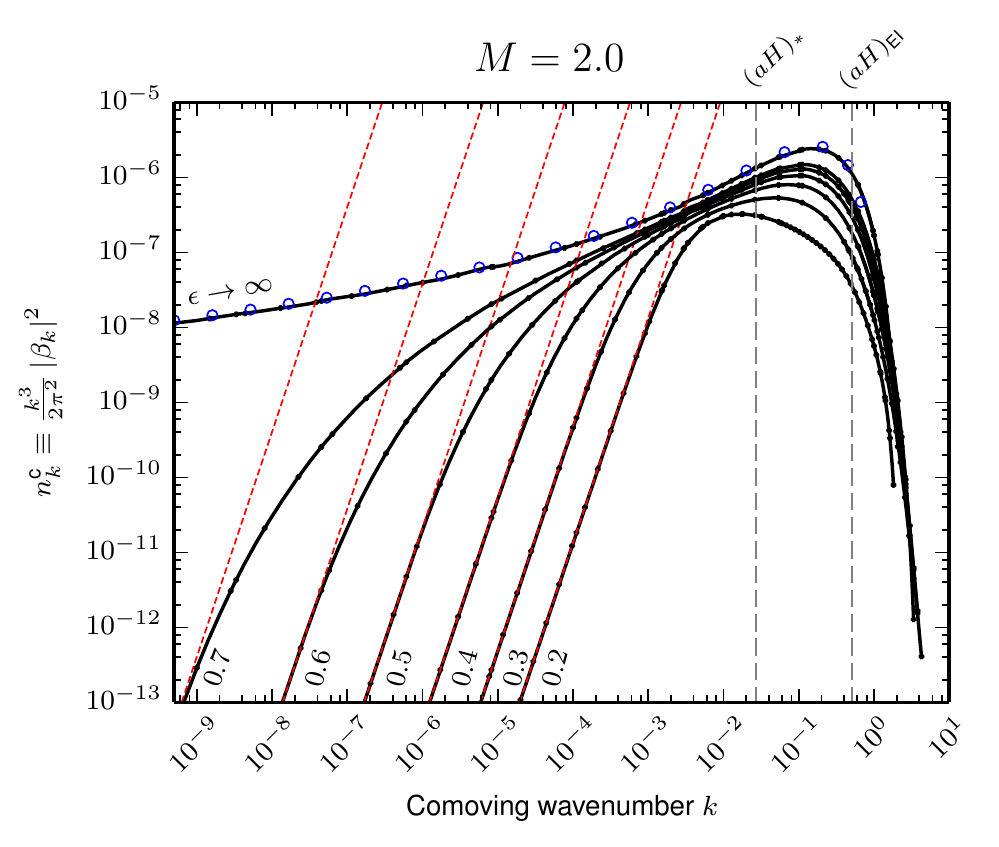}\\
\includegraphics[width = 0.49 \textwidth]{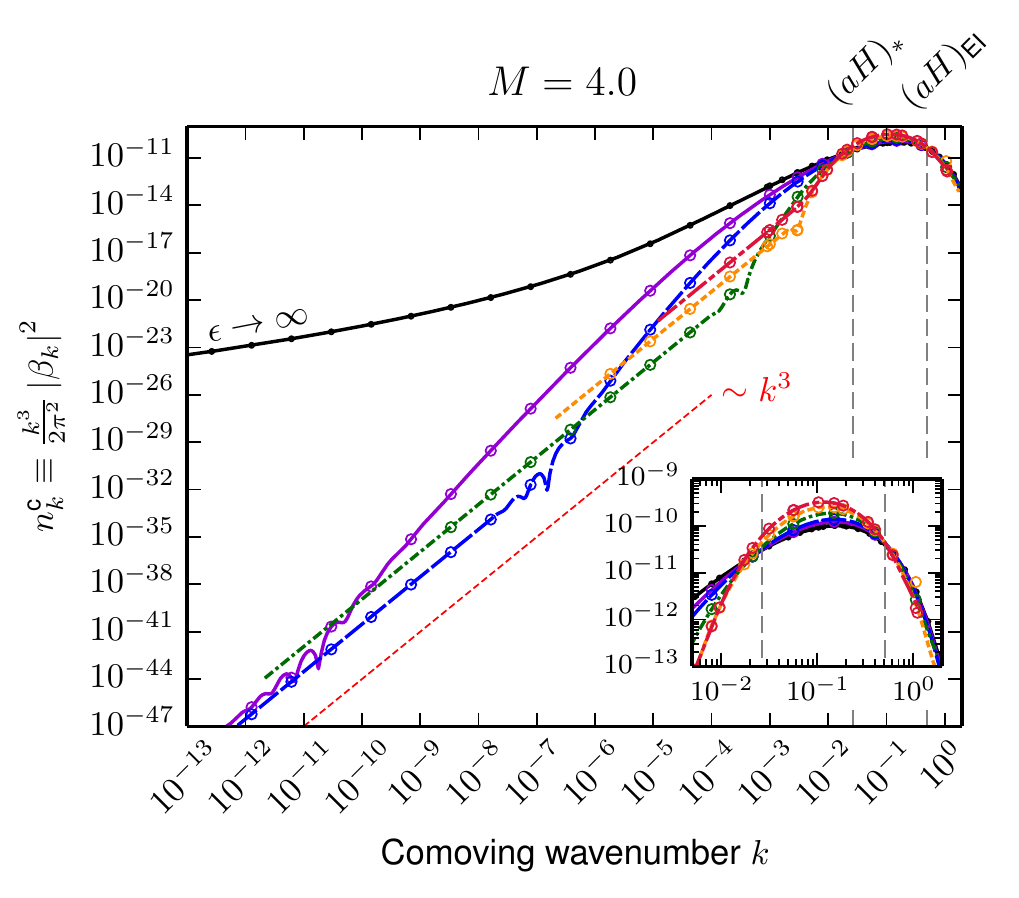}
\includegraphics[width = 0.49 \textwidth]{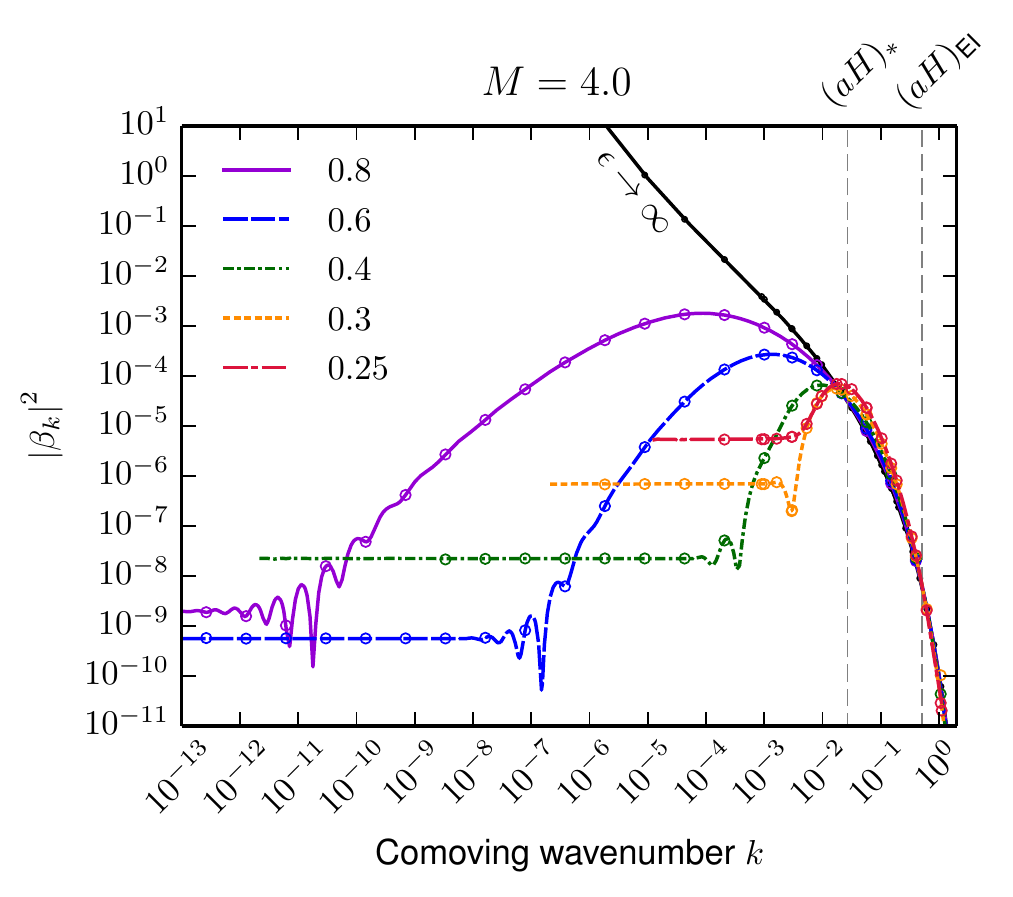}
\caption{ \label{fig:M} (Color online)  Comoving particle density spectra $\nck$ as a function of comoving $k$ (in units of $m$) for the kinetically-coupled irrupton, for various choices of $\epsilon$ with fixed $M=2$ (upper plot) and $M=4$ (lower-left plot). The thin (red) dashed lines illustrate a scaling proportional to $k^3$ (constant $|\beta_k|^2$). The (blue) circles on the line for $\epsilon \rightarrow \infty$ in the upper plot are sampled from a spectrum in Ref.\ \cite{Kuzmin:1998kk}. The inset in the lower-left plot shows detail near the peak in the spectra. Also shown for the case of $M=4$ are the values of $\bks$ (lower-right plot).  The vertical dashed lines denote $k = a_*H_*$ and $k = a_{\text{EI}}H_{\text{EI}}$, where $*$ denotes the values at the instant where $\phi = \phi_*$.  See Table \ref{tab:which_method} for a summary of which numerical methods were applied to obtain these results, and the meaning of the open circles and solid lines in the lower plots. } 
\end{figure*}

Representative particle spectra for the potentially-coupled irrupton as a function of $M_g$ are given in Fig.\ \ref{fig:potential}, while those for the kinetically-coupled irrupton as a function of $M$ are given for fixed $\epsilon = 0.6$ in Fig.\ \ref{fig:eps_0.60}, and as a function of $\epsilon$ for two fixed values of $M$ ($M=2$ and $M=4$) in Fig.\ \ref{fig:M}.   These figures capture all the important features we have observed in our numerical work.   Indeed, for the specific potential and kinetic couplings we have considered for the irrupton, many interesting features are present. Before explaining the causes of these features in the spectra, we qualitatively describe the scalings of the spectra with $k$ and $M_g$ for the potentially-coupled irrupton, or with $k$, $M$, and $\epsilon$ for the kinetically-coupled irrupton.  

Consider first the potentially-coupled irruption spectra shown in Fig.\ \ref{fig:potential}. In all these cases with $M_g \gtrsim 5$, in the infrared region the spectra scale as $\nck \sim k^{3}$, which implies $\bks \sim k^0$. Provided also that $M_g\lesssim 10$, this behavior is valid until a threshold value $k=k_*$, beyond which the spectra increase more slowly than $k^3$, but have no simple power-law scaling. The spectra then peak, more sharply for larger $M_g$, before showing a steep drop-off in the UV region; for larger $M_g$, the spectra instead roll off exponentially fast directly from the $\nck \sim k^3$ regime.  For smaller $M_g$ the IR behavior does not enter an approximate $k^3$ scaling regime (at least for the numerically sampled range of $k \gtrsim 10^{-16}$) and based on arguments we present below we do not necessarily expect that such a regime would exist for smaller $k$ when $M_g \lesssim 3$. We note that for $M_g \lesssim 60$ in the results we have presented there is a strict ordering in the size of $\nck$: at fixed $k$, $\nck$ is smaller the larger $M_g$ becomes. This behavior is modified at very large $M_g\gtrsim 200$, where the IR spectrum stops decreasing with increasing $M_g$ and instead approaches from above the limit $\nck = k^3/2\pi^2$ (which implies $\bks = 1$); on the other hand, the spectrum near the peak begins to \emph{increase} again with increasing $M_g$ approaching the same limit from below, while the peak itself shifts further to the UV.

For the kinetically-coupled irrupton, first consider the case of fixed $\epsilon$  and fairly small $M$ in the left plot of Fig.\ \ref{fig:eps_0.60}.  In the infrared region the spectra scale as $\nck \sim k^{3}$, which implies $\bks \sim k^0$.  This scaling behavior is valid up until an ``elbow'' in the spectra at a value of $k$ that depends on $M$ (and $\epsilon$). After the elbow, for increasing $k$ there is a slow decrease in $\nck$ with $k$ for $M\lesssim 1$, or a slow increase in $\nck$ for $M\gtrsim1$. For $M\lesssim 2.5$, the spectra scale as a power law in $k$ in this intermediate-$k$ regime, with an $M$-dependent power which is less than 3. This behavior continues until $k\sim 10^{-1}$, and thereafter there is a steep decrease in $\nck$. There is, however, a pronounced bump in the $\nck$ spectra at small $M$ ($M=0.2$) and $k\sim 1$; this does not occur for $M\gtrsim 0.2$.  Just as in the model of a minimally-coupled scalar field with constant $M$, there is a general trend of decreasing particle production with increasing $M$.

Consider now the impact of varying $\epsilon$ when $M$ is fairly small, starting with the case of $M=2$ shown in the upper plot in Fig.\ \ref{fig:M}. First of all, we note the good agreement of the $\epsilon \rightarrow \infty$ limit of our results with those from Ref.\ \cite{Kuzmin:1998kk} as shown by the blue circles. The next feature to note is that, just as in Fig.\ \ref{fig:eps_0.60}, in the infrared $\nck\propto k^3$ for finite $\epsilon$.  Again, at some value of $k$ there is an ``elbow'' after which the spectrum grows (for $\epsilon \gtrsim 0.4$, as a power law) more slowly than $k^3$, and there is again a steep drop in $\nck$ for large $k$.  As $\epsilon$ decreases the spectra become more peaked, and decrease in magnitude.

Finally consider the large-$M$ cases: $M=4$ in the lower plots of Fig.\ \ref{fig:M}, and large $M$ at fixed $\epsilon = 0.6$ in the right plot of Fig.\ \ref{fig:eps_0.60}. For these values of $M$ the $\nck$ spectra scale as $k^3$ in the deep infrared (constant $\bks$), show oscillations (provided that $\epsilon$ is not too small) when transitioning from this behavior to a scaling steeper than $k^3$ in an intermediate region of $k$ (corresponding to a bump in $\bks$), then peak at $k\sim 10^{-1}$ and finally decrease rapidly at larger $k$.  Particularly in the vicinity of the peak in the $\nck$ spectra in Fig.\ \ref{fig:M} (see the inset plot), there is only a weak dependence on $\epsilon$.

Now that we have described the spectra, we turn to an explanation for their behavior.   The presence or absence of the tachyonic phase, along with its duration, is crucial for understanding the spectra. For modes which can run tachyonic, the behavior of $\omega_k^2$ clearly dictates the duration of the tachyonic phase and hence the amount of particle production that can occur as the exponential increase in the mode functions during this phase is the dominant effect. For modes which never run tachyonic, we will show that a good understanding of the behavior of the adiabaticity parameters leads to a good understanding of the characteristics of the spectrum.

\begin{figure*}
\includegraphics[width = \textwidth]{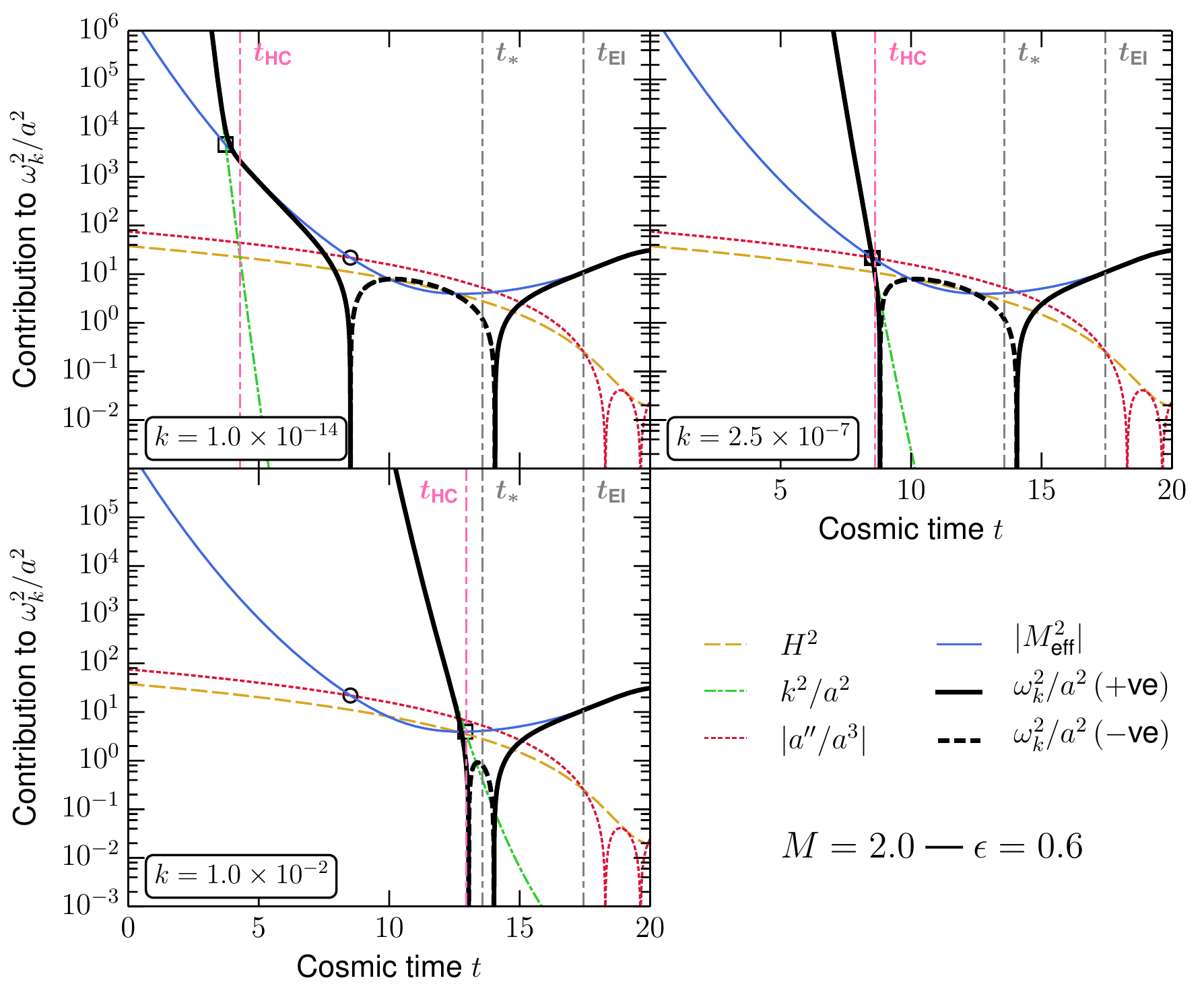}
\caption{ \label{fig:mass_plot_1} (Color online) The oscillation frequency $\omega_k^2 / a^2$ and its three contributions for the kinetically-coupled irrupton [see Eq.\ (\ref{eq:omegak2})] plotted as a function of cosmic time. The open circle in each plot indicates the point where two of the contributions to $\omega_k^2 / a^2$, namely $M_{\text{eff}}^2$ and $|a''/a^3|$, are approximately equal. The point where $k^2/a^2$ and $M_{\text{eff}}^2$ first intersect is denoted by the open square.  The value of $k$ at the ``elbow'' break-point, $k_*$, is the value of $k$ for which $k_*^2/a^2 \approx |a''/a^3| \approx M_\mathrm{eff}^2$; i.e., the open circle and the square coincide.  From the figure we see that this occurs for $k=k_*=2.5\times10^{-7}$.  For $k \ll k_*$ (illustrated by the case $k=10^{-14}$) the onset of the tachyonic phase is determined by $M_\textrm{eff}^2$ and independent of $k$.  Since the duration of the tachyonic phase is independent of $k$, $|\beta_k|^2$ should be independent of $k$, and $\nck\propto k^3$.  For $k \gg k_*$ (illustrated by the case $k=10^{-2}$) the onset of the tachyonic phase is determined by $k^2/a^2$ and the duration of the tachyonic phase is reduced, which implies $|\beta_k|^2$ will decrease with $k$ and $\nck$ will grow more slowly than $k^3$. It is clear that for even larger $k$, there may be no tachyonic phase at all, which exponentially suppresses the particle number produced. Note that larger values of $\epsilon$ flatten $M^2_{\text{eff}}$, giving a longer maximum duration of the tachyonic phase, while larger values of the $M$ move the minimum of $M^2_{\text{eff}}$ upward, which suppresses particle production by reducing the duration of the tachyonic phase (which argument holds until such a phase ceases to exist; see \figref{mass_plot_2} for further consideration of this case).} 
\end{figure*}

\begin{figure*}
\includegraphics[width = \textwidth]{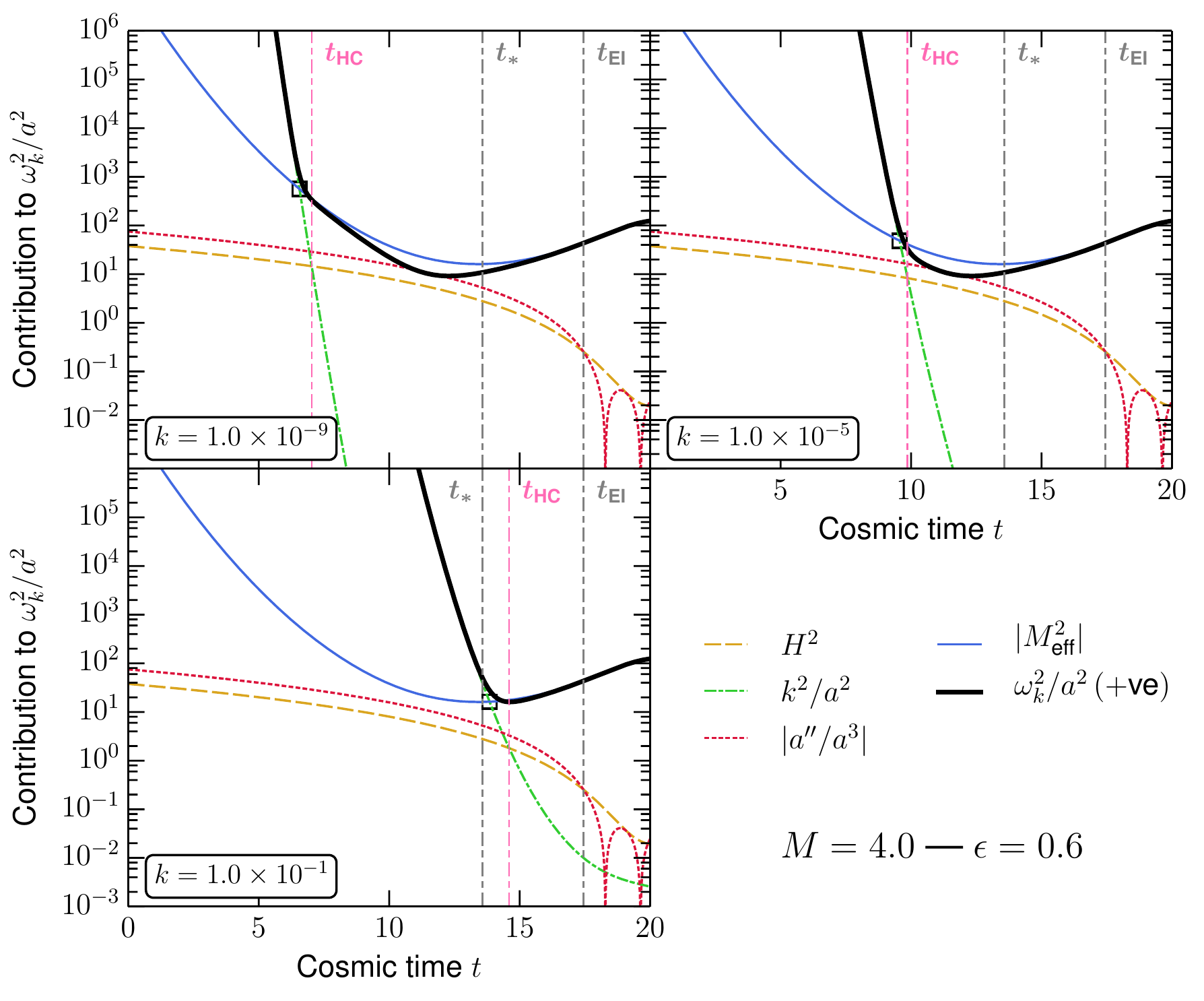}
\caption{ \label{fig:mass_plot_2} (Color online) As for \figref{mass_plot_1}, except this series of plots of the various contributions to $\omega_k^2/a^2$ for varying $k$ is shown for a case where no tachyonic phase is present. Nevertheless, the minimum value of $\omega_k^2/a^2$ still becomes $k$-independent at small $k$, which leads to a $\nck \sim k^3$ infrared behavior in the particle spectrum. As $k$ becomes very large (lower plot), the minimum value of $\omega_k^2/a^2$ is clearly shifted upward, which leads to the large-$k$ suppression in the spectra. }
\end{figure*}

We begin with a discussion of the IR behavior of the spectra. In the case where $M$ is constant (i.e., the $\epsilon\rightarrow\infty$ limit for the kinetically-coupled irrupton) and not much larger than $H$ (Fig.\ \ref{fig:eps_inf}), a mode of comoving momentum $k$ enters the period of tachyonicity almost immediately after crossing outside the comoving Hubble radius $R_H^{\text{c}}(t) = \lb[a(t)H(t)\rb]^{-1}$ since at this time, the sum $M^2 + k^2/a^2$ becomes comparable to the $-a''/a^3$ term in Eq.\ \eqref{eq:deSitter_omega} for $\omega_k^2/a^2$. (Note that $k^2/a^2$ is a rapidly falling function of $t$ since $a$ is growing exponentially; at the equality point, all three terms are of roughly the same size in our parameter region of interest. After equality, $k/a$ rapidly becomes completely negligible for all subsequent evolution. See Figs.\ \ref{fig:mass_plot_1} and \ref{fig:mass_plot_2}.) This means that the Hubble-crossing time is a reliable indicator of the onset of the tachyonic phase, and that the exit time from the tachyonic phase is independent of the value of $k$; modes of smaller $k$ thus spend much longer in such a tachyonic phase than modes of larger $k$, which implies that $\bks$ grows without bound at small $k$, explaining the absence of a $\nck \sim k^3$ scaling regime in the infrared region for the constant-$M$ case. 

However, in the case of the potentially- or kinetically-coupled irruptons with a running $M_{\text{eff}}$, the situation is different. We discuss first the scaling with $k$ for the kinetically-coupled case. In \figref{mass_plot_1} we plot $\omega_k^2/a^2$ as given in Eq.\ (\ref{eq:omegak2}) along with the magnitude of the three terms contributing to it: $k^2/a^2$, $|a''/a^3|$, and $M_\textrm{eff}^2$.  The heavy dashed curve is where the given momentum mode is tachyonic, which requires $k^2/a^2+M_\textrm{eff}^2$ to be less than the magnitude of $a''/a^3$ (recall that during inflation $a''>0$).  In the far IR (illustrated by $k=10^{-14}$) $k$ is sufficiently small that $k^2/a^2$ drops below $M_\textrm{eff}^2$ early in the evolution and the onset of tachyonicity is determined by when $M_\textrm{eff}^2$ drops below $|a''/a^3|$, implying that the onset of tachyonicity no longer closely tracks Hubble-crossing (see also  the left plot of Fig.\ \ref{fig:beta_k_example}).  As $k$ increases, eventually it, and not $M_\textrm{eff}$, will determine the onset of the tachyonic phase.  Let us call the crossover point $k_*$.  The value of $k_*$ will be the value of $k$ for which $\ok^2/a^2 \approx k_*^2/a^2 \approx M_\textrm{eff}^2 \approx |a''/a^3|$.\footnote{The value of $k_*$ is exponentially sensitive to the value of $\epsilon$ since the flatness of $M^2_{\text{eff}}$, which is clearly (see \figref{mass_plot_1}) the most important factor for deciding where $M_{\text{eff}}^2$ and $k^2/a^2$ become of roughly the same size, is directly set by this parameter.} From \figref{mass_plot_1} we see that occurs at $k=2.5\times10^{-7}$ for $M=2,\epsilon=0.6$, which agrees well with the cross-over point (i.e., the ``elbow'') in the spectrum shown in the Fig.\ \ref{fig:eps_0.60}.  For larger $k>k_*$ (illustrated by $k=10^{-2}$) the duration of the tachyonic phase is shorter. 

To understand the resultant scaling with $k$, we make the crude approximation (reasonable for $\epsilon \gtrsim 0.4$) that $\omega_k^2/a^2$ is approximately constant during the tachyonic phase: $\omega_k^2/a^2 \approx - \Omega^2$ where $\Omega^2>0$ is a $k$-independent constant (see \figref{mass_plot_1}).  As $\omega_k^2$ is negative ($i\omega_k \in \mathbb{R}$), the expression for $|\beta_k|^2$ in Eq.\ \eqref{eq:nk} is inapplicable during the tachyonic phase itself; but the expression can be used immediately before and after the tachyonic phase when $\omega_k^2>0$.  Since the value of $\ok^2/a^2$ passes through zero, it will be equal to the \emph{same} small positive value at times both immediately before, and immediately after, the tachyonic phase (see \figref{mass_plot_1}). Consider then the ratio of the values of $|\beta_k|^2$ at those times, which we denote ``before'' and ``after'',
\begin{align}
\frac{|\beta_k|^2_\textrm{after}}{|\beta_k|^2_\textrm{before}} &\sim 
\frac{(\ok)_\textrm{after}}{(\ok)_\textrm{before}}
\frac{|\mu_k|^2_\textrm{after}}{|\mu_k|^2_\textrm{before}} \nonumber \\
&= \frac{a_\textrm{after}}{a_\textrm{before}}
\frac{|\mu_k|^2_\textrm{after}}{|\mu_k|^2_\textrm{before}},
\end{align}
where we have used that $| \dot\mu_k|^2 = |a^{-1} \mu_k'|^2 \propto |\mu_k|^2$ (which will be obvious from the form of the solution shown below) and have neglected the constant term in Eq.\ \eqref{eq:nk}.  We must now estimate $|\mu_k|_\textrm{after}/|\mu_k|_\textrm{before}$.  The mode function satisfies $\mu_k''+\ok^2\mu_k=0$, which under our assumption of constant negative $\omega_k^2/a^2 = -\Omega^2$ becomes $\mu_k'' - a^2\Omega^2\mu_k=0$. Since $H$ does not change very much over the short duration of the tachyonic phase, we will use the de Sitter result $\eta = -1/aH$, so the mode equation becomes $\eta^2\mu_k''- ( \Omega^2/H^2 ) \, \mu_k=0$, whose growing mode solution is $\mu_k=(-\eta)^{\lb(1-\sqrt{1+4\Omega^2/H^2}\rb)/2}=(aH)^{\lb(\sqrt{1+4\Omega^2/H^2}-1\rb)/2}$. We thus estimate that
\begin{align}
&\frac{|\beta_k|^2_\textrm{after}}{|\beta_k|^2_\textrm{before}} \sim \left(\frac{a_\textrm{after}}{a_\textrm{before}}\right)^\zeta \nonumber \\[2ex] &\qquad\qquad\qquad \text{where} \quad\ \zeta = \sqrt{1+4\Omega^2/H^2} > 0 .
\end{align}

Now in the deep IR ($k<k_*$), we have already noted that the values of $a$ when entering and leaving the tachyonic phase are independent of $k$, so we expect $|\beta_k|^2_\textrm{after}/|\beta_k|^2_\textrm{before}$ to be independent of $k$.  Since $\nck\sim k^3\left|\beta_k\right|^2$, $\nck$ will be proportional to $k^3$. 

But for $k>k_*$, the onset of the tachyonic phase is determined by $k^2/a^2=|a''/a^3|\sim 2H^2$.  So now $a_\textrm{before}\sim k$ whereas $a_{\textrm{after}}$ is still $k$-independent, so that $|\beta_k|^2_\textrm{after}/|\beta_k|^2_\textrm{before} \sim k^{-\zeta}$ which implies $\nck\sim k^{3-\zeta}$. This correctly captures the observed behavior of power-law scaling of $\nck$ with a power less than 3 above the ``elbow'' for the cases $M\lesssim 2.5$ (i.e., those with a tachyonic phase) in \figref{eps_0.60}.\footnote{A quantitative estimate of the scaling power $(3 - \zeta)$ requires an estimate for $\Omega^2/H^2$. A well-motivated approximation is to take it equal to the largest value of $| ( \omega_k^2/a^2)/H^2 |$ attained during the tachyonic phase. There is no simple closed-form expression for this value, but it can be easily (and accurately) estimated assuming that $a,\nu$ and $ \dot\nu$ take their slow-roll values. With such an estimate, we find scaling powers offset systematically high by about 0.15 compared to the values extracted from linear fits to the power-law section of the spectra above the elbow in \figref{eps_0.60} for all values of $M$ from $0.2$ to $2.5$  ($\epsilon = 0.6$). The offset notwithstanding, we capture the $M$-dependence very well. There is also some weak $\epsilon$-dependence in the power law above the elbow (see \figref{M}, upper plot) which arises from the curvature of $\omega_k^2/a^2$ during the tachyonic phase; our approach here is manifestly inadequate to capture this.}  Note, however, that this conclusion is predicated on a period of approximate constancy of $\omega_k^2/a^2$, which is not a very good approximation when $\epsilon$ is small; therefore, we would not expect a power-law intermediate regime for $\nck$ at small $\epsilon$, but rather a spectrum with a more constantly evolving slope, as is becoming evident in the $\epsilon=0.2,\, 0.3$ spectra at $M=2$ in the upper plot of Fig.\ \ref{fig:M}.

Eventually as $k$ increases the tachyonic phase disappears altogether and the evolution becomes more adiabatic, suppressing particle production even further.

The deep-IR scaling $\nck \sim k^3$ for the potentially-coupled irrupton arises for the exact same reason as for the kinetically-coupled irrupton provided $M_g$ is not too large: the onset of tachonicity is governed by the $k$-independent condition $M_{\text{eff}}^2 \approx a''/a^3$ for small enough $k$, provided that $M_g$ is sufficiently large. Since $k=10^{-12}$ is the smallest comoving momentum value shown in Fig.\ \ref{fig:potential}, we would only expect such a regime to be manifest in the results presented if $M_g \gtrsim 4.6$ (obtained from solving $M_{\text{eff}}^2 \approx k^2/a^2 \approx a''/a^3$ for $M_g$ at $k = 10^{-12}$). For smaller $M_g$, one would need to probe smaller $k$ to enter the $\nck \sim k^3$ regime; however, once $M_g \lesssim 3$ it is always the case that $M_{\text{eff}}^2 < a''/a^3$ for all $\nu > 0.8$ (even if $\nu$ is allowed to run much larger than 3; i.e., more $e$-foldings of inflation are allowed) and so $M_{\text{eff}}^2$ cannot come to dominate in setting the onset of tachyonicity for any value of $k$ and we would thus not necessarily expect an $\nck \sim k^3$ scaling regime to exist in such cases (i.e., we do not expect an irruption of limited duration; the production duration and characteristics are more similar to the constant-$M$ case). The intermediate scaling where $\nck$ increases more slowly than $k^3$, if present, once again occurs once the onset of tachyonicity becomes governed by when $k^2/a^2 \approx a''/a^3$. Although in the kinetically-coupled case this regime gave rise to a simple intermediate power-law scaling, such a regime does not manifest itself for the potentially-coupled case. 

On the other hand, at very large $M_g$, the duration of any possible tachyonic phase becomes too small to result in an exponentially large increase in the size of the mode function.\footnote{In the IR, we expect this regime to be entered roughly when the duration of the tachyonic phase is short enough that the mode function cannot increase in size by much more than an $e$-fold during this phase. We assume for the sake of this argument that $\omega_k^2/a^2 \approx -a''/a^3 \approx - \frac{8\pi}{3}\nu_*^2 \approx \text{constant}$ during the tachyonic phase and take the duration of the tachyonic phase, $\Delta t$, to be limited both before and after $t=t_*$ by where $M_{\text{eff}}^2=a''/a^3$ giving $M_g^2\dot{\nu}_*^2 (\Delta t / 2)^2\approx \frac{8\pi}{3}\nu_*^2$ which implies $\Delta t = 4\sqrt{2\pi/3}\ |\nu_*/\dot{\nu}_*|\, M_g^{-1}$. Then requiring no more than $n$ $e$-folds of increase for $\mu_k$ during the tachyonic phase demands we set $|\omega_k/a|\Delta t \leq n$ leading to $M_g \geq (16\pi/3)\ \nu_*^2\, /\, |\dot{\nu}_*| n \approx 67n$. Taking $n\approx1-2$ gives good qualitative agreement with the value of $M_g$ for which there is a quantitative behavior change in the IR in Fig.\ \ref{fig:potential}.\label{footnote:crossover} } Instead, it begins to looks more like an impulsive ``kick'' to the mode function localized very sharply around $t=t_*$ as $\omega_k^2/a^2$ very rapidly falls from a very large value to some small negative value before rapidly increasing again. This results in $\bks$ jumping from essentially zero to its late-time value almost instantaneously; we develop this argument further into a quantitative analytical prediction for the shape of the spectrum in Appendix \ref{app:potential_analytics}. However, since this process is still $k$-independent when $k$ is sufficiently small to not significantly modify $\omega_k^2/a^2$ at $t=t_*$, it also results in a $\nck \sim k^3$ scaling in the IR. This is quantitatively confirmed by Eq.\ \eqref{eq:spectrum_analytics} which shows that in the large-$M_g$ limit, $\nck \rightarrow k^3/2\pi^2$ ($|\beta_k|^2\rightarrow 1$) for small $k$.

\begin{figure*}
\includegraphics[width = \textwidth]{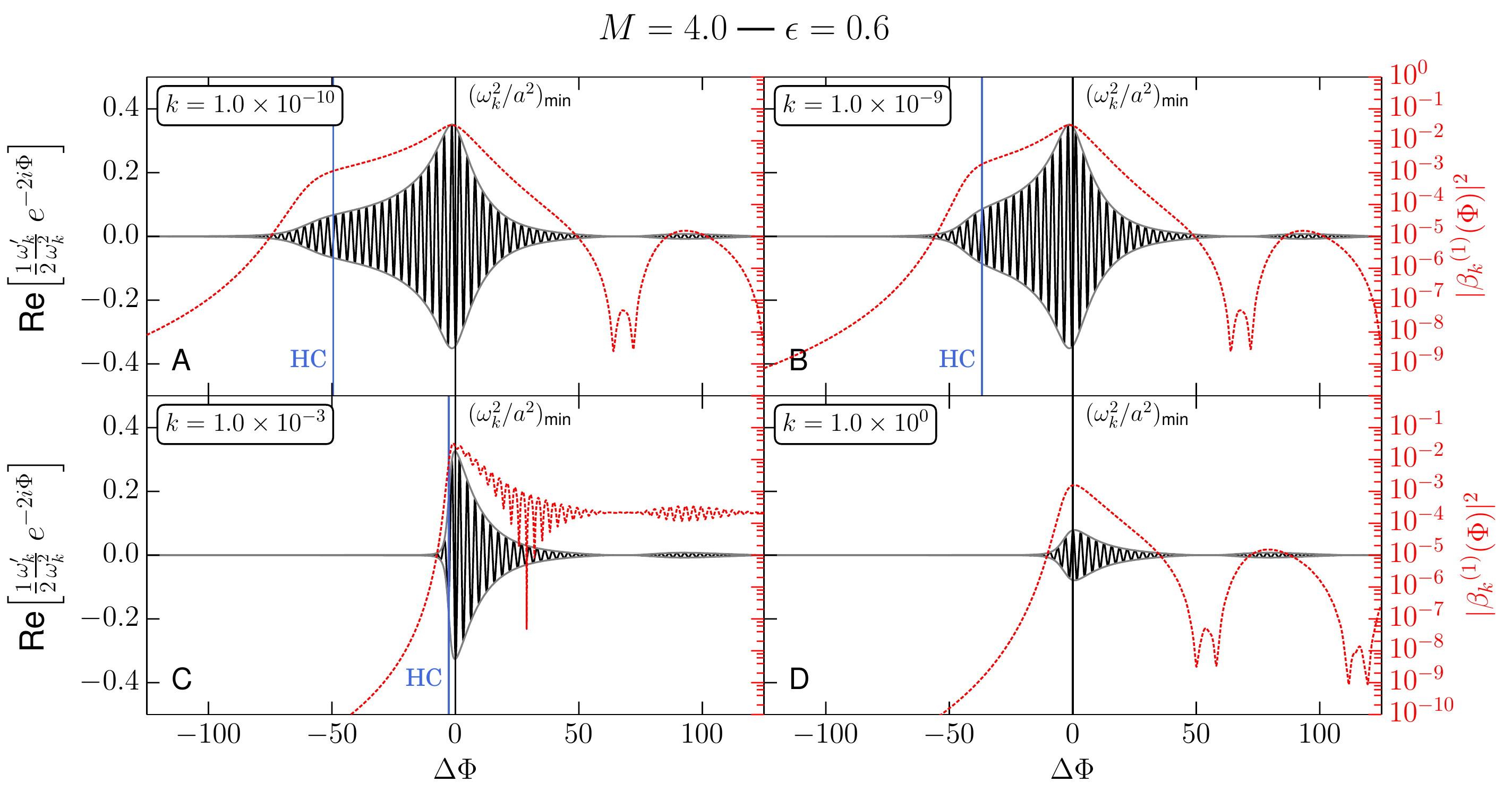}
\caption{ \label{fig:envelope} (Color online) The envelope $\pm (1/2) \lb| \omega_k' / \omega_k^2 \rb|$ (grey) bounding the real part of the rapidly oscillating integrand $(1/2) ( \omega_k' / \omega_k^2 ) e^{-2i\Phi}$ in Eq.\ \eqref{eq:beta1_phase} (black) for the first iterate ${\beta_k}^{(1)}$, for a variety of values of $k$ for the case $M=4$ and $\epsilon = 0.6$ for the kinetically-coupled irrupton, plotted as a function of the phase deviation $\Delta\Phi$ from when $\omega_k^2/a^2$ is minimized. The imaginary part of the integrand shows similar behavior; we omit it for clarity. We also show $|{\beta_k}^{(1)}(\Phi)|^2$ per Eq.\ \eqref{eq:beta1_phase} (right-scale on each set of axes; red dotted line) for illustrative purposes to indicate the impact of the shape of the envelope on the evolution of $|\beta_k|^2$. ``HC'' denotes Hubble-radius crossing; the mode with $k=1$ is always sub-Hubble-radius sized. The relevant comparisons we intend the reader to make from this series of plots are between plots A and B, and between plot C and plots A or D;  in particular, we caution the reader that the approximate order-of-magnitude equality of the value of $|\beta_k|^2$ at large positive $\Delta \Phi$ of $|\beta_k|^2$ in plot D (large $k$), and the values of the same quantity in plot A and B (small $k$), is a coincidental consequence of the values of $k$ we have chosen to display (see right plot of Fig.\ \ref{fig:eps_0.60}), so no deep significance should be attached to that approximate equality.}
\end{figure*}

Even if no tachyonic phase or impulsive kick is present owing to the relative sizes of the contributions to $\omega_k^2/a^2$, the $\nck \sim k^3$ infrared scaling can still obtain for the kinetically-coupled irrupton; for instance, the cases with $M>3$ in \figref{eps_0.60} or the $M=4$ cases in Fig.\ \ref{fig:M}. Although the explanation of this particular behavior does not require the full machinery we are about to develop, we will nevertheless have recourse to the same ideas to explain other features in the non-tachyonic cases, so we pause to carefully develop the arguments here. Firstly, we note that can obtain a good qualitative understanding of the behavior of the spectra for non-tachyonic cases by examining the lowest-order solution ${\beta_k}^{(1)}$ in the iterated-solution method, particularly in the form as given in Eq.\ \eqref{eq:beta1_phase}: we see that the size of the (square root of) the adiabaticity parameter $|\omega_k'/\omega_k^2|^2$ defines the envelope bounding the rapidly oscillating phase factor $e^{2i\Phi}$. Provided that this envelope varies only slowly as the phase advances by $\pi$, neighboring excursions in the positive and negative directions, of both the real and imaginary parts of the integrand, cancel nearly completely when integrated over. On the other hand, if the envelope varies rapidly as the phase advances by $\pi$, neighboring excursions cancel incompletely. Although in either case it is possible to obtain transient values of $|{\beta_k}^{(1)}|^2$ which are large, a non-zero late-time asymptotic value of $|{\beta_k}^{(1)}|^2$ occurs as a result of the accumulated incomplete cancellations between neighboring excursions over the full evolution out to $\Phi \rightarrow \infty$ (see Fig.\ \ref{fig:envelope}). To illustrate the point explicitly, consider a toy model in which $\omega_k'/\omega_k^2$ were an exact Gaussian with standard deviation $\sigma = \pi n$; we then find that the late-time asymptotic value of $|{\beta_k}^{(1)}|^2 \propto e^{-4\pi^2n^2}$. Clearly, the wider the Gaussian (i.e., the slower the envelope varies), the smaller the late-time asymptotic value of $|{\beta_k}^{(1)}|^2$. Although these arguments can be formalized, our goal here is simply to build qualitative intuition for the behavior of the spectrum.

In order to apply this intuition to understand the $\nck \sim k^3$ IR scaling of the spectra, we note firstly that the adiabaticity parameters are larger and evolving rapidly with increasing $\Phi$ when $\omega_k^2/a^2$ is near its minimum; for sufficiently small $k$ they are generally smaller and (provided that $t \gtrsim t_{\text{HC}}$) evolving more slowly with increasing $\Phi$ when $\omega_k^2/a^2$ is large; for $t\lesssim t_{\text{HC}}$ the adiabaticity parameters evolve fairly rapidly but are damped to very small values exponentially quickly as the phase $\Phi$ is \emph{de}creased since $\omega_k^2/a^2$ generally increases much more rapidly with decreasing phase $\Phi$ when dominated by $k^2/a^2$ than by $M_{\text{eff}}^2$ (see Fig.\ \ref{fig:mass_plot_2}). Therefore, for small values of $k$ and viewed as a function of increasing phase $\Phi$ (see Fig.\ \ref{fig:envelope}), the envelope modulating the rapid phase oscillation starts exponentially small, fairly rapidly rises to some small value around the time of Hubble-radius crossing, evolves fairly slowly for some duration of increasing phase $\Phi$, grows in size and evolves more rapidly as $\omega_k^2/a^2$ goes through its minimum, then decreases in size again as $\omega_k^2/a^2$ increases in size and finally executes small amplitude oscillations with the same period at which $\omega_k^2/a^2$ oscillates around a constant in the MD era. Crucially however, most of that evolution is completely $k$-independent, and if $k$ is further decreased, the only modification to the envelope is to add in an additional duration of fairly slow evolution of the envelope at early time by shifting to earlier times the point around $t_{\text{HC}}$ where the envelope rises from its initial very small value (compare the upper plots of Fig.\ \ref{fig:envelope}). As such, for sufficiently small $k$, we do not expect that significant \emph{additional} incomplete cancellation between neighboring excursions in the additional oscillations can occur as $k$ is further decreased. As a result we expect qualitatively that $|\beta_k|^2$ should become $k$-independent at sufficient small $k$, leading to $\nck \sim k^3$.

Furthermore, by virtue of the fact that neighboring oscillations near the maximum of the envelope become more rapidly incommensurate in size if the envelope decreases more rapidly from its maximum with changing phase, such as occurs for cases of smaller $\epsilon$, we expect that the plateau value of $\bks$ should be larger, as observed (at least for $\epsilon \leq 0.6$) in Fig.\ \ref{fig:M}. We can mock up the plateau behavior in the same toy model discussed above by supposing that the Gaussian envelope is simply cut off sharply to zero at a point $\zeta$ standard deviations before its maximum, which leads to a late-time asymptotic value of $|{\beta_k}^{(1)}|^2 \propto e^{-4\pi^2n^2} \lb| 1 - \text{Erf} \lb(- \sqrt{2} \lb( \zeta - i\pi n \rb) \rb) \rb|^2$, which for $\zeta>\pi n = \sigma$ is essentially independent of $\zeta$, and is larger if $n$ is smaller. Translating back to the language of our actual model we see that this captures all the salient features: for $k$ sufficiently small that $t_{\text{HC}}$ occurs a number of phase oscillations before $\omega_k^2/a^2$ goes through its minimum, the spectrum would be $t_{\text{HC}}$- and thus $k$-independent, and this plateau would be at a higher value if $\epsilon$ were smaller.

In almost all cases of either constant or running effective masses, as the effective mass term increases in size,\footnote{To be concrete: for the minimally-coupled (constant mass) case of \secref{simplemodel} (in the chaotic inflation background), we mean increasing $M$; for the potentially-coupled of \secref{potentialirruption}, we mean increasing $M_g$; and for the kinetically-coupled case of \secref{kineticirruption}, we mean increasing $M$ and/or decreasing $\epsilon$.}  fewer particles are produced both at fixed $k$ and overall. Qualitatively, the reason is clear: heavier modes experience changes in the background spacetime ``more adiabatically.'' We mean one of two things here: either (a) the duration of any possible tachyonic phase is shortened as the effective mass grows (see, e.g., \figref{mass_plot_1}), or (b) in the more extreme case where the effective mass is so large that no tachyonic phase at all is present (i.e., $M_{\text{eff}}^2> |a''/a^3|$ at all times during inflation; see, e.g., \figref{mass_plot_2}), the minimum value of $\omega_k^2/a^2$ reached during inflation increases in size as the effective mass increases, which generally correlates with a decrease in the maximum size of the adiabaticity parameters. This not only collapses the envelope modulating the phase factor $e^{2i\Phi}$ in Eq.\ \eqref{eq:beta1_phase} (see the lower-right plot of Fig.\ \ref{fig:envelope}) but also results in the envelope evolving more slowly with phase since larger $\omega_k/a$ results in a smaller $\Delta t$ to get the same $\Delta \Phi \approx (\omega_k/a)\Delta t$. Therefore, by arguments similar to those just advanced, production is suppressed in the latter case. 

The obvious exception in our results to this general rule of decreasing $\nck$ with increasing mass parameters occurs in the large-$M_g$ results for the potentially-coupled case in Fig.\ \ref{fig:potential}. For $k<k_{\text{th}} \approx 4\eten{-2}$, as $M_g$ increases, the spectrum goes to a limiting value of $\bks = 1$ from above, while for larger (fixed) $k>k_{\text{th}}$, the spectrum goes to the same limit from \emph{below}. The comoving momentum $k_{\text{peak}}$ where the $\nck$ spectrum peaks shifts over to the UV roughly as $k_{\text{peak}}\propto (M_g)^{0.500(5)}$ (from fits to the largest-$M_g$ numerical results we have). We already argued in footnote \ref{footnote:crossover} that once $M_g \gtrsim 70$, the duration-of-tachyonicity argument for understanding the results breaks down and we enter the regime in which the analytical expressions we develop in Appendix \ref{app:potential_analytics} apply. Indeed, examining Eq.\ \eqref{eq:spectrum_analytics} immediately explains many of the observed ``anomalous'' features at large-$M_g$: at fixed $k$, as $M_g$ gets larger the exponent in Eq.\ \eqref{eq:spectrum_analytics} goes to zero, and $\bks\rightarrow1$. Since the exponent is, for the parameter values we give in Appendix \ref{app:potential_analytics}, positive for $k \lesssim 4\eten{-2}$, and negative for larger $k$, it is also clear that the limit should indeed be approached from above for $k<k_{\text{th}}$ and from below for larger $k$. Finally, once $k \gg H_*^2,\dot{H}_*$, the exponent scales proportional to $(-k^2/M_g)$, and so for equal exponential fall-offs from $\bks=1$, $k$ has to increase as $\sqrt{M_g}$ consistent with the peak shift to the UV seen in the numerical results.

The remaining point we wish to clarify in connection with this discussion is how this specific analytical understanding of the scaling of the spectra at large $M_g$ is consistent, in the regime where no tachyonic phase exists, with the more general picture outlined just above; i.e., that the amount of production in this regime is limited by the maximum size of the adiabaticity parameters. If we were discussing the kinetically-coupled case, it is obvious that increasing $M$ necessarily monotonically increases the minimum value of $M_{\text{eff}}$ and, hence, could only increase the minimum value of $\omega_k^2/a^2$, suppressing production. However, for the potentially-coupled case, $M_{\text{eff}}$ \emph{always} goes to zero at $\nu=\nu_*$, and the resulting $M_g$ dependence of the minimum value of $\omega_k^2/a^2$ (and hence the maximum value of the adiabaticity parameters) is not immediately clear. In particular, if our general arguments are to hold, the behavior of the adiabaticity parameters with increasing $M_g$ must necessarily be non-monotonic, because we observe the spectrum to first decrease and then increase again at large $k$ as $M_g$ is increased. 

We begin by making the important observation that the threshold value $k_{\text{th}}$ is additionally the boundary, in the limit of large $M_g$, between the cases where a (brief, impulsive) tachyonic phase exists ($k<k_{\text{th}}$), and when one does not exist  ($k>k_{\text{th}}$): to see this, note that in the $M_g \rightarrow \infty$ limit, the minimization of $\omega_k^2/a^2$ occurs at $\nu = \nu_*$, so in this limit  $k_{\text{th}}$ is simply estimated by requiring that $\omega_{k_{\text{th}}}^2/a^2|_{\nu=\nu_*}=0$, which implies that $k_{\text{th}} \approx \sqrt{ 4\pi a_*^2/3 \times (2\nu_*^2-1/12\pi)} \approx 3.7\eten{-2}$ where we have used $a''/a^3=4\pi/3\times(2\nu^2-\dot{\nu}^2)$, the slow-roll value $\dot{\nu}\approx-1/\sqrt{12\pi}$, and have taken $a_* \equiv a(\nu = \nu_*) = 1.6\eten{-2}$ from our full numerical solutions.

\begin{figure*}
\includegraphics[width = 0.70 \textwidth]{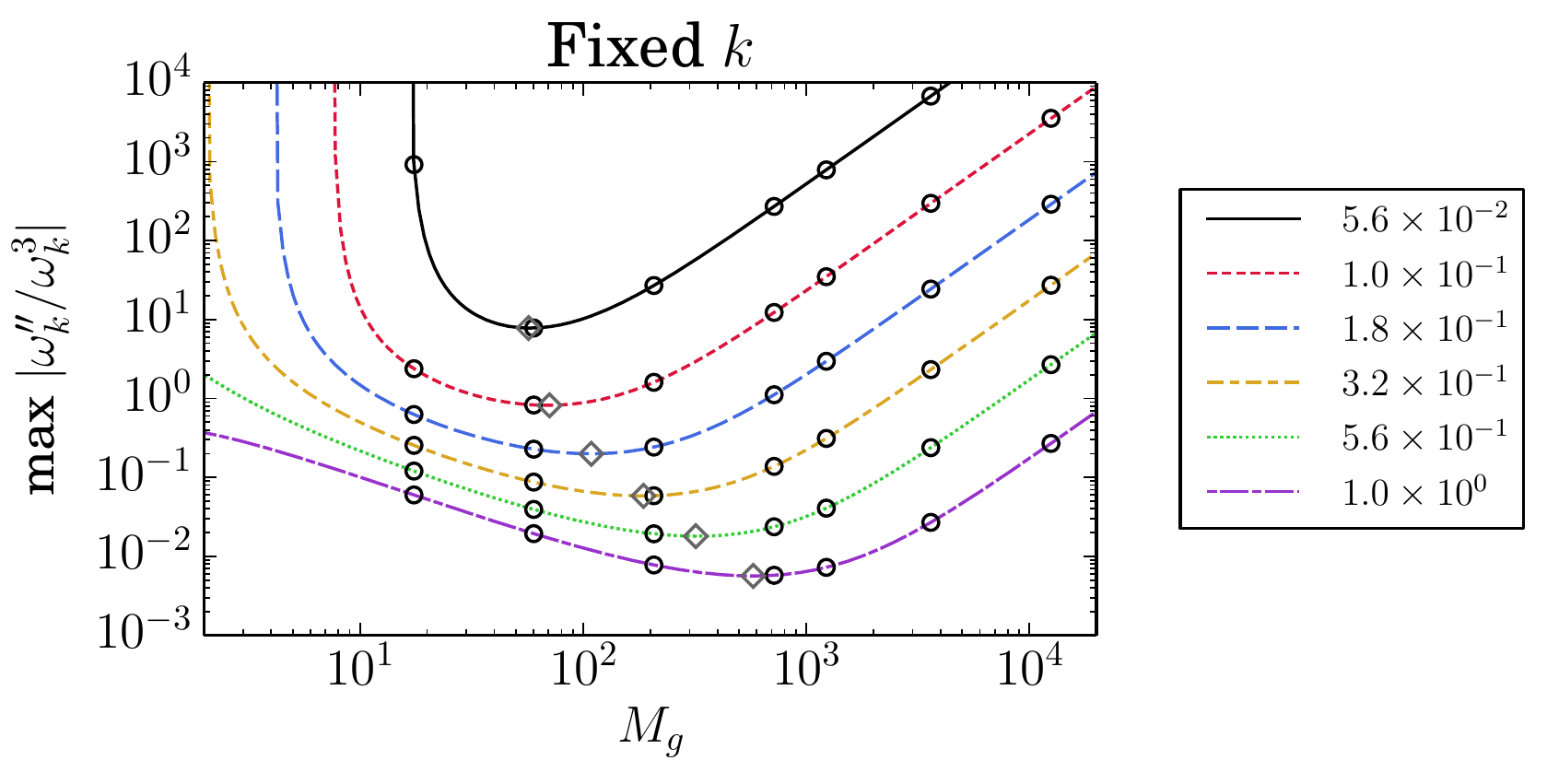}
\caption{ \label{fig:max_adiab_params}  (Color online) The maximum values attained during inflation of the adiabaticity parameter $|\omega_k''/\omega_k^3|$ (the behavior of the other adiabaticity parameter $|\omega_k'/\omega_k^2|^2$ is similar) in the potentially-coupled irrupton model at representative fixed $k$ (see legend) as a function of $M_g$ (in units of $m$). The black circles indicate the relevant values $M_g = \{\ 17,\ 60,\ 2.1\eten{2},\ 7.2\eten{2},\ 1.2\eten{3},\ 3.7\eten{3},\ 1.2\eten{4} \}$ (see Fig.\ \ref{fig:potential}). The divergences evident at small $M_g$ for some of the smaller-$k$ curves indicate that these modes allow for a tachyonic phase. The minimum of each curve is indicated by a grey diamond.}
\end{figure*}

Therefore, for $k>k_{\text{th}}$, no tachyonic phase exists for the large $M_g$ cases and the amount of particle production should indeed be governed by the size of the adiabaticity parameters $|\omega_k'/\omega_k^2|^2$ and $|\omega_k''/\omega_k^3|$ (as well as phase cancellations as we have previously argued) and so it is necessary to demonstrate non-monotonic behavior of those parameters with increasing $M_g$ for all $k>k_{\text{th}}$ if our arguments are to be consistent.  In Fig.\ \ref{fig:max_adiab_params} we plot the maximum value of the adiabaticity parameter $|\omega_k''/\omega_k^3|$ ($\textrm{max}\, \{ |\omega_k'/\omega_k^2|^2 \}$ behaves similarly) as a function $M_g$ at fixed $k$ as evaluated in in our numerical work. It is clear that the requisite non-monotonic behavior is indeed present.

It is instructive to understand the origin of this non-monotonic $M_g$ dependence of $\textrm{max}\, \{ |\omega_k''/\omega_k^3| \}$. We observe that it is usually true that the the maximum value of $|\omega_k''/\omega_k^3|$ is reached very near to the point where $\omega_k^2/a^2$ reaches a minimum. As we have already noted, for large $M_g$ the latter function is minimized at $\nu\approx\nu_*$; imposing this condition kills all $M_g$-dependence in $|\omega_k''/\omega_k^3|$ except for a term which scales as $M_g^2$ in the numerator [i.e., the surviving term arising from twice differentiating $M_{\text{eff}}^2$ as given in Eq.\ \eqref{eq:Meffpotential}], which gives rise to the $M_g^2$ scaling. 

On the other hand, once $M_g$ gets small enough, the minimum of $\omega_k^2/a^2$ (and hence the maximum of $|\omega_k''/\omega_k^3|$) is no longer reached at $\nu \approx \nu_*$. Assuming for the sake of this argument that $k$ is sufficiently large that, even though $M_g$ is small, the $-a''/a^3$ term in $\omega_k^2/a^2$ always remains negligible, the minimum is reached at a later time when $\nu \approx \tilde{\nu}$ ($\tilde{\nu}<\nu_*$), where $\tilde{\nu}$ is defined to be the point where the decreasing function $k^2/a^2$ is equal to the increasing (on $\nu < \nu_*$) function $M_{\text{eff}}^2$, yielding $\omega_k^2(\nu=\tilde{\nu}) \approx 2k^2$. This requires $\nu_*-\tilde{\nu} = k/(\tilde{a}M_g) >0$; taking the slow-roll expression $\tilde{a} \equiv a(\nu = \tilde{\nu}) \approx a_* \exp\lb[-2\pi(\tilde{\nu}^2-\nu_*^2)\rb]$ as a good approximation results in a transcendental equation for $\tilde{\nu}$, which we solve numerically. If in evaluating the adiabaticity parameters $|\omega_k''/\omega_k^3|$ and $|\omega_k'/\omega_k^2|^2$ we additionally take $\dot{\nu} \equiv -1/\sqrt{12\pi}$ and drop all $\ddot{\nu}$ terms (consistent with the slow-roll approximation for $a$), we find the adiabaticity parameters scale approximately as
\begin{align}
\lb\{ |\omega_k''/\omega_k^3|,\, |\omega_k'/\omega_k^2|^2 \rb\} \propto \frac{1}{M_g^2} \frac{\tilde{\nu}^2}{(\nu_*-\tilde{\nu})^2} \lb[ 1 + \frac{1}{2\pi\nu(\nu_*-\nu)} \rb].
\end{align}
Using the numerically obtained values of $\tilde{\nu}$, we find that the factor multiplying $M_g^{-2}$ scales roughly proportional to $M_g$ for large enough $k$, leading to an approximate scaling of $\lb\{ |\omega_k''/\omega_k^3|,\, |\omega_k'/\omega_k^2|^2 \rb\} \propto M_g^{-1}$ in the regime of large $k$ and small $M_g$. This compares favorably with the large-$k$ results in Fig.\ \ref{fig:max_adiab_params}. The behavior for smaller $k$ at small $M_g$ would be obtained from a similar argument, in which one did not neglect the $-a''/a^3$ term, but still assumed that $k$ and $M_g$ were sufficiently large to prevent a tachyonic phase.\footnote{Indeed, the divergences evident in Fig. \ref{fig:max_adiab_params} for small $k$ and $M_g$ are precisely due to these cases allowing for a tachyonic phase. For small $M_g$, since $\omega_k^2/a^2$ is not minimized at $\nu = \nu_*$, our earlier argument that $k_{\text{th}} = 4\eten{-2}$ is the threshold value beyond which no tachyonic phase is present is not applicable. Instead, the threshold value $k_{\text{th}}$ should be obtained from the full condition $\textrm{min}_\nu\, \{ \omega_{k_{\text{th}}}^2 \} = 0$, and this typically requires a larger threshold value than $k_{\text{th}} = 4\eten{-2}$ if $\omega_k^2/a^2$ is minimized at $\nu < \nu_*$. The appearance of divergences in Fig.\ \ref{fig:max_adiab_params} related to the existence of a tachyonic phase is thus not in contradiction with our earlier statements.} 

The question then naturally arises as to where the cross-over point between the two regimes occurs as this should give a good estimate of when $\textrm{max}\, \{ |\omega_k''/\omega_k^3| \}$ goes through its turning point which indicates at what value of $M_g$ one would expect the spectrum, at a fixed value of $k$, to stop decreasing and instead enter the regime where it increases again back to the limiting value of unity as we discussed above. Since we expect the $g^2$ scaling to obtain whenever $\omega_k^2/a^2$ is minimized around $\nu \approx \nu_*$, we can estimate that once $\nu_* - \tilde{\nu} \gtrsim 0.1\nu_*$, this scaling may begin to break down. Again throwing away the $-a''/a$ term in $\omega_k^2/a^2$, and solving for the value of $M_g^\text{min}$ required to achieve such a deviation assuming $a\approx a_*$, we find that $M_g^\text{min} \approx \sqrt{ 4\pi k^2 / 0.1 a_*^2 } \approx 7\eten{2}k$; this compares fairly well with the values of $M_g$ for which $\textrm{max}\!\lb\{ |\omega_k''/\omega_k^3|\rb\}$ reaches its minimum in Fig.\ \ref{fig:max_adiab_params}. Of course, the fairly good numerical agreement here is sensitive to the exact value of $a$ and the assumed small deviation from $\nu_*$ which is used in this argument; however, the fact that $M_g^{\text{min}} \propto k$ is a fairly robust prediction, and is consistent with the scaling with $k$ of the minima of the curves shown in Fig.\ \ref{fig:max_adiab_params} for sufficiently large $k$.

We now turn attention to the far-ultraviolet (UV) behavior of the spectra. To understand the rapid drop-off in the particle spectra once $k \gtrsim 10^{-1}$ (see Figs.\ \ref{fig:eps_inf}--\ref{fig:M}) one must consider three qualitatively distinct cases: (a) for small enough $k$ there exists a broad region of tachyonic behavior of the mode function (e.g., for $M \lesssim 3$ and $\epsilon = 0.6$ for the kinetically-coupled irrupton, and the potentially-coupled irrupton results for $M_g \lesssim 60$ in Fig.\ \ref{fig:potential}), (b) for small enough $k$ there is a very narrow tachyonic region which looks more like the impulsive kick we discussed above (e.g., for the $M_g\gtrsim200$ results in Fig.\ \ref{fig:potential}), and (c) there is never any tachyonic behavior of the mode functions for any value of $k$ (e.g., for $M\gtrsim 3$ for the kinetically-coupled irrupton). 

We begin with case (a) by noting again that a temporary but fairly broad tachyonic instability in the mode function, such as occurs in this case at smaller $k$, leads naturally to exponentially more particle production than in a case where no such instability exists. The important observation is that once $k^2 > \text{max}\lb\{a''/a \rb\} $, it is impossible for the mode function to become tachyonic at any point during its evolution, regardless of the value of $M_{\text{eff}}^2$. Therefore, for large enough $k$, we naturally expect the spectrum (or more exactly, $\bks$) to show a rapid fall-off compared to the values it obtains at smaller $k$. A simple-minded estimate for when this criterion is satisfied, obtained by assuming that the slow-roll regime is always valid, yields $k\gtrsim 0.4$, which is of the same order of magnitude of the point beyond which the $\nck$ spectra are observed to drop-off exponentially fast in Fig.\ 5 and at small $M_g$ in Fig.\ \ref{fig:potential}.\footnote{This estimate is about 30\% too large compared to the same criterion evaluated in our full numerical solution. Also, adding a (positive) effective mass term only helps to make $\omega_k^2/a^2$ more positive, so the estimate shown is really an upper bound which we do not expect to be saturated.} 

Case (b) is handled by our analytical treatment in Appendix \ref{app:potential_analytics}: once the value of $k$ is large enough so that the $k^2/a^2$ term in $\omega_k^2/a^2$ is significant, the spectrum rolls off exponential quickly as $\exp\!\lb[- \lb( \pi / M_g|\dot{\nu}_*|a_*^2  \rb)k^2\rb]$ from $\bks \approx 1$. We have already noted above that the quantitative prediction arising from Eq.\ \eqref{eq:spectrum_analytics} for the location of the peak in the $\nck$ spectra, which coincides with the drop-off of $\bks$, does indeed match well with the numerical results shown in Fig.\ \ref{fig:potential}.

Case (c) requires a little more care, since the evolution can never be tachyonic for any value of $k$; however, by increasing $k$ beyond the point where Hubble-crossing is occurring around $\nu = \nu_*$, the minimum value of $\omega_k^2/a^2$ achieved during inflation becomes a monotonically increasing function of $k$ (see, e.g., \figref{mass_plot_2}). We have already argued that the minimum value of $\omega_k^2/a^2$ anti-correlates with the maximum value of the adiabaticity parameters, so as this minimum value increases the particle production is suppressed as the envelope bounding the phase factor $e^{2i\Phi}$ in Eq.\ \eqref{eq:beta1_phase} collapses. The simple-minded estimate here for when the spectrum starts to drop is thus obtained from setting $k = aH$ at $\nu = \nu_*$; assuming that slow-roll is valid, this gives $k \gtrsim  0.1$, which is again in good agreement with the observed behavior in Figs.\ \ref{fig:eps_0.60} and \ref{fig:M}.

The foregoing comments about the UV spectra do have a clear exception in the slower drop-off / small bump around $k \sim 5$ in the constant-$M$ spectra for $M \lesssim 1$ (\figref{eps_inf}) and the pronounced bump around $k \sim 5$ in the kinetically-coupled irrupton spectrum for small $M$ ($M=0.2$) and $\epsilon =0.6$ (\figref{eps_0.60}). This bump clearly has a different physical origin to the rest of the spectrum; an examination of the time-evolution of $|\beta_k|^2$ and the behavior of $\omega_k^2/a^2$ indicates that it arises from modes which become light during the first few coherent inflaton oscillations immediately after the end of inflation, leading to their excitation either in the usual matter-dominated Friedmann expansion phase, or in the transition out of the inflationary phase. Again, once $k$ is sufficiently large, even this effect is suppressed.

We now discuss the some features the large-$M$ kinetically-coupled irrupton spectra (\figref{M}) to which we drew attention above: the bump in $\bks$ at intermediate $k$ (corresponding to $\nck \sim k^{3+x}$ for some $x>0$) and the nearly-universal behavior of the spectra near the peak. 

Consider first the bump in $\bks$ at intermediate $k$, which occurs provided that $\epsilon$ is not too small; see \figref{M}. The reason for this feature can be traced back to the behavior of (the square root of) the adiabatic parameter $|\omega_k'/\omega_k^2|^2$ per our earlier argument about the shape of the envelope of the rapid oscillations in the integrand in Eq.\ \eqref{eq:beta1_phase}. For such intermediate $k$ cases, $\omega_k^2/a^2$ is dominated by $k^2/a^2$ until it is very near its minimum, and from our general observation of an anti-correlation in the sizes of $\omega_k^2/a^2$ and the adiabaticity parameters it follows that this envelope is very small until very close to its maximum, making it highly asymmetric near that maximum (see the lower-left plot in Fig.\ \ref{fig:envelope}); this leads to a significantly larger accumulated incomplete cancellation between neighboring oscillations when viewed in the late-time regime. The appearance of oscillations in the spectra in the transition from the IR $k^3$ scaling to this ``bump'' regime can also be qualitatively understood as the effect of first the positive, and then the negative, excursions just before the maximum being alternately larger as the envelope opens up on the low-phase $\Phi$ side as $k$ is decreased from, e.g., the situation pictured in the lower-left plot of Fig.\ \ref{fig:envelope}. Furthermore, the fact (which we noted above) that neighboring excursions become more rapidly incommensurate in size as the envelope decreases more rapidly in size from its maximum with changing phase, such as happens in the cases of smaller $\epsilon$, makes this argument more marginal in such cases, explaining why the oscillations in smaller-$\epsilon$ cases are less pronounced (see Fig.\ \ref{fig:M}). Again, our toy model captures the essential elements of this behavior: $|{\beta_k}^{(1)}|^2 \propto e^{-4\pi^2n^2} \lb| 1 - \text{Erf} \lb(- \sqrt{2} \lb( \zeta - i\pi n \rb) \rb) \rb|^2$ increases rapidly from its plateau value as soon as $\zeta < n\pi$ (i.e., as $t_{\text{HC}}$ approaches the time at which $\omega_k^2/a^2$ is minimized), and also demonstrates some dips and wiggles, more pronounced for larger $n$ (i.e., for larger $\epsilon$) provided it is not too large, in the transition from the plateau to this regime of growth.  

The second feature of interest is the fairly universal behavior (i.e., approximately independent of changing $\epsilon$) for $k \gtrsim 10^{-2}$ for $M = 4$ in \figref{M}. Since the $\nck$ spectra are strongly peaked near $k \sim 10^{-1}$, it follows that to a fairly good approximation the total number of particles produced, $n^{\text{p}}$, thus also becomes independent of $\epsilon$. This is clearly an important qualitative feature of this result, and can be explained by appealing to arguments similar to those already advanced about the maximum size of, and behavior near the maximum of, the adiabaticity parameters in this regime via their role of defining the envelope of oscillations in Eq.\ \eqref{eq:beta1_phase}. Briefly, since at large $k$, $\omega_k^2/a^2$ is dominated by $k^2/a^2$ until quite late during inflation, $\omega_k^2/a^2$ has no $\epsilon$-dependence until very late during inflation. This implies that the evolution of the adiabaticity parameters is nearly identical for different values of $\epsilon$ until the $k^2/a^2$ term red-shifts away sufficiently to expose the $\epsilon$-dependence in $M_{\text{eff}}$, and hence in the adiabaticity parameters. However, the resulting $\epsilon$-dependence of the maximum value of the adiabaticity parameters, and their behavior near that maximum, in the $k\gtrsim 10^{-2}$ regime is very mild for all the values of $\epsilon$ we have examined, which implies that as a gross approximation, the amount of particle production should be approximately the same for each $\epsilon$ value, roughly as observed in \figref{M}. Going beyond this gross approximation, we note that the small excess in the production at $\epsilon = 0.25$ versus that at $\epsilon = 0.8$ in the regime between $k \sim 10^{-2}$ and $k \sim 5 \times 10^{-1}$ (see the inset in the lower-left plot in Fig.\ \ref{fig:M}) is borne out in the slightly larger maximum value of the adiabaticity parameter attained for the former case compared to the latter.

\section{Discussion and Applications \label{sec:applications}}

While inflation is a phenomenological success, the particle-physics foundations upon which a complete theory of inflation can be built are yet to be set. Even assuming that the dynamics of inflation may be described in terms of a scalar field (the inflaton), we do not know whether the inflaton is a ``fundamental'' scalar field describable in terms of a ultraviolet-complete theory, or whether it should be considered within the framework of an effective field theory.  We do not yet understand how inflation began (e.g., whether inflation is eternal), or how inflation ended (preheating, reheating, etc.).  We also do not know how the inflaton couples to other fields. 

If the Universe did undergo an early phase of inflation, then one probe of the dynamics results from particle production during inflation.  One relic of particle production during inflation is the ``scalar'' curvature fluctuations resulting from creation of inflaton quanta during inflation.  A second relic is the ``tensor'' perturbations (gravitational waves) resulting from creation of the transverse, traceless component of the gravitational field (viz., gravitons).  Measurements of the scalar and (especially) the tensor perturbation spectra and possible non-Gaussian signatures will go a long way toward untangling the foundations of inflation.

In addition to the inflaton and the graviton, other fields will be produced during inflation if conformal symmetry is broken through either a mass term or a non-conformal coupling to gravity.  We considered such a model in \secref{simplemodel}.  This model has been considered before (e.g., Refs.\ \cite{Chung:1998zb,Kuzmin:1998kk}); here, we extend the range of numerical results to larger mass and larger comoving wavenumber and develop a clearer physical understanding of the expected results in ranges of parameters not accessible to numerical techniques.

Another avenue of exploration is the possibility that the inflaton may couple to massive particles, where again ``massive'' is with respect to the expansion rate during inflation.  Of particular interest is the possibility that the additional field may become massless (or at least light compared to $H$) during inflation as a result of its coupling to the inflaton.  In this case there may be a ``resonant'' production of the particle species at a particular value of the inflaton field.  This is what we call an irruption of the massive particle species, and why we refer to the additional scalar field as the irrupton.

We study two such models.  While it is impossible to imagine all possible inflaton--irrupton couplings, the two models we consider exhibit a range of final spectra that should encompass a wide range of possibilities. 

The first model is a potentially-coupled irrupton in which the field couples through a term in the potential that couples the inflaton and the irrupton.  This possibility was first proposed in a similar model by \citet{Chung:1999ve}, and explored in Refs.\ \cite{Elgaroy:2003hp,Romano:2008rr,Barnaby:2009mc,Barnaby:2010sq}.  Here, we develop numerical and analytic techniques to allow us to extend the numerical range of study and to understand the behavior of the resulting spectra with parameters where a numerical calculation is problematic.

We also consider a new model for inflaton--irrupton coupling: specifically, a non-canonical kinetic-term coupling between a heavy scalar-field with a canonical mass parameter  and the scalar (inflaton) field which drives inflation. By canonically normalizing the heavy scalar kinetic term, for our choice of coupling, we find a time-dependent exponential enhancement of the effective mass of the canonically-normalized heavy scalar which allows it to briefly become as light (or lighter than) $H$, but otherwise to be much heavier at both early and late times. By using the method of Bogoliubov coefficients, and numerically solving the equations of motion for the inflaton field, the scale factor, and the mode equations for the heavy scalar field, we determine the number of these heavy scalar particles produced gravitationally by the non-adiabatic expansion of the background space-time during inflation acting on quantum fluctuations of the heavy scalar field.  From this we extract the final irrupton particle spectra.

The particle spectra in the two models are found to exhibit a variety of complex behaviors attributable to the time dependence of the effective mass, with the most generic feature being an infrared cutoff in the spectra compared to the the minimally-coupled WIMPzilla which has been previously extensively studied in the literature, in addition to the well-known usual UV fall-off. These spectra become increasingly peaked toward scales which cross the comoving Hubble radius near the end of inflation as the mass parameter $M$ or $M_g$ increases. For the second model, they additionally become more peaked as the non-canonical kinetic term increasingly singles out a specific inflaton field value as important (i.e., the parameter $\epsilon$ decreases in size). 

\begin{figure*}[p]
\includegraphics[width = 0.8 \textwidth]{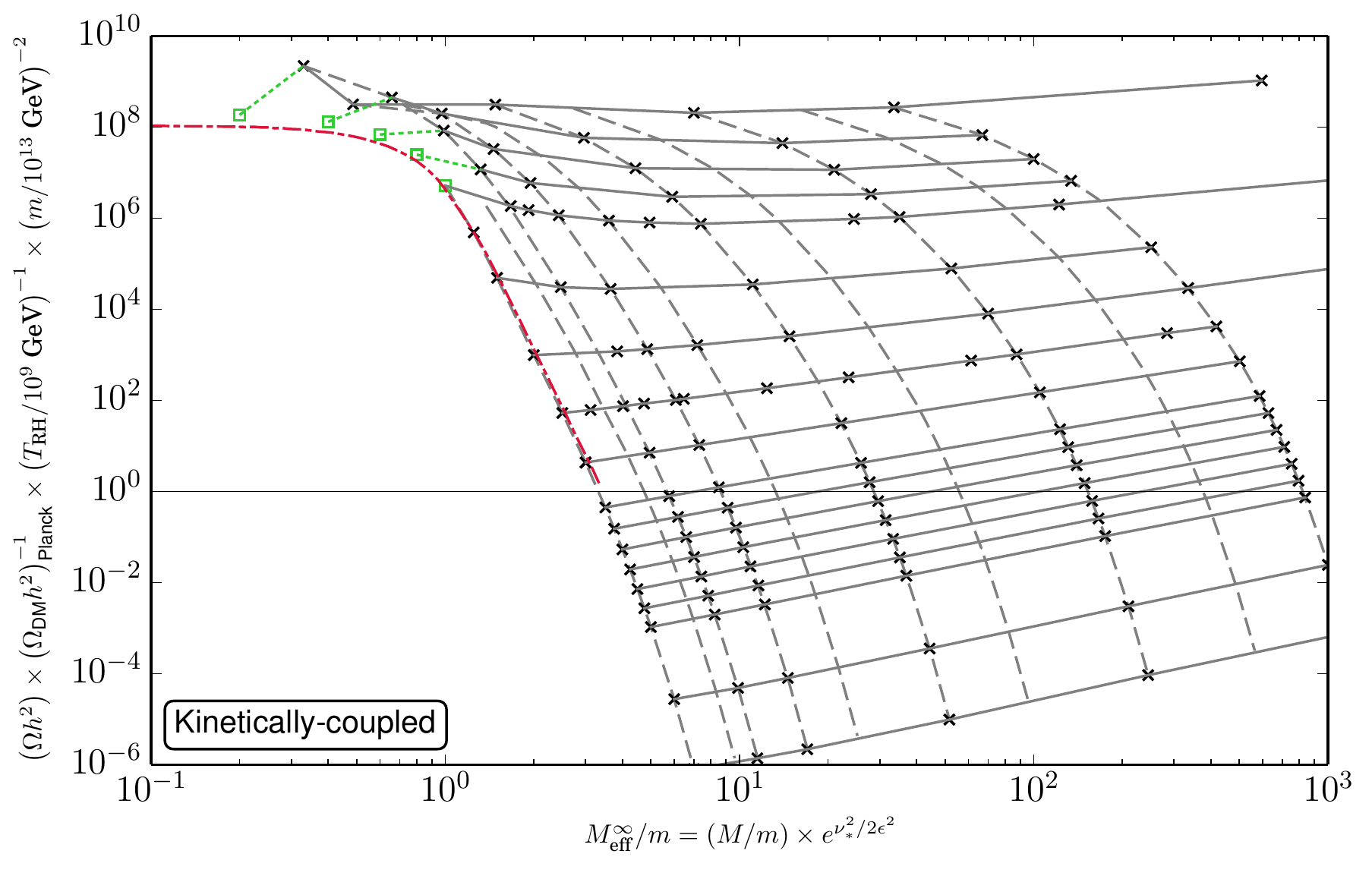}\\
\includegraphics[width = 0.8 \textwidth]{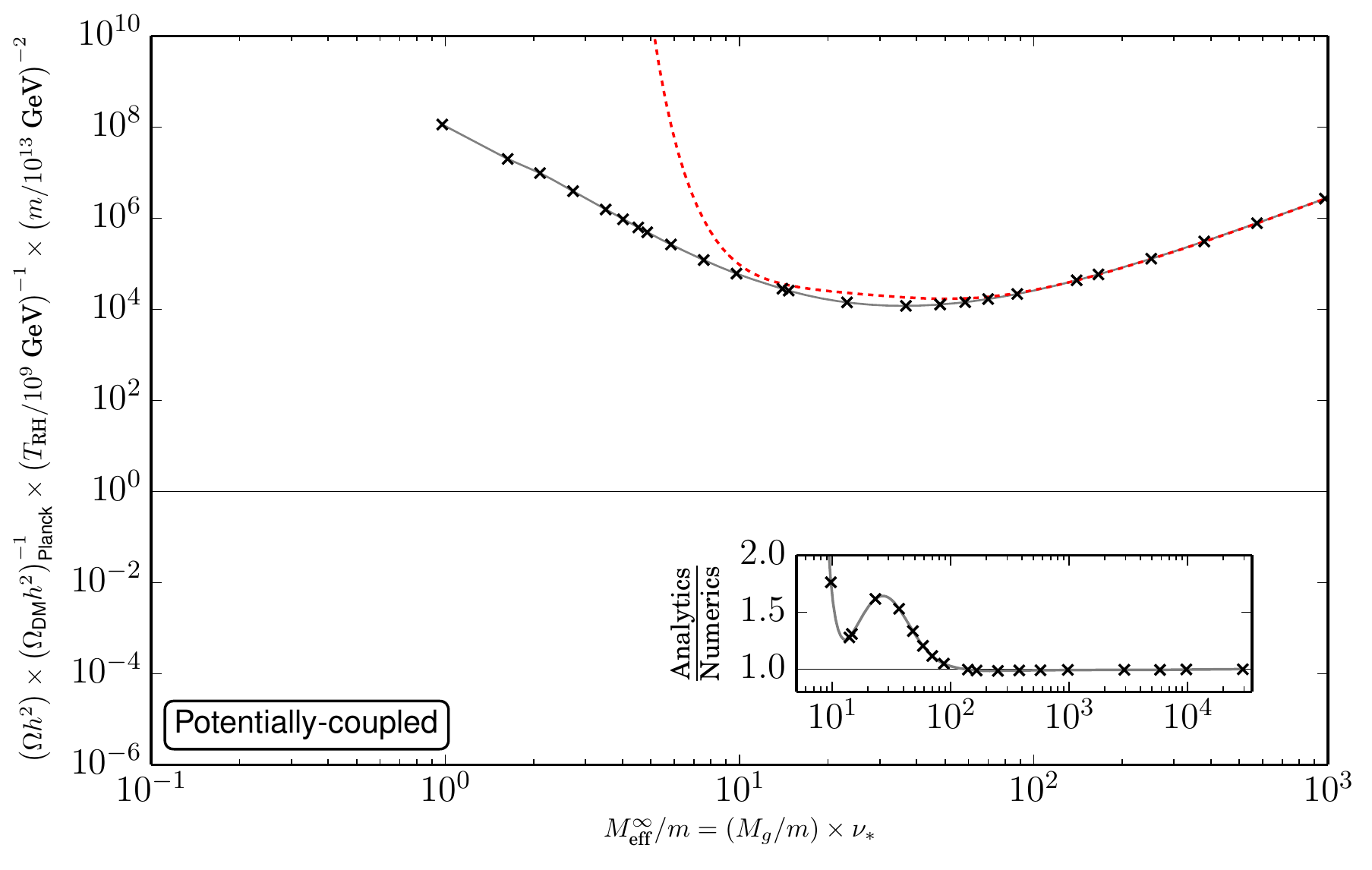}
\caption{\label{fig:relic} (Color online) The present-day relic mass-density of stable irruptons for the kinetically-coupled model (upper plot) and the potentially-coupled model (lower plot) in units of $\lb(\Omega_{\text{DM}} h^2\rb)_{\text{Planck}} \times (T_{\text{RH}} / 10^9 \text{GeV} ) \times ( m / 10^{13} \text{GeV} )^2 $ where $ \lb(\Omega_{\text{DM}} h^2\rb)_{\text{Planck}} = 0.1186$ \cite{Ade:2013zuv}, as a function of the late-time effective mass $M_{\text{eff}}^\infty$. Crosses (black) represent points we have explicitly sampled in our numerical computations. In the upper plot, solid (grey) lines join points at constant $M/m$ (from top to bottom $M=$ 0.2, 0.4, 0.6, 0.8, 1.0, 1.5, 2.0, 2.5, 3.0, 3.5, 3.75, 4.0, 4.25, 4.5, 4.75, 5.0, 6.0, 7.0), and dashed (grey) lines join points at constant $\epsilon$ (from left to right $\epsilon=\infty$, 1.0, 0.8, 0.6, 0.5, 0.4, 0.35, 0.3, 0.27, 0.25), where interpolation between sampled points has been performed. The (green) square points in the upper plot are models where it was necessary to cut the spectrum off in the infrared to obtain a finite value; while the dash-dot (red) line is the constant-$M$ result taken from \citet{Kuzmin:1998kk}, which our $\epsilon\rightarrow\infty$ results recover well except at small $M$; see footnote (\ref{footnote:IR_div}). In the lower plot, the solid (grey) line is a (log-log) cubic spline interpolant between the sampled points while the dashed (red) line is the result of our analytical expression Eq.\ \eqref{eq:spectrum_analytics}. We show the ratio of the analytical to numerical result in the inset, including also numerically sampled points up to the mass $M_{\text{eff}}^\infty/m = 2.9\eten{4}$ which are not shown on the main axes as they lie exactly on the $\Omega h^2 \propto M_g^{5/2}$ extrapolation line.}
\end{figure*}

One important result of our work is the relic density of these heavy particles, assuming they are stable, as a function of the heavy particle effective mass (\figref{relic}).\footnote{\label{footnote:IR_div}Note that in integrating $\nck$ over $k$ to extract the total particle number $n^{\text{p}}$ via Eq.\ \eqref{eq:number}, a problem arose in some cases (i.e., the kinetically-coupled irrupton at $M\leq 1$ as $\epsilon\rightarrow \infty$), indicated in \figref{relic} by (green) squares, due to the non-convergence of the integral owing to IR-divergent behavior in $\nck$. We simply cut off the integration in the IR at $k=10^{-20}$ as this was the smallest value we sampled; this corresponds to modes that crossed the Hubble radius roughly 48 $e$-folds before the end of inflation. The marked results may very well underestimate the total particle number produced, and they should thus be interpreted with caution. These results are also larger than those from Ref.\ \cite{Kuzmin:1998kk} as we have probed smaller values of $k$.} From these results, we conclude that the effect of singling out in the non-canonical kinetic-coupling a single inflaton field value as more important (i.e., decreasing $\epsilon$) at fixed sufficiently large $M$ is to increase both the late-time effective mass of the heavy particle and the relic abundance. (For smaller $M$, the effect of decreasing $\epsilon$ is first to decrease the relic abundance while increasing the late-time effective mass of the heavy particle, but this behavior is short-lived and as $\epsilon$ gets smaller, the mass and relic abundance increase together again.) As a result, we find explicitly that we can produce heavy particles with late-time effective masses more than three orders of magnitude larger than the inflation mass ($m \sim 2 H(t_{EI})$) yet with sufficient relic abundance to saturate the Planck result for $\Omega_{\text{DM}} h^2$, which is in marked contrast to the usual minimally-coupled WIMPzilla whose mass must be around $3.3m$ to achieve the same result (see our ``$\epsilon\rightarrow\infty$'' results, which agree well with those of Ref.\ \cite{Kuzmin:1998kk}). We expect that the mechanism should remain operative for even higher effective masses, possibly even up to the Planck mass scale, for suitable parameter choices. 

We note that to obtain the ratio of the irrupton mass density to the inflaton mass density during the inflaton oscillation phase, one should multiply the quantity plotted in Fig.\ \ref{fig:relic}, $\lb( \Omega h^2 \rb) \times \lb(\Omega_{\text{DM}} h^2\rb)^{-1}_{\text{Planck}} \times (T_{\text{RH}} / 10^9 \text{GeV} )^{-1} \times ( m / 10^{13} \text{GeV} )^{-2}$, by $5.8\times10^{-19}\times( m / 10^{13} \text{GeV} )^{2}$.  This implies that the irrupton mass density is a very small fraction of the total since the maximum value of the plotted quantity is around $10^9$ for all cases we have considered, and justifies ignoring it in the dynamics of expansion.\\

We will now recapitulate some of the important results we have obtained for the various models investigated in this paper and make some additional comments. Consider first the constant-$M$ model:

\begin{enumerate}

\item The only parameter in the model is $M$ (always expressed in units of the inflaton mass, $m \simeq 10^{13}$ GeV).

\item The spectra are slightly red for $M<1$, slightly blue for $1 < M \alt 2$, and increasingly peaked around $10^{-2}\alt k \alt 1$ for $2 \alt M$.  There is a sharp drop in the spectra for $k>1$.

\item The spectra of produced particles decreases very rapidly with $M$ for $M \agt 1$; the decrease in the spectra with $M$ for wave-numbers in the range $10^{-2} \alt k \alt 10^0$ is still substantial, but less pronounced than in the infrared. 

\item If the produced particle is stable, $\Omega h^2 = 0.12$ is obtained for $M\simeq 3.3$, provided $T_{\text{RH}}=10^9$ GeV. (This can be inferred from the $\epsilon \rightarrow \infty$ results in \figref{relic}.)

\end{enumerate}

For the potentially-coupled irrupton model, some of our important results are as follows:

\begin{enumerate}

\item There are two parameters in this model, $M_g=g M_{\text{Pl}}$, and $\nu_*$.  The late-time value of the irrupton mass is $M_\text{eff}^\infty = M_g\nu_*$.  Our numerical investigation fixed $\nu_*=0.8$ which corresponds to four $e$-folds before the end of inflation.

\item For $M_g<3$ the spectrum is similar in shape to the constant-$M$ model, owing to the existence in both cases of a broad tachyonic phase whose duration depends on $k$. That is, at small $M_g$ this model has continuous particle production over extended durations rather than a localized irruption.

\item The spectrum for the potentially-coupled case is much larger than the corresponding spectrum for the constant-mass model with $M_g=M$.

\item For still larger $M_g$, say $M_g \agt 5$ or so (see \figref{potential}), the spectrum becomes increasingly peaked around $10^{-2}\alt k \alt 1$.

\item The irruption production mechanism is too efficient at producing particles to allow for their interpretation as a possible superheavy DM candidate unless the reheat temperature is fairly low: the total number of particles produced over-saturates the Planck result for $\Omega h^2$ \cite{Ade:2013zuv} by a factor of at least $10^4 \times \lb( T_{\text{RH}} / 10^9\, \text{GeV} \rb) \times \lb( m / 10^{13}\, \text{GeV} \rb)^2$ for all choices of $M_g$ which we have studied (see \figref{relic}). This conclusion is, however, rather sensitive to the value of $\nu_*$: since $\Omega h^2 \sim \nu_*\, a_*^3$ [see Eq.\ \eqref{eq:np_analytics}], had we taken $\nu_* \gtrsim 1.1$ (greater than eight $e$-folds before the end of inflation), the minimum value of  $\lb( \Omega h^2 \rb) \times \lb(\Omega_{\text{DM}} h^2\rb)^{-1}_{\text{Planck}} \times (T_{\text{RH}} / 10^9 \text{GeV} )^{-1} \times ( m / 10^{13} \text{GeV} )^{-2} $ would drop below 1, and it would be possible to attain the right relic abundance even for $T_{\text{RH}} \sim 10^9$GeV.

\item For fixed $M_g$, our understanding of how the behavior of $M_{\text{eff}}$ impacts the spectra leads us to conclude on general grounds that as $\nu_*$ is increased, the $\nck$ spectrum will have the same general shape but will (a) shift to the infrared because smaller values of $k$ are needed to allow for strongly non-adiabatic / tachyonic behavior if $M_{\text{eff}}$ has its zero earlier during inflation when $a_*$ is smaller, (b) broaden on the low-$k$ side due to an extension of the tachyonic phase to earlier times, and (c) decrease in amplitude owing to the greater dilution of an NR species if it is created earlier during inflation. The dilution effect is always dominant, leading to a exponential suppression of the abundance of particles as measured by $\Omega h^2$ after the end of inflation, as $\nu_*$ is increased linearly. These points are all explicitly borne out by our large-$M_g$ analytical expressions Eqs.\ \eqref{eq:approx_analytics} and \eqref{eq:np_analytics}, as discussed further in Appendix \ref{app:potential_analytics}. 

\item Our analytical expressions in Appendix \ref{app:potential_analytics} also indicate that at large $M_g$, this model produces \emph{more} particles with increasing $M_g$: $\np \propto M_g^{3/2}$ and $\Omega h^2 \propto M_g^{5/2}$; our numerical results agree with these scalings.

\end{enumerate}

Finally, we turn to some additional final comments on the kinetically-coupled irrupton model: 

\begin{enumerate}

\item There are three parameters in this model: $M$, $\epsilon$, and $\nu_*$.  The late-time value of the irrupton mass is $M_\text{eff}^\infty = Me^{\nu_*^2/2\epsilon^2}$.  For $\epsilon \rightarrow \infty$ the constant-$M$ model is recovered.  $M_\text{eff}^\infty$ is exponentially sensitive to $\nu_*/\epsilon$.

\item The simple large-$M$ scaling behavior with changing $\epsilon$ evident in \figref{relic} traces its origin to the insensitivity of the spectra, in the regions where they contribute dominantly to the particle number integral in Eq.\ \eqref{eq:number}, to changing values of $\epsilon$ (at fixed $M$) which we also noted with regard to our discussion of Fig.\ \ref{fig:M} in \secref{results}: for example, $n^{\text{p}}$ increases by a factor of only about $2$ between $\epsilon = 0.80$ and $\epsilon = 0.25$ for $M=4$, yet $\Omega_{\text{DM}} h^2$ increases by a factor 2 orders of magnitude larger than that. That is, the clean scaling behavior of $\Omega h^2$ with varying $\epsilon$ at fixed $M$ seen in \figref{relic} at large $M$ arises predominantly through the \emph{parametric} dependence on $\epsilon$ of the horizontally- and vertically-plotted quantities [both proportional to $e^{\nu_*^2/2\epsilon^2}$; see Eqs.\ \eqref{eq:relic_abundance} and \eqref{eq:Meff_inf}] rather than through the impact on the particle spectra of varying $\epsilon$ \emph{per s\'e}. 

\item The simple large-$M$ scaling with increasing $M$ arises jointly from parametric dependence causing $M_{\text{eff}}^\infty$ to increase linearly, while the spectra themselves undergo an exponentially fast decrease, fairly uniform over a fairly wide range of $M$, in normalization causing the relic abundance (proportional to $n^{\text{p}})$ to drop, notwithstanding its parametric scaling proportional to $M$. The integrated spectrum itself can be very well fit over the range from $M=4$ to 8 by a function of the form $n^{\text{p}} \sim e^{-a M + b M^2 }$ where $a$ and $b$ are some positive real numbers and $b\ll a$.

\item The more complex behavior at smaller $M$ is due to the more non-trivial dependence of the particle spectra in the vicinity of their maxima (e.g., Fig.\ \ref{fig:M}) on $M$ and $\epsilon$, in addition to the parametric dependence just discussed. As an example, consider the case of varying $\epsilon$ at small $M$. Two effects compete against one another: with decreasing $\epsilon$ the normalization of the particle spectrum drops at fixed small $M$ (e.g., the upper plot of \figref{M}), but the late-time effective mass rises proportional to $e^{\nu_*^2/2\epsilon^2}$. Since $\Omega h^2$ is proportional to $ M_{\text{eff}}^\infty\, n^{\text{p}}$, this gives rise to the turning-point behavior around $\epsilon = 0.4$ seen in, e.g., the $M = 1$ results because at small $\epsilon$, the latter effect wins whereas at larger $\epsilon$, the former effect does.

\item Our general understanding of the behavior of $M_{\text{eff}}$ leads us to conclude that if a tachyonic phase is already present, further increasing $\nu_*$ will increase the duration of that phase, and will shift it to slightly earlier times. While the former effect would argue for an increase in the $\nck$ spectra, the latter effect leads to a greater dilution of the NR species as it is produced earlier. Simulations indicate that the latter effect wins, causing the amplitude of the $\nck$ spectra to decrease with increasing $\nu_*$. The extension of the tachyonic phase to earlier times causes the spectrum to broaden on the low-$k$ side, while its truncation at later times causes the UV drop-off to happen for slightly smaller $k$. In other words, with increasing $\nu_*$, the $\nck$ spectrum gets smaller at fixed $k$, and broadens as a function of $k$, and moves to the IR. \label{pt:tachy}

\item In contrast, if no tachyonic phase exists (e.g., at large $M$), increasing $\nu_*$ can cause more complex changes to the spectrum. This is perhaps easiest discussed by way of an example; for instance, consider the case of $M=4$. As $\nu_*$ is increased from 0.4 to 0.8, the $\nck$ spectrum \emph{increases} in amplitude, broadens and the peak moves to the IR due to the fact that $\text{min}\!\lb\{ \omega_k^2/a^2 \rb\}$ decreases and is attained at slightly earlier times (the increase in the production from the former effect swamps the dilution implied by the latter effect). As $\nu_*$ is increased further from 0.8 to 1.2, the spectrum continues to move to the IR, but its amplitude drops as the dilution effect becomes the dominant. Eventually for large enough $\nu_*$, a tachyonic phase develops, and point (\ref{pt:tachy}) begins to apply.

\end{enumerate}

All the models we consider are capable of producing massive particles during inflation.  For the constant mass model the efficacy of particle creation drops precipitously for $M \agt H$.  Also, for constant mass models with mass light enough for appreciable particle production, the spectrum of produced particles tends to be flat.  We considered two models with varying mass: a ``potentially-coupled'' model and a ``kinetically-coupled'' model.  Both of these models are capable of producing an irruption of a particle species as the effective mass of the species vanishes or becomes small compared to $H$.  Both of these models are able to produce particles with mass (after inflation) much larger than $H$.  They are also capable of producing highly-peaked spectra. 

While the models we considered do not exhaust the space of models with varying particle mass due to the coupling of the particle to the inflaton, they do encompass a remarkable range of phenomena and results.  

While a complete study of the applications of massive particle production is outside of the scope of this work, we conclude by remarking on some possible implications and applications.

\begin{enumerate}
\item \emph{Backreaction on the Inflaton Field:} The original motivation for studying irrupton of massive particles during inflation was the backreaction of particle production on the inflaton field, which can lead to features in the scalar density spectrum \cite{Chung:1999ve}. It was later demonstrated that there are additional effects on the density spectrum due to the scattering of the produced particles with the inflaton field \cite{Barnaby:2009mc,Barnaby:2010sq}.  All of these studies assumed a potentially-coupled irrupton.  

Both of the aforementioned effects are sensitive to the magnitude and the duration of particle irruption.  As we have shown, a wide range of possibilities for magnitude and duration are possible just in the two irrupton models we have considered.  We have shown that for a same mass ($M_\text{eff}^\infty$) particle species, the spectrum and amplitude of produced particles may differ greatly between the potentially-coupled and the kinetically-coupled models.  This will have a large effect on the calculation of the effect of the backreaction on the inflaton field as well as irrupton--inflaton scattering after production.

\item \emph{Superheavy Dark Matter (``WIMPzilla'') production:}
The concept of dark matter produced by particle creation during inflation was proposed within the framework of the constant-mass model \cite{Chung:1998zb,Kuzmin:1998kk}.  The idea is that the particle coupling to the inflaton is stable and is produced gravitationally during inflation in the correct abundance to be the relic dark matter.  As we can see from \figref{relic}, the correct relic mass density in this model is obtained if $M \simeq 3.3 m$.  This is in broad agreement with previous analyses.

We can now extend this idea to models with species irruption.  The results for a potentially-coupled irrupton model is also shown in \figref{relic}.  The interesting result is that for fairly small values of $\nu_*$ there is \emph{no} value of the model parameters that do not overproduce the particle, provided the reheat temperature is not quite low.  For this model, $\Omega h^2$ actually has a minimum around $M^\infty_{\text{eff}} \approx 30$ as a result of the fact that $n^{\text{p}}$ does not continue to monotonically decrease with increasing $M_g$ when the latter is large (Fig.\ \ref{fig:potential}): once $\partial n^{\text{p}}/ \partial M_g \sim M_g^{-1}$, the relic density (proportional to $M_g n^{\text{p}}$) will go through a minimum. Once $\nu_*\gtrsim 1.1$, however, the correct relic abundance can be obtained (even for $T_{\text{RH}} \approx 10^9$GeV) due to the additional dilution of the NR species owing to it being produced earlier during inflation.

For a kinetically-coupled irrupton, $M_\text{eff}^\infty$ can be (many) orders of magnitude larger than $3.3m$ and still give rise to the right abundance of particles (see \figref{relic}); in fact, it could na\"ively be as large as the Planck mass (i.e., it can be a Planckon).  We will remark briefly on the possibility of super-Planckian particle production below.

\item \emph{Isocurvature modes:} Isocurvature modes are produced in the WIMPzilla scenario \cite{Chung:2004nh,Chung:2011xd} because the dark matter is produced by the dynamics of the coupled irrupton-inflaton system and the curvature perturbations are produced by the inflaton dynamics alone.  

Again, the calculation of amplitude and spectrum of the isocurvature component will differ in the constant-mass case and either of the kinetically- or potentially- coupled irrupton models. This is particularly important as the limits to the isocurvature model become more stringent \cite{Ade:2013uln}.

\item \emph{Non-Gaussian Features:}  Non-Gaussian features in the scalar perturbation spectrum will occur in all of the models we study.  Isocurvature perturbations were considered in the WIMPzilla scenario in Ref.\ \cite{Chung:2011xd} assuming a constant-$M$ model.  If there is an irruption of massive particles during inflation there is another source of non-Gaussianity called infrared cascading in \citet{Barnaby:2009mc}  (see Ref.\ \cite{Barnaby:2010sq} for a review).  The model used in that study is the potentially-coupled model of this paper.  The calculations for both of these effects (WIMPzilla and infrared cascading) will be modified in the kinetically-coupled scenario. 

\item \emph{Planck-mass particle production:}  As we noted above, the efficacy of the either the potentially- or kinetically- coupled irrupton production mechanism with regard to large-mass particles is expected to extend beyond the regime we have explicitly investigated numerically. This raises the prospect that by dialing the parameters $M, \epsilon$ and $\nu_*$ (kinetically-coupled case) or $M_g$ and $\nu_*$ (potentially-coupled case), we could na\"ively raise the late-time (as well as early-time) effective mass of the irruptons to $M_{\text{eff}}^\infty \gtrsim M_{\text{Pl}}$ while still maintaining a non-negligible abundance.  

This raises the possibility that inflation could be sensitive to particles\footnote{\citet{Dvali:2010bf} argue that these states are not properly quantum particles, but black holes.} with mass larger than the Planck mass.  In this scenario \emph{except} for an extremely narrow region near $\phi=\phi_*$ the Planckian state can be integrated out of the effective field theory describing inflation.  But it is exactly near $\phi=\phi_*$ that the particles will be produced, and when they are produced they are light.  Only after the inflaton continues to evolve past $\phi=\phi_*$ will the particle regain its proper Planckian mass.  Presumably, then the ``particle'' becomes a classical black hole, and description of its dynamics in terms of single-particle excitations of a fundamental scalar field is then inappropriate. Furthermore, with Planck-mass particles present, the local space-time will be subject to significant back-reaction, and if a black hole forms, it would look locally like Schwarzschild, rather than homogeneous quasi--de Sitter, so it is clear that our results in this regime are on shaky footing. At some point our simple model must break down and it is therefore unclear whether it does actually allow for the production of trans/super-Planckian particles. This is an interesting open question.

\end{enumerate}

\clearpage
\appendix
\onecolumngrid
\section{Late-time asymptotics \label{app:late_time_asymptotics}}

It is instructive to develop a further analytical understanding of the late-time behavior of $\bks$, building from our discussion in Sec.\ \ref{sec:MD}. We examine here the expression for the first-order iteration, $\beta^{(1)}_k(t)$, for $t \gg t_{\text{ref}}$.  Starting with Eq.\ \eqref{eq:beta_late} for $n=1$, and using $\alpha_k^{(0)}(t)=1$, we have
\begin{align}
\beta^{(1)}_k(t) = \beta^{(1)}_{k,\text{ref}}+\frac{1}{2} e^{-2i\Phi_\text{ref}} \int_{t_\text{ref}}^t \left[H(t_1)-2\delta\, \dot{\nu}(t_1)\right]e^{-2iM_\infty(t_1-t_\text{ref})}\ dt_1.
\label{eq:beta1late}
\end{align}
Applying Eq.\ \eqref{eq:nudot}, the integral in Eq.\ \eqref{eq:beta1late} can be written in terms of exponential integral functions $\text{Ei}(y)$.  Two standard integrals will enter:
\begin{align}
\mathcal{I}_1(y;\tilde{y},\xi) &\equiv \int_0^y \frac{e^{-i\xi (y_1+\tilde{y})}}{y_1+\tilde{y}} dy_1 = \text{Ei}( -i \xi ( y + \tilde{y} ) ) - \text{Ei}( -i \xi \tilde{y} ),\\
\mathcal{I}_2(y;\tilde{y},\xi) &\equiv \int_0^y \frac{e^{-i\xi (y_1+\tilde{y})}}{(y_1+\tilde{y})^2} dy_1 = e^{-i\xi \tilde{y}} \left[ \frac{1}{\tilde{y}} - \frac{e^{-i\xi y}}{y+\tilde{y}} \right] -i\xi \left[ \text{Ei}( -i \xi ( y + \tilde{y} ) ) - \text{Ei}( -i \xi \tilde{y} ) \right] \ .
\end{align}
Here, we will have either $\xi = 2M_\infty$ or $\xi = 2M_\infty\pm1$, and $y \equiv t-t_\text{ref}$, $\tilde{y} \equiv t_\text{ref}-\tau$, and $y+\tilde{y}=t-\tau$  (all are positive, provided $M_\infty> 1/2$).  Finally, we can write the desired late-time expression for $\beta_k(t)$:
\begin{align}
\beta^{(1)}_k(t) &\approx \beta^{(1)}_{\text{ref}} + \frac{1}{2} e^{-2i(\Phi_{\text{ref}}-M_\infty \tilde{y})} 
\left\{ \frac{2}{3}  \mathcal{I}_1(y;\tilde{y},2M_\infty)  \right. \nonumber \\
&  - \delta \Big[ (A+iB)\, \mathcal{I}_1(y;\tilde{y},2M_\infty-1) 
+(A-iB)\, \mathcal{I}_1(y;\tilde{y},2M_\infty+1) \Big] \nonumber \\
& \left. \phantom{\frac{2}{3}} -i \delta \Big[ (A+iB)\, \mathcal{I}_2(y;\tilde{y},2M_\infty-1) 
-(A-iB)\, \mathcal{I}_2(y;\tilde{y},2M_\infty+1) \Big] \right\} \ .
\label{eq:phew}
\end{align}

We are interested in the late-time asymptotic limit of Eq.\ \eqref{eq:phew}.  Since the exponential integral function has a branch cut one must be careful when making asymptotic expansions:
\begin{align}
\text{Ei}(\pm i a) &\rightarrow \pm i \pi  + e^{\pm i a} \lb[ \mp \frac{i}{a} - \frac{1}{a^2} + \cdots \rb] \quad \text{as} \quad  a\rightarrow +\infty \quad ( a\in \mathbb{R} ).
\end{align}
Thus, the late-time asymptotic expansions for the two standard integrals are
\begin{align}
\mathcal{I}_1(y\rightarrow \infty;\tilde{y},\xi)  &\sim  -\lb[ i\pi + \text{Ei}( -i \xi \tilde{y} )\rb] + e^{-i\xi (y+\tilde{y} ) } \lb[  \frac{i}{\xi y} - \frac{1+i\xi \tilde{y}}{\xi^2 y^2} + \cdots  \rb]\\
\mathcal{I}_2(y\rightarrow \infty;\tilde{y},\xi) &\sim \frac{e^{-i\xi \tilde{y}}}{\tilde{y}} + i\xi \left[ i\pi + \text{Ei}( -i \xi \tilde{y} ) \right] + \frac{e^{-i\xi(y+\tilde{y})}}{y}\lb[ \frac{i}{\xi y} - 2\frac{1+i\xi \tilde{y}}{\xi^2 y^2} + \cdots \rb] \ .
\end{align}
Finally we may expand this late-time asymptotic solution in the large-mass limit ($\xi\rightarrow\infty$), yielding
\begin{align}
\mathcal{I}_1(y\rightarrow\infty;\tilde{y},\xi \rightarrow \infty)  &\sim - \frac{i}{\xi} \lb[ e^{-i\xi \tilde{y}} \frac{1}{\tilde{y}} - e^{-i\xi(y+\tilde{y})} \lb( \frac{1}{y} - \frac{\tilde{y}}{y^2} \rb) + \cdots \rb]  + \frac{1}{\xi^2} \lb[ e^{-i\xi\tilde{y}} \frac{1}{\tilde{y}^2} - e^{-i\xi(y+\tilde{y})} \frac{1}{y^2} +\cdots \rb] + \cdots\\
\mathcal{I}_2(y\rightarrow \infty;\tilde{y},\xi \rightarrow \infty) &\sim  -\frac{i}{\xi} \lb[ e^{-i\xi \tilde{y}} \frac{1}{\tilde{y}^2} - e^{-i\xi (y+\tilde{y}) } \lb( \frac{1}{y^2} - \frac{2\tilde{y}}{y^3} \rb) + \cdots \rb] + \frac{2}{\xi^2} \lb[ e^{-i\xi\tilde{y}} \frac{1}{\tilde{y}^3} - e^{-i\xi(y+\tilde{y})}\frac{1}{y^3} + \cdots \rb]+\cdots 
\end{align}

After a bit of manipulation, the late-time, large-mass asymptotic value of $\beta_k^{(1)}$ can be written as 
\begin{align}
\beta^{(1)}_k(t\rightarrow \infty) &\approx \beta^{(1)}_{k,\text{ref}} 
-i\, \frac {e^{-2i\Phi_\text{ref}}} {6M_\infty} \left\{ \frac{1}{t_\text{ref}-t_0} 
 \lb[ 1 - 3\delta \,  (t_{\text{ref}}-t_0 )\, \dot{\nu}_{\text{ref}}  \rb] \right. \nonumber \\
& \left. + \frac{e^{-2iM_\infty(t-t_\text{ref})}}{t-t_\text{ref}}
\bigg[ 1 - 3\delta\bigg( \left( \nu_{\text{ref}} + ( t_{\text{ref}} - t_0 ) \dot{\nu}_\text{ref} \right) \cos( t-t_\text{ref} ) - \nu_\text{ref}  \left( t_\text{ref} - t_0 \right) \sin(t-t_\text{ref}) \bigg) \bigg] 
\right\} \ .
\label{eq:LTLM}
\end{align} \newpage
\twocolumngrid

Armed with this expression, we can immediately see the qualitative features of the late-time, large-mass solution:
\begin{enumerate}
\item A non-zero (in general) late-time value of 
\begin{align} 
& |\beta^{(1)}_k(t\rightarrow\infty)|^2 \nonumber \\
&= \lb| \beta^{(1)}_{k,\text{ref}} - i\, \frac{e^{-2i\Phi_\text{ref}}}{6M_\infty} \frac{1}{t_\text{ref}-t_0} \lb[ 1 - 3\delta \,  (t_{\text{ref}}-t_0 )\, \dot{\nu}_{\text{ref}}  \rb]   \rb|^2 
	\end{align}
arising from the constant term squared.
\item A fast oscillation at frequency $2M_\infty$, whose amplitude is modulated at frequency $m$, which damps out as $1/t$. This arises from the cross-term between the constant term and the damped term. Schematically, this contribution takes the form $\dfrac{1}{t} \cos(2M_\infty t + \phi ) \Big[ 1 + \zeta \cos(m t + \phi') \Big]$. Although we have not displayed it here, if we expanded $\beta_k^{(1)}$ itself up to $1/t^2$, there would also be a similar term damped at $1/t^2$, but this will be subdominant.
\item A slow oscillation at frequency $m$ on top of a constant $1/t^2$ contribution. This arises from the damped term squared. Schematically, this contribution takes the form $\dfrac{1}{t^2}  \Big[ 1 + \zeta' \cos(mt + \phi'') \Big]^2$.
\end{enumerate}
We see all these behaviors in the numerical solutions; in particular, we have performed Fourier analysis on selected late-time numerical solutions and found strong peaks in the Fourier power spectrum at all the expected frequencies (including sidebands generated by the modulations).

\section{Analytics for the potentially-coupled case \label{app:potential_analytics} }
This appendix is based on ideas and methods to be found in Section VII B of \citet{Kofman:1997yn}.

We consider the potentially-coupled case at large $M_g$ (larger than any other scale in the problem, except possibly $k$). In this case, the duration of production is extremely short around $t=t_*$, and we need only consider the solution of the mode equation in this short interval, taking $|\beta_k|^2=0$ identically at some short time just before $t=t_*$, and extracting the asymptotic value of $\bks$ shortly after $t=t_*$ (``short'' will be made precise below). As a result of this observation, we may Taylor expand $a(t), H(t),$ etc.\ in the mode equation around the point $t=t_*$ and will keep only terms up to $t^2$. 

Rather than solve the mode equation Eq.\ \eqref{eq:timemode} for $\chi_k$, we solve the equation for $f_k(t) \equiv a^{1/2}\chi_k(t)$ \cite{Barnaby:2010sq}:
\begin{align}
\ddot{f}_k + \lb[ M_g^2 (\nu - \nu_*)^2  + \frac{k^2}{a^2} - \frac{9}{4} H^2 - \frac{3}{2} \dot{H} \rb] f_k = 0,
\end{align}
where we used the fact that $a''/a^3=2H^2+\dot{H}$. 

We may now Taylor expand the relevant functions to second-order in powers of $t$, for the present purposes re-zeroing $t$ at $\nu=\nu_*$:
\begin{align}
M_g^2 (\nu - \nu_*)^2 &= M_g^2 \dot{\nu}_*^2 t^2 \equiv k_\star^4 t^2 \nonumber 
\\
\frac{k^2}{a^2} &= \frac{k^2}{a_*^2} \lb[ 1 - 2H_* t + \lb( 2H_*^2 - \dot{H}_* \rb) t^2 \rb] \nonumber \\
H^2 &= H_*^2 + 2H_* \dot{H}_* + ( \dot{H}_*^2 + H_* \ddot{H}_* ) t^2 \nonumber \\
\dot{H} &= \dot{H}_* + \ddot{H}_* t + \frac{1}{2}\! \lb.\frac{d^3H}{dt^3}\rb|_* t^2 \ ,
\end{align}
where we defined $k_\star \equiv \sqrt{ M_g | \dot{\nu}_* | }$. To evaluate the higher derivatives here in terms of $\nu$ and $\dot{\nu}$, one can exploit Eq.\ \eqref{eq:Hubble_nu}, $\dot{H}= - 4\pi\dot{\nu}^2$ and $\ddot{\nu}=-3H\dot{\nu}-\nu$, all of which are exact relations when the back-reaction is ignored and the irrupton energy density is assumed to be a very small fraction of the inflaton energy density.  We thus have
\begin{align}
\ddot{f}_k + \lb[ p t^2+ qt+r  \rb]f_k = 0,
\end{align}
where we have defined
\begin{align}
p &= k_\star^4 + \frac{k^2}{a_*^2} \lb[ 2H_* - \dot{H}_* \rb] - \frac{9}{4} \lb( \dot{H}_*^2 + H_* \ddot{H}_* \rb) - \frac{3}{2} \lb.\frac{d^3H}{dt^3}\rb|_* \nonumber \\
q &= - 2H_* \frac{k^2}{a_*^2} - \frac{9}{2} H_*\dot{H}_* - \frac{3}{2} \ddot{H}_* \nonumber \\
r &= \frac{k^2}{a_*^2} - \frac{9}{4} H_*^2 - \frac{3}{2} \dot{H}_*.
\end{align}
Defining
\begin{align}
z &\equiv \sqrt{ 2 } p^{1/4}  \lb( t + \frac{q}{2p} \rb) \nonumber \\
c &\equiv \frac{1}{2\sqrt{p}} \lb( \frac{q^2}{4p} - r \rb),
\end{align}
the equation for $f_k$ can be written as
\begin{align}
\frac{d^2f_k}{dz^2} + \lb( \frac{1}{4} z^2 - c \rb) f_k = 0.
\end{align}
The solutions to this equation can be written in terms of confluent hypergeometric functions ${}_1\hspace{-0.2mm}F_{1}$ (see \S 9.2 of Ref.\ \cite{gradshteyn2007} or Ref.\ \cite{wolfram}):
\begin{align}
f_k &= e^{-iz^2/4} \Big[ A\ {}_1\hspace{-0.2mm}F_1\!\lb( \tfrac{1}{4} - \tfrac{i}{2}c , \tfrac{1}{2} , \tfrac{i}{2}z^2 \rb)  \nonumber \\&\qquad\qquad \quad + B e^{i\pi/4}\,z\ {}_1\hspace{-0.2mm}F_1\!\lb( \tfrac{3}{4} - \tfrac{i}{2}c , \tfrac{3}{2} , \tfrac{i}{2} z^2\rb) \Big],
\end{align}
where $A$ and $B$ are integration constants to be chosen to specify the adiabatic in-vacuum. 

Recall that $f_k = a^{1/2} \chi_k$ and the in-vacuum solution for $\chi_k$ must reduce to $\chi_k \rightarrow (1/2\omega_k)^{1/2} \exp\left(-i\int^t dt' \ \omega_k/a \right)$ as $t\rightarrow -\infty$. Therefore, we must have $f_k \rightarrow (a/2\omega_k)^{1/2} \exp\left(-i\int^t dt' \ \omega_k/a \right)$ as $t\rightarrow -\infty$. We now Taylor expand $\omega_k^2$ and look sufficiently far from $t=0$: $|t| \gg k_*^{-1}$. Assuming that the mass term is the largest contribution to $\omega_k^2/a^2$, the latter is dominated by the $t^2$ term, and up to small corrections like $\dot{H}_* / k_\star^4$, etc., we then have that $\omega_k / a \approx \sqrt{p}\ t \approx  p^{1/4}z/\sqrt{2}$. Therefore, the in-vacuum asymptotic early-time form of $f_k$ must be $f_k \rightarrow \exp(+iz^2/4)/\left(2^{1/4} p^{1/8}\sqrt{-z}\right)$ as $z\rightarrow-\infty$, where we have ignored the presence of a possible irrelevant overall phase and have dropped small terms. Note the opposite sign to the na\"ive expectation appears in the exponent since the $t<0$ form of $\omega_k$ must be used.

By demanding this early-time asymptotic form, we recover the values of $A$ and $B$ for the correctly normalized modes which give an in-vacuum; although these expressions are not particularly enlightening, we present them here for completeness:
\begin{align}
A &= -\frac{(-1)^{7/8} 2^{i c/2}}{4\pi  p^{1/8}} e^{-3 \pi  c / 4} \left(1 + e^{2 \pi  c}\right) \nonumber \\ &\qquad \times \Gamma \left(\tfrac{i
   c}{2}+\tfrac{1}{4}\right) \Gamma \left(\tfrac{1}{2}-i c\right), \\[2ex]
B &=-\frac{(-1)^{3/8} 2^{ ic/2}}{2^{3/2}\pi  p^{1/8}} e^{- 3 \pi  c / 4} \left(1 + e^{2 \pi  c}\right) \nonumber \\ &\qquad \times \Gamma \left(\tfrac{i
   c}{2}+\tfrac{3}{4}\right) \Gamma \left(\tfrac{1}{2}-i c\right).
\end{align}

To extract $\alpha_k$ and $\beta_k$ we simply look at the late-time behavior of $f(z)$ as $z\rightarrow\infty$: $\alpha_k$ is the coefficient of the term $\exp(-iz^2/4)/\left(2^{1/4}\, p^{1/8}\, \sqrt{z}\right)$ and $\beta_k$ is the coefficient of the term $\exp(+iz^2/4)/\left(2^{1/4}\, p^{1/8}\, \sqrt{z}\right)$. The results are
\begin{align}
|\alpha_k|^2 &= 1 + e^{2\pi c} \nonumber \\
|\beta_k|^2 &= e^{2 \pi  c} = \exp\lb[ \frac{\pi}{\sqrt{p}} \lb( \frac{q^2}{4p} - r \rb) \rb], \label{eq:spectrum_analytics}
\end{align}
which clearly satisfy the Wronskian condition $|\alpha_k|^2 - |\beta_k|^2 = 1$.

The (red) dashed line in the lower plot of Fig.\ \ref{fig:relic} is obtained via numerical integration of Eq.\ \eqref{eq:spectrum_analytics} using Eq.\ \eqref{eq:relic_abundance}: $\Omega h^2 \times \lb( \Omega h^2 \rb)_{\text{Planck}}^{-1} \times ( T_{\text{RH}} / 10^9 \text{GeV} )^{-1} \times ( m / 10^{13} \text{GeV} )^{-2} \approx \lb( 2.7043\times10^6 \rb) \times  M_g \nu_* \times \int_0^\infty k^2 |\beta_k|^2\, dk $. The relevant parameters in terms of which $p$, $q$, and $r$ can be evaluated must be taken from our numerical background solutions: when $\nu = \nu_* = 0.8$, we have $H_* = 1.6697,\ \dot{\nu}_* = - 0.1599$ and $a_* = 0.0160$. 

At very large $M_g$ further simplifications are possible: 
\begin{align}
p & \approx k_\star^4 = M_g^2 \dot{\nu}_*^2 \nonumber \\
c & \approx -\frac{r}{2\sqrt{p}} \approx \frac{1}{2M_g|\dot{\nu}_*|} \lb[ \frac{9}{4} H_*^2 + \frac{3}{2} \dot{H}_* - \frac{k^2}{a_*^2} \rb] ,
\end{align}
so that using $\dot{H}_* = -4\pi\dot{\nu}_*^2$,
\begin{align}
|\beta_k|^2 &\approx \exp\lb[ \frac{\pi}{M_g|\dot{\nu}_*|} \lb( \frac{9}{4} H_*^2 + \frac{3}{2} \dot{H}_* - \frac{k^2}{a_*^2}  \rb) \rb] \nonumber \\
&\approx \exp\lb[ \frac{\pi}{M_g|\dot{\nu}_*|} \lb( \frac{9}{4} H_*^2 - 6\pi \dot{\nu}_*^2 - \frac{k^2}{a_*^2}  \rb) \rb]. \label{eq:approx_analytics}
\end{align}
Expressing $H^2$ in terms of $V(\phi)$ and $\dot{\phi}$, and using $\dot{H} = - (4\pi/M_{\text{Pl}}^2) \dot{\phi}^2$, we have
\begin{equation}
\frac{9}{4} H_*^2 + \frac{3}{2}\dot{H}_* = \frac{9}{4}H_*^2 \lb[ \frac{V(\phi_*) - \tfrac{1}{2}\dot{\phi}_*^2}{V(\phi_*) + \tfrac{1}{2} \dot{\phi}_*^2} \rb] = -\frac{9}{4}H_*^2 w_* \ ,
\end{equation} 
which is always positive during inflation (the equation of state at any point $\nu=\nu_*$ during inflation satisfies $w_*<-1/3$). Therefore, it is clear that $|\beta_k|^2$ is flat for small $k$ and approaches unity from above for large $M_g$. It also drops rapidly once the $k^2$ term drives the exponent negative, but at fixed large $k$ approaches unity from below as $M_g$ is further increased. Granted, this expression is invalid when $k$ dominates $\omega_k$, but since the rapid drop-off sets in by this point for large enough $M_g$, we can simply neglect this regime. 

Furthermore, if we confine ourselves to consideration of values of $\nu_*$ such that slow-roll inflation is still a very good approximation around $t=t_*$, it follows that $\dot{\nu}_*$ is very nearly independent of $\nu_*$, so $H_*^2$ as given by Eq.\ \eqref{eq:Hubble_nu} is proportional to $\nu_*^2$. We also have $a_* \propto e^{-2\pi\nu_*^2}$. Therefore, the deep-IR value $|\beta_{k=0}|^2 \propto \exp\lb( 3\pi^2 \nu_*^2 / M_g |\dot{\nu}_*| \rb)$ increases exponentially quickly as $\nu_*$ is increased linearly. However, the value $k= k_1$ required to cause a one--$e$-fold drop-off in $\bks$ from this IR value decreases exponentially quickly roughly as $k_1\propto a_*H_* \propto \nu_*e^{-2\pi \nu_*^2}$. The net result is that as $\nu_*$ is increased linearly, the $\nck$ spectrum broadens on the low-$k$ side (since $|\beta_{k=0}|^2$ is larger), yet peaks at a much smaller value of $k$ and as a result has a much smaller maximum value.

Integrating the approximate very-large-$M_g$ spectrum \eqref{eq:approx_analytics} over all $k$ yields
\begin{align}
& \frac{1}{2\pi^2} \int_0^\infty k^2 |\beta_k|^2\, dk \nonumber \\ 
&= \frac{1}{2\pi^2} \int_{-\infty}^\infty k^3 |\beta_k|^2\, d\ln k \nonumber \\ & =   \frac{a_*^3}{8\pi^3} \lb( M_g |\dot{\nu}_*| \rb)^{3/2} \exp\lb[  \frac{3\pi}{4} \frac{ 3H_*^2 - 8\pi \dot{\nu}_*^2 }{ M_g |\dot{\nu}_*| } \rb] \nonumber \\ 
&\approx\lb( \frac{a_*\sqrt{|\dot{\nu}_*|}}{2\pi} \rb)^3 M_g ^{3/2}.\label{eq:np_analytics}
\end{align}
This shows us that in the large $M_g$ limit, the total number of produced particles goes like $n^p \propto M_g^{3/2}$, and $\Omega h^2 \propto M_g^{5/2}$, both of which \emph{increase} with increasing $M_g$. Also, as $\nu_*$ increases, both $\np$ and $\Omega h^2$ (measured at the same fixed time $t$ in the MD era following the end of inflation) drop exponentially quickly as $a_*^{+3}$, as expected for the dilution of an NR species being produced earlier during inflation.


\section*{Acknowledgements}
We would like to thank Jennifer Lin for discussions in the early stage of this project. This work was supported in part by the Kavli Institute for Cosmological Physics at the University of Chicago through grant NSF PHY-1125897 and an endowment from the Kavli Foundation and its founder Fred Kavli. The work of E.W.K.\ was supported by the Department of Energy through grant DE-FG02-13ER41958.  E.W.K.\ would like to acknowledge the hospitality of the Institute of Theoretical Physics of the University of Heidelberg. M.W.\ thanks the James Arthur Postdoctoral Fellowship at New York University for support during part of the time in which this work was completed.


\bibliographystyle{apsrev4-1}
\bibliography{arxiv_v2}

\end{document}